\documentclass[12pt,draft]{ucthesis}

\usepackage{amsthm,amssymb,makeidx,amscd}

\usepackage[all]{xy}

\makeindex

\newcommand{\A}{\mathbf{A}}
\newcommand{\Act}{\mathcal{S}}

\newcommand{\bra}[1]{\left\langle #1\right\vert}
\newcommand{\bracket}[2]{\langle{#1}\mid{#2}\rangle}
\newcommand{\C}{\mathbb{C}}

\newcommand{\Dyn}{\mathbf{D}}
\newcommand{\dd}{\mathrm{d}}

\newcommand{\dom}{\mathop{\mathrm{dom}}}
\newcommand{\E}{\mathbf{E}}
\newcommand{\e}{\mathbf{E}}

\newcommand{\from}{\colon}

\newcommand{\g}{\mathfrak{g}}

\newcommand{\gr}{\mathop{\mathrm{Gr}}}
\newcommand{\Ham}{\mathcal{H}}
\newcommand{\Hilb}{\mathbf{H}}

\newcommand{\id}{\mathbf{1}}
\newcommand{\ie}{\textit{i.e.}}
\newcommand{\Imag}{\mathrm{Im}}

\newcommand{\K}{\mathbf{K}}
\newcommand{\ket}[1]{\left\vert #1\right\rangle}
\newcommand{\Lagr}{\mathcal{L}}
\newcommand{\maps}{\colon}
\newcommand{\matElem}[3]{\bra{#1}#2\ket{#3}}
\newcommand{\N}{\mathbb{N}}
\newcommand{\Phase}{\mathbf{P}}
\newcommand{\R}{\mathbb{R}}

\newcommand{\ran}{\mathop{\mathrm{ran}}}
\newcommand{\Real}{\mathrm{Re}}

\newcommand{\suchthat}{\colon}

\newcommand{\U}{\mathrm{U}}

\newcommand{\var}{\mathrm{Var}}

\newcommand{\vol}{\mathrm{vol}}
\newcommand{\W}{\mathcal{W}}
\newcommand{\Wick}[1]{\mathopen{:}{#1}\mathclose{:}}
\newcommand{\X}{\mathbf{X}}
\newcommand{\Y}{\mathbf{Y}}

\newtheorem{definition}{Definition}
\newtheorem{theorem}{Theorem}
\newtheorem{lemma}[theorem]{Lemma}
\newtheorem{result}[theorem]{Result}
\newtheorem{proposition}[theorem]{Proposition}
\newtheorem{corollary}[theorem]{Corollary}

\begin{document}


\title{\rm Loop Quantization\\
versus\\
Fock Quantization\\ 
of $\mathrm{p}$-Form Electromagnetism\\
on Static Spacetimes\strut}

\author{Miguel Carri\'on \'Alvarez\strut}
\degreeyear{December 2004\strut}
\degree{Doctor of Philosophy\strut}
\chair{Dr. John C. Baez\strut}
\othermembers{Dr. Michel L. Lapidus\strut\\
Dr. Xiao-Song Lin\strut}

\numberofmembers{3}
\prevdegrees{B.Sc. Physics (Complutense University of Madrid, Spain) 1998\strut\\
B.Sc. Mathematics (Complutense University of Madrid, Spain) 2000\strut\\
M.Sc. Mathematics (University of California at Riverside) 2002\strut}

\field{Mathematics\strut}
\campus{Riverside\strut}

\begin{frontmatter}

\maketitle
\copyrightpage 
\approvalpage

\begin{acknowledgements}


I am indebted in one way or another to the following people. 

Pilar \'Alvarez, porque madre no hay m\'as que una; 
Pedro Carri\'on, who taught me to count (the rest just
follows); 
Coral Duro, who introduced me to the Feynman Lectures on Physics at a
tender age;
Petra Solera, who sent me to the Math Olympiad;
Luis V\'azquez, who sent me on my Erasmus exchange;
Guillermo Garc\'ia--Alcaine, who taught me quantum mechanics;
Antonio Dobado, who taught me high-energy physics;
Lee Smolin, who suggested that I study with John Baez;
Jos\'e Gaite, who supervised me on my first serious research;
Fernando Bombal, who taught me functional analysis; 
Miguel Mart\'{\i}n D\'{\i}az, who taught me probability theory;
Ignacio Sols, who tried to teach me things about algebra that I had to
rediscover on my own years later;
John Baez, the best advisor this side of the Virgo cluster; 
Fotini Markopoulou, who was interested enough in my research to invite
me to PI;
and
Barbara Helisov\'a, who is just wonderful, and wonderfully patient
too.

Without them I would never have come this far.


\end{acknowledgements}

\begin{dedication}
\null\vfil
{
\begin{center}
A mis padres,\strut\\\vspace{12pt}
Pedro Carri\'on L\'opez y Pilar \'Alvarez Ur\'{i}a.\strut\\\vspace{12pt}
que la sabr\'an apreciar\strut\\ 
en su justa medida\strut
\end{center}}
\vfil\null
\end{dedication}

\begin{abstract}

As a warmup for studying dynamics and gravitons in loop quantum
gravity, Varadajan showed that Wilson loops give operators on the Fock
space for electromagnetism in Minkowski spacetime---but only after
regularizing the loops by smearing them with a Gaussian.
Unregularized Wilson loops are too singular to give densely defined
operators.  Here we present a rigorous treatment of unsmeared Wilson
loops for vacuum electromagnetism on an arbitrary globally hyperbolic
static spacetime.  Our Wilson loops are not operators, but
``quasioperators'': sesquilinear forms on the dense subspace of Fock
space spanned by coherent states corresponding to smooth classical
solutions.  To obtain this result we begin by carefully treating
electromagnetism on globally hyperbolic static spacetimes, addressing
various issues that are usually ignored, such as the definition of
Aharonov--Bohm modes when space is noncompact.  We then use a new
construction of Fock space based on coherent states to define Wilson
loop quasioperators.  Our results also cover ``Wilson surfaces'' in
$p$-form electromagnetism.

\end{abstract}

\tableofcontents
\listoffigures

\end{frontmatter}

\chapter{Introduction}

This work is motivated by the open problem of representing gravitons
in loop quantum gravity~\cite{rovelli98}, a proposed quantum theory of
geometry and candidate for a theory of quantum gravity. The great
virtue of loop quantum gravity is that it is manifestly
background-free and diffeomorphism-invariant. Unfortunately, because
the usual construction of the graviton Fock space depends explicitly
on a background metric, it is difficult to say precisely how the
notion of \emph{graviton} arises in this formalism. At least at the
kinematical level, in loop quantum gravity states of quantum geometry
are described not in terms of gravitons but in terms of spin
networks~\cite{baez96}, which had been invented independently by
Penrose~\cite{penrose71} and can be seen as a generalization of the
Wilson loops introduced in the 1970's for the study of non-abelian
gauge theories~\cite{wilson74}. However, describing the dynamics of
quantum gravity in terms of spin networks remains a difficult open
problem.  So, we are not yet in a position to study how this dynamics
reduces to that of gravitons in some limit, as presumably it should.

As a warmup, it is natural therefore to investigate the dynamics of
Wilson loops in a gauge theory which is better understood: vacuum
electromagnetism. However, until recently we were in the embarrassing
situation of not even knowing the precise relation between the loop
representation of electromagnetism and the usual Fock
representation. Here, of course, the theory is linear and formulated
on a fixed background metric, which drastically simplifies the
situation. The technical problem is that the loop representation is
based on a diffeomorphism-invariant vacuum, while the traditional Fock
vacuum is tied to a particular background metric, which implies that
photon (Fock) states are not part of the loop state space and Wilson
loop states are not part of the Fock state space. In particular, with
respect to the Fock vacuum, the photon 2-point correlation function blows up at
short distances at such a rate that Wilson loops are not well-defined
operators on Fock space.

Varadarajan~\cite{varadarajan00,varadarajan01} tackled this problem by
``smearing'' the loop~$\gamma$ using Gaussian convolution in Minkowski
space. Varadarajan's procedure puts photons and Wilson loops in a
common framework. Our goal in the present work is to understand
electromagnetic Wilson loops without the need for smearing, and on
general static, globally hyperbolic spacetimes. A related and
important outstanding problem in loop quantum gravity is that spin
network dynamics is poorly understood, and here we tackle the
analogous problem of electromagnetic Wilson loop dynamics in the Fock
representation.

The modern view of electromagnetism is that the electromagnetic
potential~$A$ is a connection on a~$U(1)$ or~$\R$ bundle over spacetime,
and the electromagnetic field is the curvature of this connection. A
\emph{Wilson loop observable} is what mathematicians call
the \emph{holonomy} of the connection around a closed
loop. In quantum theory, observables of a physical system are
represented by operators on a Hilbert space of states of the
system. In the case of electromagnetism in Minkowski spacetime, the
state space of the electromagnetic field is the so-called Fock
space. The main problem with the Wilson loop approach to quantum gauge
field theories is that, even in the simple case of electromagnetism,
Wilson loop operators are not defined on Fock space. Because in
quantum field theory there is a correspondence between observables and
states, this means that there are also no Wilson loop states in the
Fock space of electromagnetism.

Quantum field theory on curved spacetimes is a famously problematic
subject, as it combines the difficulties of quantum field theory,
notably ultaviolet divergences, with a lack of a well-defined vacuum
state due to the lack of global symmetries in a curved spacetime. For
a free quantum field theory on a static spacetime, such as we are
studying, these problems go away as there are no divegent interactions
and there is a unique time-invariant vacuum state. Because of this,
most physicist would say that vacuum electromagnetism on a static
spacetime is well-understood. This is more or less true for scalar
fields~\cite{wald94}, but then despite it being known~\cite[\S
  4.7]{wald94} that
\begin{quote}
the requirement that the classical field equations have a well-posed
initial value formulation in curved spacetime is a highly nontrivial
restriction: the straightforward generalization to curved spacetime of
the standard spin-$s$ field equations in flat spacetime do not admit a
well posed initial value formulation for~$s>1$
\end{quote}
even researchers concerned only with electromagnetism and not with
scalar fields work on the assumption that the mathematical theorems on
scalar fields apply without modification to other fields~\cite{dimock92}.
\begin{quote}For globally hyperbolic manifolds, the usual classical
  linear field equations will have global solutions if they are
  well-behaved locally. We quote the result for scalar fields.\end{quote} 

Part of the point of this thesis is to show that things are not so
simple: there are subtleties involved due to gauge invariance and
noncompact spacetimes which interact in unexpected ways. Our first
goal is to clear this up and give a rigorous general treatment of
vacuum electromagnetism on a static, globally hyperbolic
spacetime. The subtleties arise mainly from the difference between the
usual de~Rham cohomology and a certain \emph{twisted} $L^2$ cohomology
arising from gravitational time-dilation. Indeed, in a careful
treatment the electromagnetic vector potential is not a
smooth~$1$-form modulo exact smooth~$1$-forms, but a
normalizable~$1$-form modulo exact normalizable~$1$-forms. Similarly,
the Aharonov--Bohm effect arises not from closed smooth modulo exact
smooth vector potentials, but from closed normalizable modulo exact
normalizable ones. This distinction would be inconsequential if space
were compact, but this is not believed to be the case in physically
realistic models of spacetime.

In Chapter~\ref{sec:harmonic} we present a rogues' gallery of
pathologies and counterexamples which illustrate how these subtleties
can manifest themselves as physical effects, including the photon
acquiring a mass due to the interaction of gravitational time dilation
and the asymptotic geometry at spatial infinity.

When we quantize electromagnetism in Chapter~\ref{chap:qed}, we will
actually exclude the Aharonov--Bohm modes from our analysis.
Chapter~\ref{chap:linear} describes our quantization
procedure---essentially just Fock quantization, but done in a way that
emphasizes the role of coherent states. The reason for this is that
Wilson loop ``operators''
$$
{\textstyle\oint_\gamma\hat A}
\qquad\hbox{or}\quad
\Wick{e^{i\oint_\gamma\hat A}}
$$
are \emph{not} densely-defined operators on Fock space, but their
matrix elements
$$
\matElem{\phi}{\textstyle\oint_\gamma\hat A}{\psi}
\qquad\hbox{or}\quad
\matElem{\phi}{\Wick{e^{i\oint_\gamma\hat A}}}{\psi}
$$
exist when~$\phi,\psi$ are linear combinations of regular coherent
states---that is, coherent states corresponding to sufficiently smooth
classical solutions of Maxwell's equations. Such regular
coherent states span a dense subspace of Fock space, so they are
sufficiently general to study Wilson loop dynamics. We are then able
to prove formulas such as
$$
{\dd\over\dd t}{\textstyle\oint_\gamma\hat
    A}={\textstyle\oint_\gamma\hat E}
$$
and
$$
{\dd\over\dd t}{\matElem{X'}{\Wick{e^{i\oint_\gamma\hat 
A}}}{X}\over\langle X'\mid
X\rangle}=i{\matElem{X'}{{\textstyle\oint_\gamma\hat
    E}}{X}\over\langle X'\mid X\rangle}\exp i{\matElem{X'}{{\textstyle\oint_\gamma\hat
    A}}{X}\over\langle X'\mid X\rangle},
$$
where~$\ket{X},\ket{X'}$ are regular coherent states.

The plan of this dissertation is as follows: in Part~$I$ we study
classical vacuum electromagnetism, and in Part~$II$ the quantization
of vacuum electromagnetism. Part~$I$ consists of three chapters. In
Chapter~$2$ we study ordinary vacuum electromagnetism in
a~$(3+1)$-dimensional static, globally hyperbolic spacetime. In
Chapter~$3$ we generalize our results to~$(n+1)$-dimensional
spacetimes and also consider theories where the electromagnetic
potential is not a~$1$-form but any~$p$-form, including the massless
scalar field ($p = 0$) and the Kalb-Ramond field ($p = 2$), which
plays a role in string theory.  Finally, in Chapter~$4$ we survey the
theory of~$L^2$ cohomology and suggest physical interpretations of
some of its main results. Part~$II$ consists of two
chapters. Chapter~$5$ is where we describe our coherent-state
quantization of linear dynamical systems and develop the concept of a
quasioperator. Lastly, in Chapter~$6$ this quantization method is
applied to vacuum electromagnetism and used to make sense of
unregularized Wilson loop quasioperators.


\part{Classical electromagnetism}

In this part we lay the classical groundwork for a a rigorous
quantization of the vacuum Maxwell equations and the analogous
equations for $p$-form electromagnetism with gauge
group~$\R$\index{$\R$!gauge group} on an arbitrary static, globally
hyperbolic, $(n+1)$\index{$n$!dimension of space}-dimensional
spacetime. In other words, we assume that spacetime is invariant under
time evolution and time reversal, and that the time evolution of
fields in spacetime is completely determined by initial data. In fact,
any such spacetime is topologically $M = \R \times
S$\index{$M$!spacetime}\index{$S$!space}, and has a metric of the form
$$
g_M = e^{2\Phi} (-\dd t^2 + g)
\index{$g_M$!spacetime metric}
\index{$\Phi$!Newtonian potential}
\index{$\dd$!exterior derivative}
\index{$t$!time coordinate}
\index{$g$!optical metric on space}
$$ 
where~$g$\index{$g_M$!spacetime metric} is a complete metric
on~$S$\index{$S$!space}, so that no lightlike geodesics run off to
spatial infinity in a finite amount of their affine parameter.
\index{quantization} 
\index{vacuum Maxwell equations} 
\index{electromagnetism!$p$-form} 
\index{gauge group}
\index{static spacetime}
\index{globally hyperbolic spacetime}
\index{spacetime}
\index{time evolution}
\index{time reversal}
\index{initial data}
\index{metric}
\index{complete metric}
\index{lightlike geodesic}
\index{spatial infinity}
\index{affine parameter}

Because the Lie algebras of~$\R$\index{$\R$!Lie
group} and~$U(1)$\index{$U(1)$!Lie group} are canonically
isomorphic, there is no difference between the versions of
electromagnetism with
either gauge group as far as the local formulation
of the Maxwell equations is
concerned. Globally there is a difference, though, because
all~$\R$\index{$\R$!Lie group}-bundles are
trivializable
whereas~$U(1)$\index{$U(1)$!Lie group}-bundles may not
be. In~$3+1$ 
dimensions, the second Chern class
 of a
nontrivial gauge bundle
manifests itself as a topological magnetic
charge whose field can be gauged away
locally, but not globally. While topological
charges are interesting, our primary goal is
to study the effects of spatial non-compactness on
quantization, and so we choose the gauge
group~$\R$\index{$\R$!Lie group} to eliminate the
possibility of nontrivial bundles. When a principal~$\R$\index{$\R$!gauge group}-bundle is trivialized,
connections on it are
ordinary~$1$-forms. 
\index{Lie algebra} 
\index{canonical isomorphism}
\index{electromagnetism}
\index{gauge group}
\index{Maxwell equations}
\index{bundle}
\index{trivializable bundle}
\index{bundle}
\index{second Chern class}
\index{nontrivial bundle}
\index{gauge bundle}
\index{topological charge}
\index{topological charge}
\index{topological charge}
\index{characteristic class}
\index{topological charge}
\index{non-compact space}
\index{quantization}
\index{gauge group}
\index{nontrivial bundle}
\index{gauge bundle}
\index{trivialization}
\index{connection}
\index{differential $1$-form}
\index{gauge group}
\index{electromagnetic potential}
\index{connection}

Technically, the subtlest aspects of our work arise from the
function~$\Phi$\index{$\Phi$!Newtonian potential} appearing in the
spacetime metric. This function measures the time dilation due to the
gravitational field, and reduces to the Newtonian gravitational
potential in the limit $\Phi \to 0$\index{$\Phi$!Newtonian potential}.
When $\Phi = 0$\index{$\Phi$!Newtonian potential}, $p$-form
electromagnetism uses rather familiar mathematics, mainly this portion
of the~$L^2$ de~Rham cohomology complex:
$$
\begin{CD}
{L^2 \Omega^{p-1}_S}@>{\dd_{p-1}}>>{L^2 \Omega^p_S}@>{\dd_p}>>{L^2 \Omega^{p+1}_S}
\end{CD}
\index{$L^2\Omega^p_S$!square-integrable $p$-forms on~$S$}
\index{$\dd_p$!exterior derivative on~$p$-forms}
$$ 
where~$L^2 \Omega^p_S$\index{$L^2\Omega^p_S$!square-integrable
$p$-forms on~$S$} stands for the Hilbert space of
square-integrable $p$-forms
on~$S$\index{$S$!space}.  The case $\Phi \ne 0$\index{$\Phi$!Newtonian
potential} requires some less familiar mathematics---except when~$p+1$
is half the dimension of spacetime\index{spacetime}, in which case
$p$-form electromagnetism is conformally invariant, allowing us to
eliminate~$\Phi$\index{$\Phi$!Newtonian potential} by an appropriate
rescaling of the fields. Even in the absence of
conformal invariance, the most elegant
approach is still to hide all the factors
involving~$\Phi$\index{$\Phi$!Newtonian potential} by a field
redefinition, and replacing the exterior
derivative with the `twisted'
differential
$$ 
D_k = e^{{1\over 2}(n-2p-1)\Phi} \dd_k e^{-{1\over 2}(n-2p-1)\Phi}
\index{$D_k$!twisted exterior derivative on~$k$-forms}
\index{$\Phi$!Newtonian potential}
$$ 
obtained by conjugating the ordinary differential by the rescaling
factor. This gives rise to a `twisted' version
of~$L^2$ cohomology which, on a
noncompact space, can differ from the usual~$L^2$
cohomology which, in turn, can
differ from the smooth de~Rham cohomology.
\index{spacetime}
\index{metric}
\index{time dilation}
\index{gravitational field}
\index{Newtonian gravitational potential}
\index{electromagnetism!$p$-form}
\index{de~Rham cohomology}
\index{cohomology complex}
\index{Hilbert space}
\index{square-integrable}
\index{differential $p$-form}
\index{conformal invariance}
\index{electromagnetism!$p$-form}
\index{field rescaling}
\index{conformal invariance}
\index{field redefinition}
\index{exterior derivative}
\index{twisted differential}
\index{rescaling factor}
\index{twisted~$L^2$ cohomology}
\index{square-integrable cohomology}
\index{de~Rham cohomology}

With this machinery in place we model the phase space\index{phase
space} of classical~$p$-form
electromagnetism\index{electromagnetism!$p$-form}
on~$(n+1)$-dimensional spacetime\index{spacetime} as a real Hilbert
space\index{real Hilbert space} with continuous
Hamiltonian\index{Hamiltonian} and symplectic
stucture\index{symplectic structure}. In the process, we address the
Aharonov--Bohm effect\index{Aharonov--Bohm effect} in situations where
the twisted~$L^2$ cohomology\index{twisted~$L^2$ cohomology} differs
from the usual de~Rham cohomology\index{de~Rham cohomology}, a subtle
issue that is largely neglected in the literature.

Among the most rigorous published treatments of Maxwell's
equations\index{Maxwell's equations} on a fairly generic manifold
stands that of Dimock~\cite{dimock92}, which however is restricted to
$(3+1)$-dimensional spacetimes\index{spacetime} with compact Cauchy
surfaces\index{Cauchy surface}. At the time of his writing, he said
``nothing that follows is particularly new, but it seems that the
various pieces have not been put together''. A later paper reviewing
the canonical\index{canonical mechanics} and covariant\index{covariant
mechanics} formulations of the classical Maxwell theory\index{Maxwell
theory} on a generic globally hyperbolic spacetime\index{globally
hyperbolic spacetime} is the one by Corichi~\cite{corichi}, again
``intended to fill an existing gap in the literature''.

Dimock constructs the classical phase space\index{phase space} from
gauge equivalence classes\index{gauge equivalence class} of Cauchy
data\index{Cauchy data} and the symplectic structure\index{symplectic
structure} obtained from the Noether current\index{Noether
current}. Gauge fixing\index{gauge fixing} appears as a technical step
used to show that Maxwell's equations\index{Maxwell's equations} are
strictly hyperbolic\index{hyperbolic equations}, so that solutions are
determined by their Cauchy data\index{Cauchy data}. Dimock uses
``fundamental solutions''\index{fundamental solutions} (essentially
Green's functions\index{Green's functions}) to parameterize the phase
space\index{phase space}, a technique that only works for linear field
equations\index{linear field equations}. Time evolution\index{time
evolution} enters the picture through symplectic
transformations\index{symplectic transformations} induced on phase
space\index{phase space} by changes in the choice of Cauchy
surface\index{Cauchy surface}. In fact, Dimock makes ``no choice of
Hamiltonian\index{Hamiltonian} or special time coordinate'', following
the covariant canonical formalism\index{covariant canonical mechanics}
of~\cite{crnkovic87}. Dimock points out how the field
strength\index{field strength} does not provide a complete set of
observables\index{complete set of observables} when the first homology
class\index{first homology class} of the Cauchy surfaces\index{Cauchy
surface} is nontrivial. In Chapter~\ref{sec:3+1} we relate this
phenomenon to the Aharonov--Bohm effect\index{Aharonov--Bohm effect}
and in Chapter~\ref{sec:harmonic} we present a thorough overview of
the situation in the non-compact case. Dimock assumes a
trivial~$U(1)$-bundle saying ``presumably our results can be extended
to non-trivial bundles for which~$A$ is only defined locally'', while
we take the more drastic step of assuming an~$\R$-bundle.

For the purposes of this Part, Dimock's presentation of Maxwell's
equations\index{Maxwell's equations} does have a couple of important
limitations. First, the restriction to compact Cauchy
surfaces\index{compact Cauchy surface} may be unphysical, and
certainly excludes many cases of theoretical interest. We address the
thorny analytic issues associated to allowing noncompact Cauchy
surfaces\index{noncompact Cauchy surface} in Chapter~\ref{sec:3+1},
albeit with the additional assumption that spacetime is
static\index{static spacetime}, which Dimock does not need. The
topological implications of noncompactness\index{noncompactness} are
discussed in Chapter~\ref{sec:harmonic}. Dimock's use of compact
Cauchy surfaces\index{compact Cauchy surface} allows his to bring
Hodge's theorem\index{Hodge's theorem} to bear on the Cauchy
data\index{Cauchy data} and, using the Kodaira
decomposition\index{Kodaira decomposition}, to show that the
symplectic structure\index{symplectic structure} is
non-degenerate. Although Hodge's theorem\index{Hodge's theorem} does
not hold on a noncompact space\index{noncompact space} (see
Chapter~\ref{sec:harmonic}), we are nevertheless able to prove a form
of Kodaira's decomposition\index{Kodaira decomposition} in
Chapter~\ref{sec:3+1}.

Dimock also states without proof or reference that ``for globally
hyperbolic manifolds, the usual classical linear field equations will
have global solutions if they are well-behaved locally. We quote the
result for scalar fields''. We repaired this defect by reference to
Chernoff's work in Chapter~\ref{sec:harmonic}. In the proof of existence of
solutions with given Cauchy data Dimock states ``The equation [above]
has principal part~$g^{\mu\nu}\partial_\mu\partial_\nu$ and thus is
strictly hyperbolic''; hyperbolicity easily follows from Chernoff's
work. Finally, the phase space constructed by Dimock does not have a topology
other than that induced by imposing the continuity of the symplectic
structure. Therefore, it is not a real inner-product space like ours
is. 

While not assuming compact Cauchy surfaces\index{compact Cauchy
surface}, Corichi's paper is ``not very precise about
functional-analytic issues'' in the author's own words. The
covariant\index{covariant mechanics} formulation is, like Dimock's,
based on the formalism of~\cite{crnkovic87}, and differs mostly in the
notation. The canonical\index{canonical mechanics} formulation is
written in a manifestly covariant way, in terms of the
foliation\index{foliation} generated by an arbitrary time coordinate
function. Both formulations of classical
electromagnetism\index{electromagnetism} are more general than ours,
and the relationship between Corichi's covariant and canonical
descriptions of phase space\index{phase space} is equivalent to
Dimock's treatment of Cauchy data\index{Cauchy data} in the
covariant\index{covariant mechanics} formalism.

The plan of this Part is as follows.  We begin in
Chapter~\ref{sec:3+1} by setting up classical
electromagnetism\index{classical electromagnetism} with gauge
group\index{gauge group}~$\R$\index{$\R$!Lie group}, leading up to Theorems~\ref{thm:3+1}
and~\ref{thm:3+1phys}, in which we make the phase space\index{phase
space} for this theory into a real Hilbert space\index{real Hilbert
space} on which the classical Hamiltonian\index{Hamiltonian} is a
continuous nonnegative quadratic form\index{quadratic form}.  In
Chapter~\ref{sec:N+1} we generalize this work to~$p$-form
electromagnatism\index{electromagnetism!$p$-form} in~$n+1$ dimensions
using the twisted de~Rham complex\index{de~Rham cohomology complex},
leading up to the analogous Theorems~\ref{thm:N+1}
and~\ref{thm:n+1phys}. In Chapter~\ref{sec:harmonic} we survey what is
known about~$L^2$ cohomology\index{square-integrable cohomology} on
noncompact spaces, and study a number of examples illustrating some of
the associated subtleties.

\chapter{Classical vacuum electromagnetism}
\label{sec:3+1}

In this chapter we discuss the classical vacuum Maxwell equations on a
$(3+1)$-dimensional static globally hyperbolic spacetime. In
particular, we explain how the classical phase space of
electromagnetism splits into two parts, one containing the oscillatory
modes of the electromagnetic field and the other containing the
`topological' modes responsible for the `Aharonov--Bohm' effect. 

The plan of this chapter is as follows: we begin in
Section~\ref{sec:3+1-geometry} by describing in detail our assumptions
and notation concerning spacetime geometry, decompose spacetime in the
form~$M\cong\R\times S$, and confront a number of analytical issues
arising from trying to define the exterior derivative on
square-integrable differential forms. In Section~\ref{sec:maxwell} we
give an overview of the stationary action formulation of classical
mechanics, and use it to derive the Maxwell equations, Noether
current, Hamiltonian and symplectic structure, as well as kinematical,
dynamical and physical phase spaces. Finally, in
Section~\ref{physical.interpretation} we describe the splitting on the
physical phase space of classical vacuum electromagnetism into an
sector consisting of oscillating modes, and a sector consisting of
topological modes responsible for the Aharonov--Bohm effect.

After seeing that the spacetimes we are interested split in the
form~$M\cong\R\times S$, where~$S$ is \emph{space}, we define the
exterior derivative~$\dd$ and coderivative~$\dd^*$ so that they act on
square-integrable differential forms on space and satisfy
\begin{equation}
\label{eqn:adjoint}
\int_Sg(\alpha,\dd\beta)\vol=\int_Sg(\dd^*\alpha,\beta)\vol
\index{$S$!space}
\index{$g$!optical metric on space}
\index{$\dd$!exterior derivative}
\index{$\vol$!volume form of optical metric}
\index{$\dd^*$!exterior coderivative}
\end{equation}
whenever~$\alpha$ and~$\beta$ are square-integrable differential forms
of appropriate degrees. The key is to show that no `boundary terms at
infinity'\index{boundary terms at infinity} appear in the integration
by parts\index{integration by parts} implicit in
Equation~(\ref{eqn:adjoint}). This can be used to show that the
Laplacian on square-integrable differential forms is
essentially self-adjoint\index{essentially self-adjoint} and
nonnegative, properties necessary for rigorous quantization.
\index{vacuum Maxwell equations}
\index{static spacetime}
\index{globally hyperbolic spacetime}
\index{spacetime}
\index{classical phase space}
\index{oscillatory mode}
\index{electromagnetic field}
\index{topological modes}
\index{Aharonov--Bohm effect}

In the \emph{temporal gauge}\index{temporal gauge} (vanishing
electrostatic potential\index{electrostatic potential}) the
configuration space\index{configuration space} of classical
electromagnetism\index{electromagnetism} on~$M$\index{$M$!spacetime}
consists of~$\R$-connections\index{connection} on~$S$ modulo gauge
transformations\index{gauge transformation}, and so is isomorphic to a
space of $1$-forms modulo square-integrable exact
$1$-forms\index{exact~$1$-forms} on~$S$. In physics, such a~$1$-form
is called a \emph{vector potential}\index{vector potential}. We make
the configuration space\index{configuration space} into a real Hilbert
space\index{real Hilbert space} by defining it as
$$
\A={\dom\{\dd\maps L^2\Omega^1_S\to L^2\Omega^2_S\}\over\overline{\ran}\{\dd\maps
L^2\Omega^0_S\to L^2\Omega^1_S\}}
\index{$\A$!space of vector potentials}
\index{$\dd$!exterior derivative}
\index{$L^2\Omega^p_S$!square-integrable $p$-forms on~$S$}
$$
with its natural real inner product. That is,~$\A$\index{$\A$!space of
vector potentials} consists of equivalence classes of
square-integrable $1$-forms with square-integrable exterior
derivatives, modulo exact $1$-forms. This space is naturally a real
Hilbert space\index{real Hilbert space}.

The canonical conjugate\index{canonical conjugate} of the vector
potential\index{vector potential}~$[A]$\index{$[A]$!gauge equivalence
class of vector potentials} is a divergenceless\index{divergenceless}
$1$-form~$E$\index{$E$!electric field}, called the \emph{electric
field}\index{electric field}. The space of electric fields
$$
\E=\ker\{\dd^*\maps L^2\Omega^1_S\to L^2\Omega^0_S\}
\index{$\E$!space of electric fields}
\index{$\dd^*$!divergence}
\index{$L^2\Omega^p_S$!square-integrable $p$-forms on~$S$} 
$$
is also naturally a real Hilbert space. The phase space of classical
electromagnetism is, then, the real Hilbert space
$$
\Phase=\A\oplus\E.
\index{$\Phase$!phase space}
\index{$\A$!space of vector potentials}
\index{$\E$!space of electric fields}
$$ 
The spaces~$\A$\index{$\A$!space of vector potentials}
and~$\E$\index{$\E$!space of electric fields} are dual
to each other by
$$
\bigl([A],E\bigr)=\int_S g\bigl(A,E)\vol,
\index{$(~,~)$!$p$-form inner product on~$S$}
\index{$S$!space}
\index{$g$!optical metric}
\index{$[A]$!gauge equivalence class of vector potentials}
\index{$E$!electric field}
\index{$\vol$!volume form of~$g$}
$$ 
which is independent of the representative~$A$ chosen for~$[A]$
because~$E$ is divergenceless. The symplectic
structure\index{symplectic structure} on~$\Phase$\index{$\Phase$!phase
space} is constructed from this duality pairing\index{duality pairing}
by antisymmetrization:
$$
\omega\bigl([A]\oplus E,[A']\oplus E'\bigr)=\int_S\bigl[g(A,E')-g(A',E)\bigr]\vol.
$$

Because of global hyperbolicity\index{global hyperbolicity}, any point
$X=[A]\oplus E$\index{$X$!field configuration}\index{$[A]$!gauge
equivalence class of vector potentials}\index{$E$!electric field} of
the physical phase space determines a unique solution of Maxwell's
equations\index{Maxwell's equations} on all
of~$M$\index{$M$!spacetime}. Time evolution\index{time evolution} is
given by a continuous one-parameter group\index{one-parameter group}
of continuous symplectic
transformations~$T(t)\from\Phase\to\Phase$\index{$T(t)$!time
evolution}\index{$\Phase$!phase space}. Unlike the symplectic
structure\index{symplectic structure} and the
Hamiltonian\index{Hamiltonian}, the natural Hilbert space norm
on~$\Phase$\index{$\Phase$!phase space} is not preserved by this time
evolution\index{time evolution}.

As a result of gauge-fixing\index{gauge fixing}, when
restricted to the phase space\index{phase space} the
Laplacian\index{Laplacian} on~$1$-forms
is~$\Delta=\dd^*\dd$\index{$\Delta$!Laplacian}\index{$\dd$!exterior
derivative}. The assumption that spacetime is static\index{static
spacetime} then implies that time evolution\index{time evolution}
commutes with~$\Delta$\index{$\Delta$!Laplacian}, and so the phase
space\index{phase space} admits the decomposition
$$
\Phase=\Phase_o\oplus\Phase_f
\index{$\Phase$!phase space}
\index{$\Phase_o$!oscillating sector of phase space}
\index{$\Phase_f$!free sector of phase space}
$$ 
where~$\Phase_f$\index{$\Phase_f$!free sector of phase space} is
the kernel of~$\Delta$\index{$\Delta$!Laplacian}
in~$\Phase$\index{$\Phase$!phase space} and consists of generalized
Aharonov--Bohm modes\index{Aharonov--Bohm modes}. From the point of
view of dynamics, the direct
summand~$\Phase_o$\index{$\Phase_o$!oscillating sector of phase space}
consists of `oscillating modes'\index{oscillating modes}
and~$\Phase_f$\index{$\Phase_f$!free sector of phase space} of `free
modes'\index{free modes}. Specifically,
on~$\Phase_o$\index{$\Phase_o$!oscillating sector of phase space} the
Hamiltonian\index{Hamiltonian} is a positive-definite quadratic form,
and so that `sector' of the electromagnetic
field\index{electromagnetic field} has the dynamics of an
infinite-dimensional harmonic oscillator\index{harmonic
oscillator}. The free sector\index{free
sector}~$\Phase_f$\index{$\Phase_f$!free sector of phase space} has
dynamics analogous to those of a free particle\index{free particle}.
For the free sector\index{free sector} one can successfully apply the
algebraic approach to quantization\index{algebraic quantization} of
Chapter~\ref{chap:linear}, but the existence of a Hilbert-space
representation\index{Hilbert-space represetation} on which time
evolution is unitarily implementable\index{unitary time evolution} is
not guaranteed unless~$\Phase_f$\index{$\Phase_f$!free sector of phase
space} is finite-dimensional. As we shall see in
Chapter~\ref{sec:harmonic}, that may not be the case on a noncompact
space\index{noncompact space} even if it is topologically trivial.

%



\section{Geometric setting}
\label{sec:3+1-geometry}

In this section we describe the mathematical framework for our study
of classical electromagnetism\index{electromagnetism}, and explain the
mathematical reasons why various physical restrictions are imposed on
the class of spacetimes\index{spacetime} under consideration.

\subsection{Static globally hyperbolic spacetimes}

Let us begin by recalling the precise definition of a
static\index{static spacetime}, globally hyperbolic
spacetimes\index{globally hyperbolic spacetime}. In physical terms, a
spacetime\index{spacetime} is stationary\index{stationary spacetime}
if it is invariant under time translations\index{time translations}
and static\index{static spacetime} if, in addition, it is invariant
under time reversal\index{time reversal}. Our first definition casts
these intuitive concepts in the language of (pseudo-)Riemannian
geometry.

\begin{definition}[stationary and static spacetimes]
A Lorentzian manifold\index{lorentzian manifold} without timelike
loops\index{timelike loop} (also called a
\emph{spacetime})\index{spacetime} is
\emph{stationary}\index{stationary spacetime} if, and only if, it
admits a one-parameter group of isometries\index{isometry} with
smooth, timelike orbits. A stationary spacetime\index{stationary
spacetime} is \emph{static}\index{static spacetime} if, in addition,
it is foliated\index{foliation} by a family of spacelike
hypersurfaces\index{spacelike hypersurface} everywhere orthogonal to
the orbits of the isometries.
\end{definition}

\begin{proof}[Note]
Spacetimes\index{spacetime} with closed timelike loops\index{closed
timelike loop} lead to a breakdown of the ordinary
initial-value\index{initial-value problem} formulation of
dynamics\index{dynamics}, and so must be excluded from our
analysis. Diffeomorphism\index{diffeomorphism} with smooth, timelike
orbits are generated by an everywhere timelike vector field. A vector
field\index{vector field} generating isometries is called a
\emph{Killing vector field}\index{Killing vector field}, and the
isometries\index{isometry} generated by a timelike Killing
field\index{timelike Killing field} are called \emph{time
translations}\index{time translation}. A stationary\index{stationary
spacetime} spacetime~$M$\index{$M$!spacetime} is
diffeomorphic\index{diffeomorphic} to~$\R\times
S$\index{$\R$!time}\index{$S$!space} for some smooth
manifold\index{smooth manifold}~$S$\index{$S$!space} representing
`space'\index{space}; if, in addition, $M$\index{$M$!spacetime} is
static\index{static spacetime}, it admits a metric\index{metric} of the
form
$$
g_M=-e^{2\Phi}\dd t^2+g_S,
\index{$g_M$!spacetime metric}
\index{$\Phi$!Newtonian potential}
\index{$t$!time coordinate}
\index{$g_S$!space metric}
$$
where~$\Phi$\index{$\Phi$!Newtonian potential} is a time-independent
function on~$S$\index{$S$!space}, and~$g_S$\index{$g_S$!space metric}
is a time-independent Riemannian metric\index{Riemannian metric}
on~$S$\index{$S$!space}. A stationary spacetime\index{stationaty
spacetime} would require cross-terms of the form~$e^\Phi(\dd
t\otimes\alpha+\alpha\otimes\dd t)$ in the
metric\index{metric},~$\alpha$ being a nonzero
time-independent~$1$-form on~$S$. For proofs of these statements see,
for instance,~\cite{wald84}.
\end{proof}

The concept of global hyperbolicity\index{global byperbolicity} is
more subtle, but it is related to the simple idea of
causality\index{causality}: that points of spacetime\index{spacetime}
are partially ordered\index{partial order} by the
relation\index{relation} `being to the future of'. The name `global
hyperbolicity'\index{global hyperbolicity} originally referred to a
property of systems of partial differential equations\index{partial
differential equation} on Euclidean space\index{Euclidean space}. By
reinterpreting those equations as coordinate\index{coordinate}
representations of equations adapted to a curved Lorentzian
manifold\index{Lorentzian manifold}, the
hyperbolicity\index{hyperbolicity} of the system became a geometric
property of the spacetime\index{spacetime} itself (see~\cite{geroch70}
and references therein). As we shall see, global
hyperbolicity\index{global hyperbolicity} of the
spacetime\index{spacetime} implies that the evolution
equations\index{evolution equation} of massless fields\index{massless
field} are globally hyperbolic\index{globally hyperbolic} systems of
partial differential equations\index{partial differential equation}.

Hyperbolic systems of partial differential equations\index{hyperbolic
partial differential equations} have a finite propagation
velocity\index{finite propagation velocity}, meaning that
compactly-supported initial data\index{initial data} evolve into
compactly-supported solutions after a finite time. Under the
reinterpretation of hyperbolic systems\index{hyperbolic system} as
propagation equations on Lorentzian manifolds\index{Lorentzian
manifold}, the finite propagation velocity\index{finite propagation
velocity} means that solutions with compactly-supported initial
data\index{initial data} are completely contained in the light
cones\index{light cone} of the support of their initial data. This is
one of the manifestations of causality\index{causality}.

The following definition formalizes the geometric ideas of
causality\index{causality} and global hyperbolicity\index{global
hyperbolicity}.

\begin{definition}[globally hyperbolic spacetime]
A piecewise-smooth curve\index{smooth curve} in a
spacetime\index{spacetime}~$M$\index{$M$!spacetime} is \emph{causal}
if its tangent vector\index{tangent vector} is everywhere
timelike\index{timelike}. A set is \emph{achronal}\index{achronal set}
if there are no causal curves between any two of its points.  The
\emph{domain of dependence}\index{domain of dependence} of a set
consists of all points~$p\in M$\index{$p$!point of
spacetime}\index{$M$!spacetime} such that every
inextensible\index{inextensible curve} causal curve\index{causal
curve} through~$p$\index{$p$!point of spacetime} intersects the set. A
\emph{Cauchy surface}\index{Cauchy surface} in a
spacetime\index{spacetime}~$M$\index{$M$!spacetime} is a
closed\index{closed set} achronal set\index{achronal set} whose domain
of dependence\index{domain of dependence} is all
of~$M$\index{$M$!spacetime}. A spacetime\index{spacetime} is
\emph{globally hyperbolic}\index{globally hyperbolic spacetime} if,
and only if, it admits a Cauchy surface\index{Cauchy surface}.
\end{definition}

\begin{proof}[Note]
The domain of dependence\index{domain of dependence} is also called
the \emph{Cauchy development}\index{Cauchy development}.  Both names,
`domain of dependence'\index{domain of dependence} and `Cauchy
development'\index{Cauchy development}, betray their origin in the
theory of partial differential equations\index{partial differential
equations}, as does the term `Cauchy surface'\index{Cauchy surface}.
A Cauchy surface\index{Cauchy surface} in a
spacetime\index{spacetime}~$M$\index{$M$!spacetime} is an achronal
set\index{achonal} intersecting every inextensible\index{inextensible
curve} causal curve\index{causal} in~$M$. It is not hard to see that
closed timelike curves\index{closed timelike curves} cannot intersect
an achronal hypersurface\index{achronal hypersurface}, and so
spacetimes\index{spacetime} with closed timelike curves\index{closed
timelike curve} cannot be globally hyperbolic\index{globally
hyperbolic}. For a static spacetime\index{static spacetime} with
metric\index{metric}
\begin{equation}\label{eq:optical}
g_M=e^{2\Phi}(-\dd t^2+g),
\index{$g_M$!metric on spacetime}
\index{$\Phi$!Newtonian potential}
\index{$t$!time coordinate}
\index{$g$!optical metric on space}
\end{equation}
global hyperbolicity\index{global hyperbolicity} is equivalent to
completeness of the metric\index{metric}~$g=e^{-2\Phi}g_S$\index{$g$!optical metric on
space}\index{$\Phi$!Newtonian potential}\index{$g_S$!metric on space}. This
metric\index{metric}~$g$\index{$g$!optical metric} is sometimes called
\emph{optical metric}\index{optical metric} (see, for
instance,~\cite{MR1694235,MR1605624,MR776077,MR0207364}) because light
rays follow geodesics of this metric. More precisely, the
geodesics\index{geodesic} of~$g$\index{$g$!optical metric on space}
parameterized by arc length\index{arc length} lift to affinely
parameterized\index{affine parameterization} lightlike
geodesics\index{lightlike geodesic} of~$-\dd t^2+g$\index{$t$!time
coordinate}\index{$g$!optical metric}, with the
time~$t$\index{$t$!time coordinate} corresponding to the
arc-length\index{arc lenght} parameter on geodesics\index{geodesic}
of~$g$\index{$g$!optical metric}. Hence, the propagation of
light\index{light propagation} in the geometric optics\index{geometric
optics} approximation is determined by~$g$\index{$g$!optical metric}
alone. We will consistently use the optical
metric\index{optical metrix}~$g$\index{$g$!optical metric}
on~$S$\index{$S$!space} rather than~$g_S$\index{$g_S$!metric on
space}.
\end{proof}

\subsection{Spacetime geometry and topology}

We model spacetime\index{spacetime} as a static\index{static
manifold}, globally hyperbolic\index{globally hyperbolic manifold},
$(3+1)$-dimensional Lorentzian manifold\index{Lorentzian
manifold}. That is, spacetime\index{spacetime} will be represented by
a smooth $(3+1)$-dimensional manifold~$M$\index{$M$!spacetime}
diffeomorphic to~$\R\times S$\index{$\S$!space} and admitting a
Lorentzian metric\index{Lorentzian metric} of the form given in
Equation~(\ref{eq:optical}).

For convenience, we also assume~$S$\index{$S$!space} is oriented. In
that case, the metric~$g$\index{$g$!optical metric} determines a
volume form~$\vol$\index{$\vol$!volume form on space}
on~$S$. Similarly, the
spacetime\index{spacetime}~$M$\index{$M$!spacetime} acquires a volume
form\index{volume form}~$\vol_M$\index{volume form on spacetime} from
the metric~$g_M$\index{$g_M$!metric on spacetime}. The canonical
volume forms\index{canonical volume form} are related by
\begin{equation}
\vol_M =  e^{4\Phi} \vol \wedge \dd t .
\index{$\vol_M$!volume form on spacetime}
\index{$\Phi$!Newtonian potential}
\index{$\vol$!volume form on space}
\index{$t$!time coordinate}
\label{eq:vol}
\end{equation}
If~$S$\index{$S$!space} were nonorientable\index{nonorientable space},
we could still carry through our whole discussion with minor
modifications, the most important of which being
that~$\vol$\index{$\vol$!volume form on space}
and~$\vol_M$\index{$\vol_M$!volume form on spacetime} would have to be
treated as densities\index{density}.

We religiously follow the convention of writing all differential
forms\index{differential form} on spacetime\index{spacetime} with a
subscript~`$M$'\index{$M$!spacetime}. We also write the so-called
temporal part\index{temporal part} with a subscript~`$0$', and the
spatial part\index{spatial part} with no subscript. We decompose
$k$-forms\index{$k$-forms} on~$M$\index{$M$!spacetime} into
spatial\index{spatial part} and temporal parts\index{temporal part}
thus:
\begin{equation}
\alpha_M=\dd t\wedge\alpha_0+\alpha,
\index{$\alpha_M$!differential form on spacetime}
\index{$t$!time coordinate}
\index{$\alpha_0$!temporal part of~$\alpha$}
\index{$\alpha_S$!spatial part of~$\alpha$}
\label{eq:stpart}
\end{equation}
where~$\alpha_0$ is a $(k-1)$-form and~$\alpha$ is a $k$-form
on~$S$, both $t$-dependent.  

The exterior derivative\index{exterior derivative} operators on
spacetime\index{spacetime}~$\dd_M\from C_0^\infty\Omega^k_M\to
C_0^\infty\Omega^{k+1}_M$ and on space $\dd\from
C_0^\infty\Omega^k_S\to C_0^\infty\Omega^{k+1}_S$,
where~$C^\infty_0\Omega^k_S$ denotes smooth, compactly supported
$k$-forms on~$S$, are related by $\dd_M=\dd t\wedge\partial_t+\dd$; in
other words,
\begin{equation}
\label{eq:stderivative}
\dd_M\alpha_M=\dd t\wedge(\partial_t\alpha-\dd\alpha_0)+\dd\alpha.
\end{equation}
for all compactly-supported smooth~$k$-forms~$\alpha_M\in
C_0^\infty\Omega^k_M$. 

We use~$g$ and~$g_M$ to denote the respective induced metrics
on $k$-forms, satisfying
\begin{equation}
g_M(\alpha_M,\beta_M)=e^{-2k\Phi}\bigl[g(\alpha,\beta)-g(\alpha_0,\beta_0)\bigr],
\label{eq:stpart2}
\end{equation}
and define the positive-definite bilinear forms
\begin{equation}
(\alpha_M,\beta_M)_M=\int_M g_M(\alpha_M,\beta_M)\vol_M
\qquad\hbox{and}\quad
(\alpha,\beta)=\int_S g(\alpha,\beta)\vol
\label{eq:inner}
\end{equation}
on~$C^\infty_0\Omega^k_M$ and~$C^\infty_0\Omega^k_S$, which are related by
\begin{equation}
(\alpha_M,\beta_M)_M=\int_\R e^{(4-2k)\Phi}\bigl[(\alpha,\beta)-(\alpha_0,\beta_0)\bigr]\dd t. 
\label{eq:inner2}
\end{equation}

We denote by~$\delta$ the formal adjoint of~$\dd$ with respect to the
bilinear form~$(~,~)$. This means that the operator~$\delta\from
C_0^\infty\Omega^{k+1}_S\to C_0^\infty\Omega^k_S$ is defined by
\begin{equation}
(\alpha,\dd\beta)=(\delta\alpha,\beta)
\qquad\hbox{for all}\quad
\alpha\in C_0^\infty\Omega_S^{k+1}
\qquad\hbox{and}\quad
\beta\in C^\infty_0\Omega_S^k,
\label{eq:formal_adjoint}
\end{equation}
The compact support in
Equations~(\ref{eq:stderivative}) and~(\ref{eq:formal_adjoint}) has
the function of avoiding boundary terms on the implicit integration by
parts involved in the definition of~$\delta$. 

\subsection{Issues of analysis on noncompact spaces}

\label{sec:analysis}

A restatement of Equation~(\ref{eq:formal_adjoint}) is the existence
of operators
\begin{equation}
\xymatrix{C^\infty_0\Omega^k_S\ar@<.5ex>[r]^\dd &
C^\infty_0\Omega^{k+1}_S\ar@<.5ex>[l]^\delta}
\label{eq:formal_adjoint2}
\end{equation}
which are formal adjoints of each other. Our goal is to extend these
to densely defined operators between~$L^2\Omega^k$
and~$L^2\Omega^{k+1}$ which are adjoint to each other in the strict
sense of operator theory, where~$L^2\Omega^k$ denotes the space of
square-integrable~$k$-forms on~$S$. It turns out that this can be done
precisely because~$g$ is a complete metric on~$S$, which we have seen
is equivalent to global hyperbolicity of spacetime.

There are both physical and mathematical reasons for wanting to do
this. Mathematically, a mutually adjoint pair of unbounded operators
between two Hilbert spaces are much better behaved than
formally-adjoint operators between spaces of smooth diferential forms,
although the latter have more intuitive geometric appeal. From a
physical point of view, we do not wish to restrict ourselves to
compactly-supported fields in a noncompact space, but on the other
hand we need the fields to be square integrable in order for the
Hamiltonian and symplectic structure on phase space to be finite at
all times. These sorts of physical considerations demand that we
treat~$\dd$ and~$\delta$ as unbounded operators between Hilbert spaces
of square-integrable differential forms. To prove that time evolution
maps the classical phase space to itself, we will also need to
extend~$\delta\dd$ to an unbounded self-adjoint operator on
square-integrable $1$-forms. Finally, once we insist on
interpreting~$\dd$ as an operator between spaces of square-integrable
forms, the electromagnetic gauge transformations will need to have
square-integrable generators.

All this requires a short detour into functional analysis, which is
contained in this subsection. While the facts we need are well-known
to the experts, they may be unfamiliar to some readers, so we review
them in a fair amount of detail.  We omit most of the proofs, many of
which can be found in Reed and Simon's textbook~\cite{RS}. The reader
who is more interested in the physical use of these operators can skip
to Section~\ref{sec:maxwell}, with the observation that from then on
the operator~$\delta$ is denoted~$\dd^*$, as in
\begin{equation}
\xymatrix{L^2\Omega^k_S\ar@<.5ex>[r]^{\dd} &
L^2\Omega^{k+1}_S\ar@<.5ex>[l]^{\dd^*}},
\label{eq:formal_adjoint3}
\end{equation}
in order to free the
symbol~$\delta$ for use in variational calculus. Making sense of
Equation~(\ref{eq:formal_adjoint3}) is the main purpose of this
subsection. 

In going from Equation~(\ref{eq:formal_adjoint2}) to
Equation~(\ref{eq:formal_adjoint3}), the first thing we need to do is
establish that the operators~$\dd\from C^\infty_0\Omega^k_S\to
C^\infty_0\Omega^{k+1}_S$ and~$\delta\from C^\infty_0\Omega^{k+1}_S\to
C^\infty_0\Omega^k_S$ appearing in Equations~(\ref{eq:formal_adjoint})
and~(\ref{eq:formal_adjoint2}) can be interpreted as densely-defined
operators between~$L^2\Omega^k_S$ and~$L^2\Omega^{k+1}_S$. This
follows from Lemma~\ref{lem:dense}.

\begin{lemma}\label{lem:dense} 
The completion of~$C^\infty_0\Omega^k_S$ with respect to the inner product
$(~,~)$ is~$L^2\Omega^k_S$.
\end{lemma}

\begin{proof}  
We can reduce this to the well-known case where $S = \R^n$ using a
partition-of-unity argument.
\end{proof}

The domain of the operator~$\dd$ is the dense subspace 
$$ 
\dom\dd=C^\infty_0\Omega^k_S \subseteq L^2 \Omega^k_S . 
$$  
We then define the adjoint~$\dd^*$ in the usual way, as
follows. First, the domain of~$\dd^*$ consists of all~$\alpha\in
L^2\Omega^{k+1}_S$ for which there exists a~$\gamma\in L^2\Omega^k_S$
such that
$$
(\alpha,\dd \beta) =(\gamma,\beta)
$$
for all~$\beta\in C^\infty_0\Omega^k_S$. If such a~$\gamma$ exists it is
unique because~$C^\infty_0\Omega^k_S$ is dense in~$L^2\Omega^k_S$, and we
then define~$\dd^*\alpha$ to equal this~$\gamma$, so that
\begin{equation}
(\alpha,\dd \beta)=(\dd^*\alpha,\beta) \qquad \forall\beta\in
C^\infty_0\Omega^k_S. 
\label{eq:adjoint}
\end{equation}
as desired. Note that, because~$\beta$ is required to be of compact
support,~$\alpha$ is not required to have compact support.

Similarly, the dense domain of~$\delta$ is~$C^\infty_0\Omega^{k+1}_S$,
the domain of~$\delta^*$ is not restricted to compactly-supported
forms, and~$\delta^*\beta$ can be defined by
$$
(\alpha,\delta^* \beta)=(\delta \alpha,\beta) \qquad \forall\alpha\in C^\infty_0\Omega^{k+1}_S. 
$$

We can also define operators~$\overline\dd$ and~$\overline \delta$,
the respective closures of~$\dd$ and~$\delta$.  For~$\dd$ this goes as
follows.  We define the graph of~$\dd$ to be the linear subspace
$$ 
\gr(\dd)=\{\alpha\oplus\dd\alpha \mid \alpha \in C_0^\infty \Omega^k_S \}
\subseteq L^2\Omega^k_S\oplus L^2\Omega^{k+1}_S
$$
where the latter space is a Hilbert space in an obvious way. This
subspace is typically not closed, and we say that~$\dd$ is closable if
the closure of~$\gr(\dd)$ is the graph of an operator, which we then
denote~$\overline\dd$. In other words,
\begin{equation}
\alpha\in\dom\overline\dd \Leftrightarrow
\alpha={\displaystyle\lim_{n\to\infty}}\alpha_n
\qquad\mathrm{and}\quad
\dd\alpha_n\to\overline\dd\alpha
\qquad\hbox{for some}\quad
\alpha_n\in
C^\infty_0\Omega^k_S.
\label{eq:closure}
\end{equation}
We define the closure~$\overline \delta$ in essentially the same way.

Because of Equation~(\ref{eq:formal_adjoint}) both~$\dd^*$
and~$\delta^*$ are densely defined, so the following lemma applies.

\begin{lemma}\label{lem:doubledual} 
A densely defined operator $T$ is closable if, and only if, $T^*$ is
densely defined. In that case, $\overline T=T^{**}$.
\end{lemma}

Observe that~$T^*$ is automatically closed and $\overline
T^*=T^*$. As a result, $\overline\dd=\dd^{**}$ and
$\overline\delta=\delta^{**}$. We have
$\dd\subseteq\overline\dd\subseteq\delta^*$ and
$\delta\subseteq\overline\delta\subseteq\dd^*$. 

\begin{proof}  See Reed and Simon's textbook~\cite[Theorem
VIII.1]{RS}.
\end{proof}

We have argued that
$$
\dom\overline\dd \subseteq \dom\delta^* 
\qquad\hbox{and}\quad
\dom\overline\delta \subseteq \dom\dd^*,
$$
but $\overline\dd$ and $\overline \delta$ will be mutual adjoints only
if these are actually equalities.  Having them be mutual adjoints is
highly desirable, as otherwise there are at least two possible
self-adjoint extensions of the operator~$\delta\dd$,
namely~$\dd^*\overline\dd$ and~$\overline\delta\delta^*$. This means
we need to understand how the equations $\dom\overline\dd =
\dom\delta^*$ and $\dom\overline\delta = \dom\dd^*$ could fail to
hold.

The answer has to do with boundary values.  Suppose that~$S$ is a
relatively compact open subset of some larger Riemannian manifold~$X$,
and its boundary~$\partial S$ is a smooth submanifold of~$X$.  In this
case the desired equalities \emph{never} hold, and there is a
well-developed theory of boundary values which explains why
\cite{evans98}. In brief, if~$\alpha,\beta$ are compactly supported
smooth forms on~$S$, integration by parts gives
$$
(\dd\alpha,\beta) = (\alpha,\delta\beta)
\qquad\hbox{for all}\quad\alpha\in C^\infty_0\Omega^k_S
\qquad\hbox{and}\quad\beta\in C^\infty_0\Omega^{k+1}_S.
$$
From this, an approximation argument gives
$$
(\overline\dd\alpha,\beta) = (\alpha,\overline \delta\beta)
\qquad\hbox{if}\quad
\alpha \in \dom\overline\dd 
\quad\hbox{and}\quad
\beta\in\dom\overline\delta .
$$
On the other hand, if~$\alpha,\beta$ are merely smooth forms
on~$S$ that extend smoothly to~$X$, integration by parts gives
\begin{equation}
(\dd\alpha,\beta) - (\alpha,\delta\beta) = (\alpha,\beta)_{\partial S},
\qquad\hbox{for all}\quad\alpha\in C^\infty\Omega^k_S
\qquad\hbox{and}\quad\beta\in C^\infty\Omega^{k+1}_S
\label{eq:boundary}
\end{equation}
and from this, again by an approximation argument, one can show
$$
(\delta^*\alpha,\beta)=(\alpha,\dd^*\beta)+(\alpha,\beta)_{\partial S}
\qquad\hbox{if}\quad 
\alpha \in \dom \delta^* 
\quad\hbox{and}\quad 
\beta\in\dom\dd^* .
$$
Thus we cannot have $\dom\overline\dd = \dom\delta^*$ and
$\dom\overline\delta = \dom\dd^*$ in this case: the nonzero boundary
term~$(\alpha,\beta)_{\partial S}$ gets in the way.

The same sort of problem can occur even when~$S$ is not a relatively
compact open subset of some larger Riemannian manifold.  However, in
this more general situation the concept of `boundary value' needs to
be reinterpreted as `value at spacelike infinity'. In fact,
Equation~(\ref{eq:boundary}) can be used to define the notion of
boundary at infinity of~$S$. The domain of~$\overline\dd$ can be
understood as the space of square-integrable differential forms with
square-integrable exterior derivatives and vanishing `values at
infinity', while the domain of~$\delta^*$ consists of
square-integrable differential forms with square-integrable exterior
derivatives and no restriction on values at infinity.  Thus, the
desired equation~$\overline\dd = \delta^*$ fails to hold if an element
of~$\dom\delta^*$ can fail to `vanish at infinity'.  Similar remarks
apply to the equation~$\overline \delta =\dd^*$.  Simply put, the
problems arise when there are boundary terms at infinity when we
integrate by parts.

Luckily, the folowing result of Gaffney implies that these problems
never happen when~$g$ is a \emph{complete} Riemannian metric on~$S$.

\begin{proposition}[Gaffney] \label{prop:gaffney} 
If $S$ is a complete oriented Riemannian manifold, then
$$
(\delta^*\alpha,\beta)=(\alpha,\dd^*\beta)
$$
whenever $\alpha\in\dom\delta^*$ and $\beta\in\dom\dd^*$.
\end{proposition}

Gaffney calls manifolds where the conclusion of
Proposition~\ref{prop:gaffney} holds ``manifolds with negligible
boundary''.

\begin{proof}  
This can be found in Gaffney's paper \cite{gaffney54}; we will also
give a proof of a more general result in
Corollary~\ref{cor:essential}, based on work of
Chernoff~\cite{chernoff73}. 
\end{proof}

\begin{corollary}
If $S$ is a complete oriented Riemannian manifold, then
$$
\overline\dd=\delta^*
\qquad\mathrm{and}\quad
\overline\delta=\dd^*.
$$ 
\end{corollary}

This means that~$\dd$ and~$\delta$ have mutually adjoint
closures
$$
\xymatrix{L^2\Omega^k\ar@<.5ex>[r]^{\overline\dd} &
L^2\Omega^{k+1}\ar@<.5ex>[l]^{\overline\delta}}.
$$
As we pointed out above, this implies that the operators~$\delta\dd$ 
and~$\dd\delta$ have unique self-adjoint closures.

\begin{proof} We will prove that $\overline\dd=\delta^*$, as the
other equality then follows by lemma~\ref{lem:doubledual}. We already
know that $\overline\dd\subseteq\delta^*$, so we need only show that
$\delta^*\subseteq\overline\dd$. To this end, let
$\alpha\in\dom\delta^*$ and $\beta\in\dom\dd^*$. By
Lemma~\ref{lem:dense}, $\alpha\in\dom\delta^*$ is the~$L^2$ limit of a
sequence~$\alpha_n$ of compactly-supported differential
forms. Gaffney's Proposition~\ref{prop:gaffney} allows us to write
$$
(\delta^*\alpha\mid\beta)=(\alpha\mid\dd^*\beta)=\lim_{n\to\infty}(\alpha_n\mid\dd^*\beta) 
$$
By the definition of~$\dd^*$ in Equation~(\ref{eq:adjoint}),
$$
(\delta^*\alpha\mid\beta)=\lim_{n\to\infty}(\alpha_n\mid\dd^*\beta)=\lim_{n\to\infty}(\dd\alpha_n\mid\beta).  
$$
Since this holds for arbitrary~$\beta$ in the dense domain of~$\dd^*$,
not only $\alpha_n\to\alpha$ but also $\dd\alpha_n\to\delta^*\alpha$,
and so $\alpha\in\dom\overline\dd$ and
$\overline\dd\alpha=\delta^*\alpha$ by the definition of
$\overline\dd$ in Equation~(\ref{eq:closure}).
\end{proof}

As we shall see, the uniqueness of the self-adjoint closure
of~$\delta\dd$ (in other words, the essential self-adjointness
of~$\delta\dd$) is necessary to make sense of the Fock quantization of
the electromagnetic field. By Gaffney's result, the essential
self-adjointness of~$\delta\dd$ follows from completeness of~$S$
which, as we have pointed out, is equivalent to the global
hyperbolicity of the original static spacetime~$M$. Intuitively, if a
spacetime is globally hyperbolic there is no information coming from
or lost to infinity, so no boundary conditions are necessary to
uniquely determine time evolution of square-integrable differential
forms and, in fact, space has `negligible boundary' in the sense of
Gaffney. This, in retrospect, is the justification for the assumption
that spacetime is globally hyperbolic although, strictly speaking,
this is a sufficient but not a necessary condition for~$S$ to have
negligible boundary.

Because our assumption of global hyperbolicity implies
that~$\overline\dd=\delta^*$ and~$\overline\delta=\dd^*$ there is no
ambiguity in the closing of the operators~$\dd$ and~$\delta$, and from
this point on we shall assume that~$\dd$ and~$\delta$ have been closed
unless otherwise stated. We will slightly abuse notation by
writing~$\dd$ to denote the closed version of the exterior
derivative. As noted before, its adjoint will be denoted~$\dd^*$ so as
to preserve~$\delta$ for use in variational calculus.

Sometimes, as shorthand or in order to avoid confusion between
exterior derivative operators acting on different spaces, an
additional bit of notation will be necessary; namely, we will denote
by~$\dd_k$ the operator~$\dd\from L^2\Omega^k_S\to L^2\Omega^{k+1}$,
so that~$\dd_k^*$ will stand for~$\dd^*\from L^2\Omega^{k+1}_S\to
L^2\Omega^k_S$.

\section{Maxwell's theory}

\label{sec:maxwell}

In the rest of this section we derive the Maxwell equations by
applying Hamilton's principle of stationary action, and define the
phase space of the theory as the collection of gauge equivalence
classes of solutions of the equations of motion. The phase space is
constructed in three steps (see, for instance, \cite{relativistic,
partial}): a \emph{kinematical phase space} on which the Hamilton
least action principle can be formulated, but not supporting a
Hamiltonian or symplectic structure; a \emph{dynamical phase space} of
solitions of the equations of motion on which a conserved Hamiltonian
and Noether current are defined, but without a symplectic structure;
and a \emph{physical phase space} with no remaining gauge freedom,
which is a symplectic space.

For simplicity, we only consider Maxwell's equations in the case where
the electromagnetic vector potential is a connection on a
\emph{trivial} bundle over spacetime. Luckily, this is a vacuous
restriction when the gauge group is~$\R$, as we are assuming.
For~$U(1)$ electromagnetism, nontrivial bundles can be used to model
magnetic monopoles. Having a trivial bundle means we can treat the
vector potential as a 1-form~$A_M$ on spacetime; that is, the
covariant exterior derivative on~$M$ is~$\dd_M+A_M\wedge$. The field
strength is the curvature~$2$-form
$$
F_M = \dd_M A_M.
$$
and the Maxwell action is
\begin{equation}
\label{eq:actionM}
\Act[A_M] = -{1\over 2}(F_M , F_M ),
\end{equation}
which is invariant under gauge transformations of the form
$$
A_M\mapsto A_M+\dd_M\phi.
$$
The equations of motion follow from applying the Hamilton principle of
stationary action to Equation~(\ref{eq:actionM}).

To obtain a Hamiltonian formulation of the equations of motion one
needs to use an explicit foliation of spacetime into a family of
Cauchy surfaces related by a time translation symmetry. We can do this
because we have assumed that spacetime is globally hyperbolic and
static. We use Equation~(\ref{eq:stpart}) to split~$A_M$ and~$F_M$
into spatial and temporal parts:
$$
A_M=\dd t\wedge A_0+A
\qquad\hbox{and}\quad
F_M=\dd t\wedge F_0+F,
$$
whose physical interpretation is that~$F_0$ is the electic field and~$F$
the magnetic field, as we shall see below. By
Equation~(\ref{eq:stderivative}) 
$$
F_0=\partial_t A-\dd A_0
\qquad\hbox{and}\quad
F=\dd A.
$$ 
Gauge transformations leave~$F_M$ unchanged, but their effect
on~$A_0$ and~$A$ is
\begin{equation}
\label{eq:stgauge}
A\mapsto A+\dd\phi
\qquad\hbox{and}\quad
A_0\mapsto A_0+\partial_t\phi.
\end{equation}

Using Equation~(\ref{eq:inner2}), Equation~(\ref{eq:actionM}) can be
rewritten as
\begin{eqnarray}
\Act[A,A_0]=
{1\over 2}\int_\R
\bigl[(\partial_t A -\dd A_0,\partial_t A -\dd A_0) -
(\dd A,\dd A)\bigr] \dd t.
\label{eq:action}
\end{eqnarray} 
Note that a factor of~$e^{-4\Phi}$ in the metric on 2-forms from
Equation~(\ref{eq:stpart2}) has cancelled the factor of~$e^{4\Phi}$ in
the volume form on spacetime from Equation~(\ref{eq:vol}). This makes
the $3+1$-dimensional case of Maxwell's theory special, and it is
intimately related to the fact that Maxwell's equations are
conformally invariant in this dimension. Conformal invariance is
another reason why the decomposition~$g_M=e^{2\Phi}(-\dd t^2+g)$ is
preferable to~$g_M=-e^{2\Phi}\dd t^2+g_S$, at least in this case. The
action of Equation~(\ref{eq:action}) is the time-integral of the
Lagrangian
\begin{equation}
\Lagr[A,A_0]={1\over 2}\bigl[(\dot A-\dd A_0,\dot A-\dd A_0)-(\dd
  A,\dd A)\bigr],
\label{eq:Lagr}
\end{equation}
where~$\dot A=\partial_t A$. 

Because of energy conservation, the integral of
Equation~(\ref{eq:action}) is likely to diverge unless it is restricted
to a finite interval of~$t$. This restiction is, in any case,
necessary to use the action principle to study time evolution between
two given instants of time. In addition to evaluating the action
integral over a finite interval of time, sufficient conditions for
Equations~(\ref{eq:stgauge})--(\ref{eq:Lagr}) to make sense include
that
$$
\phi(t),A_0(t)\in\dom\{\dd\from L^2\Omega^0_S\to L^2\Omega^1_S\}
\qquad\hbox{and}\quad
A(t)\in\dom\{\dd\from L^2\Omega^1_S\to L^2\Omega^2_S\},
$$ 
for almost, with~$t$ with all the~$L^2$ norms being square-integrable
over any compact interval of~$t$; and that their respective time
derivatives are in the same spaces. This imposes nontrivial smoothness
and decay restrictions on the electromagnetic potentials~$A_0,A$, and
also on the allowed generators~$\phi$ of gauge transformations. In the
case when space is compact, any smooth gauge generator will
automatically be bounded and square-integrable, but in the noncompact
case we are forced to exclude some gauge transformations which are too
large at infinity but would otherwise na\"{\i}vely be allowed. This restriction on the gauge generators
cannot manifest itself in physical effects on any bounded region of
spacetime.

\subsection{Overview of covariant mechanics}

Hamilton's principle states that physically allowed field
configurations~$X$ in a region~$R$ of a spacetime~$M$ are critical
points (not necessarily minima) of an action
functional~$\Act_R[X]$. We assume that the action is \emph{local},
that is, that~$\Act_R[X]$ is the integral over the spacetime
region~$R$ of a Lagrangian density~$\Lagr[X]$ which, at each point of
spacetime, depends only on~$X$ and a finite number of its derivatives
(usually just the first) at that point. The action functional is often
calculated by evaluating the integral in Equation~(\ref{eq:action})
over a bounded region of spacetime, and almost always over a finite
interval of time. In fact the action calculated over all of time may
be infinite, and the variation of the action might also be ill-defined
unless restricted to be compactly supported in time, which amounts to
evaluating the action integral over a finite interval of time in the
first place. Hamilton's principle is formulated on a \emph{kinematical
phase space}~$\X_R$ large enough to contain all plausible field
configurations and small enough that~$\Act_R[X]=\int_R\Lagr[X]$ is
well-defined.
\begin{figure}
$$
\begin{xy}
(30,0)*{};
(25,15)**\crv{(20,5)&(25,10)};
(-5,15)**\crv{(15,10)&(5,20)};
(0,0)**\crv{(-5,10)&(-10,5)};
(30,0)**\crv{(20,5)&(10,-5)}?(.5)+(-2,-1)*{\scriptstyle S};
(0,10)*{\scriptstyle M};
(12,8)*\xycircle<14pt,7pt>{};
(12,8)*{\scriptstyle R};
\end{xy}
$$
\caption{Schematic representation of spacetime, space and the domain
  of integration for the action functional}
\end{figure}
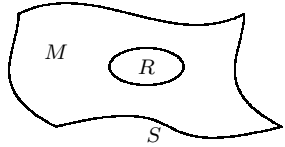

The stationary action principle implies the vanishing of the first
variation of the action on any region~$R$:
$$
0=\delta\Act_R[X]=\int_R\delta\Lagr[X]=-\oint_{\partial R}\theta[X]+\int_R E[X].
$$ It has been shown~\cite{zuckerman87,crnkovic87} that it is possible
and advantageous to choose~$\X$ to be an infinite-dimensional manifold
(possibly even a vector space) and interpret the variational
derivative~$\delta$ as an exterior derivative on~$\X$. The Lagrangian
density~$\Lagr$ is then an~$(n+1)$-form on~$\X\times M$ proportional
to~$\vol_M$. The condition that~$\Lagr$ be a \emph{local Lagrangian}
means that, at any point~$p\in M$,~$\Lagr$ depends on~$X$ only through
the values of~$X$ and finitely many ot its derivatives at~$p$. The
exterior derivative on~$\X\times M$ is~$\delta+\dd_M$, which implies
the anticommutation relation~$\delta\dd_M+\dd_M\delta=0$. The
quantity~$E$ is an~$(n+2)$-form on~$X\times M$ which is a~$1$-form
with respect to~$\X$ and proportional to the~$(n+1)$-form~$\vol_M$;
similarly,~$\theta$ is an~$(n+1)$-form which is a~$1$-form with
respect to~$\X$ and an~$n$-form with respect to~$M$. Tangent vectors
to~$\X$ are variations of field configurations. We denote a typical
such tangent vector by~$\partial_X$.

If~$\delta\Act_R[X]$ is evaluated at a stationary field
configuration~$X$, on variations~$\partial_X$ vanishing on the
boundary~$\partial R$, the stationary action condition implies the
\emph{Euler--Lagrange equations of motion}~$E[X](\partial_X)=0$. We define the
\emph{dynamical phase space} associated to the region~$R$ as the variety
$$
\Dyn_R=\{X\in\X\suchthat
E[X](\partial_X)=0\quad\hbox{on}\quad
R\qquad\hbox{if}\quad\partial_X=0\quad\hbox{on}\quad\partial R\}.
$$

The so-called \emph{Noether current}~$\theta[X]$ is defined only up to
an exterior derivative, and can be interpreted as a generator of
conserved quantities associated to continuous symmetries of solutions
to the Euler--Lagrange equations of motion. To see this, consider a
tangent vector to~$\Dyn_R$, which is a variation of solutions to the
Euler--Lagrange equations of motion. Because the Euler-Lagrange
equations are satisfied throughout, we have
$$
\oint_{\partial M}\theta[X](\partial_X)=0.
$$
Suppose now that~$R\cong [0,1]\times T$. Then,
$$
\int_{T_0}\theta[X](\partial_X)-\int_{T_1}\theta[X](\partial_X)=\int\limits_{\rlap{$\scriptstyle[0,1]\times\partial
T$}}\,\theta[X](\partial_X), 
$$ 
where the right-hand side represents the time integral of the flux
of the conserved quantity through~$\partial T$. In the case where~$R$
is a globally hyperbolic region with Cauchy surface~$T$, the latter
has negligible boundary in the sense of Gaffney, and
$$
\int_{T_0}\theta[X](\partial_X)=\int_{T_1}\theta[X](\partial_X),
$$
so~$\int_T\theta[X](\partial_X)$ is a conserved quantity of the
motion. 
\begin{figure}
$$
\begin{xy}
(30,0)*{};
(0,0)**\crv~lc{(15,15)};
(30,0)**\crv{(12,11)};
(0,0)**\crv{(12,6)};
(30,0)**\crv{(18,-4)};
(0,0)**\crv{(18,-9)};
(30,0)**\crv~lc{(15,-15)};
(0,0)**\dir{-}?(.5)+(0,1.5)*{\scriptstyle{T}};
(11,-9)*{\scriptstyle R};
\end{xy}
$$
\caption{Schematic representation of a globally hyperbolic region
  foliated by a family of Cauchy surfaces.}
\end{figure}
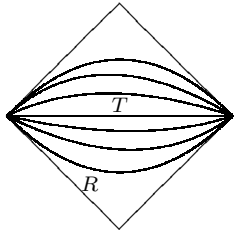
For instance, in the case where~$M\cong\R\times S$ is static
and~$R=[t_0,t_1]\times S$, the variation~$\partial_X$ might represent
the generator of a one-parameter group of isometries of~$S$ (a
translation or rotation) on the field configuration~$X$, and the
associated conserved quantity would be the corresponding momentum (linear or angular) of~$X$. If~$\partial_X$ represented the action of an internal symmetry
of the field variables at each point (a gauge transformation), the
conserved quantity would be the conserved charge associated to the
gauge symmetry.

The variational derivative of the Noether current is a skew-symmetric
$2$-form on~$\Dyn_R$,
$$
\omega_T[X]=\int_T\delta\theta[X].
$$
Given two variations of solutions, 
$$
\omega_T[X](\partial_X,\partial_X')
$$
is a conserved quantity of the solution~$X$. It is possible
that~$\omega_T[X]$ is degenerate, admitting variations of
solutions~$\partial_X$ such that
$$
\omega_T[X](\partial_X,-)=0.
$$
Each such degenerate direction~$\partial_X$ generates a gauge
transformation of the dynamical phase space. The space of gauge orbits
of~$\Dyn_R$ is the \emph{physical phase space}~$\Phase_R$. As we have
pointed out, it is in
general not a manifold, but an `infinite-dimensional variety
with singularities'. By
construction,~$\omega_S$ would project to a non-degenerate symplectic 
structure on~$\Phase_R$. 

A more cogent approach to the physical phase space~$\Phase_R$ would be
as follows. Let~$N$ denote the space of degenerate directions
of~$\omega_T$. The smooth functions~$f$ on~$\Dyn_R$ such
that~$\partial_Xf=0$ whenever~$\partial_X\in N$ constitute a
subalgebra of~$C^\infty(\Dyn_R)$, the so-called gauge-invariant
observables on~$\Dyn_R$. The spectrum of homomorphisms of this algebra
would be~$\Phase_R$, and we can map the algebra of gauge-invariant
observables homeomorphically to~$C^\infty(\Phase_R)$. Whether or
not~$\Phase_R$ turns out to be a manifold that can support a
symplectic structure, the algebra of gauge-invariant supports the
canonical Poisson structure
$$
\{f,g\}=\omega_T(\partial_f+N,\partial_g+N)
\qquad\hbox{for all}\quad f\in C^\infty(\Phase_R),
$$ 
where~$\partial_f$ is a tangent vector to~$\Dyn_R$ such
that~$\omega_T(\partial_f,\partial_Y)=\delta f(\partial_Y)$ for all
tangent vectors to~$\Dyn_R$. Conveniently,~$\partial_f$ is 
defined precisely up to addition of elements of~$N$, so one can associate a
unique equivalence class in~$T\Dyn_R/N$ to it,
namely~$\partial_f+N$. Since~$\omega_T$ is, in fact,
non-degenerate on~$\Dyn_R$, the algebra of gauge-invariant observables
is a Poisson algebra, whose spectrum is the physical phase space. 

This construction simplifies considerably when the action functional
is quadratic, as in that case the equations of motion and the Noether
current are linear, and all the spaces involved are vector spaces. In
addition, in a stationary, globally hyperbolic spacetime~$M$ there is a
preferred foliation~$M\simeq\R\times S$ by Cauchy surfaces isometric to~$S$. When
there is a single timelike Killing field, there is a canonical
identification of the different Cauchy surfaces, and time evolution
can be represented as a transformation of the field configuration on a
single Cauchy surface. It is then possible to define a Hamiltonian
function. 

In the next few sections we construct the phase space of
electromagnetism using this method. First, the \emph{kinematical phase space} is a space~$\X$ of
\emph{field configurations} on which the Maxwell action can be
defined, or on which the Maxwell equations can be written. The precise
definition of the kinematical phase space is somewhat arbitrary, as
long as it is large enough to contain all the actual solutions of the
equations of motion. In the next section we shall see three acceptable
formulations of the least action principle on different kinematical
phase spaces before settling on one of them.

Next, setting the first variation of the action to zero yields the
Maxwell equations of motion, whose space of solutions if the
\emph{dynamical phase space}~$\Dyn$ and is a linear subspace of the
kinematical phase space (in more general cases,~$\Dyn$ is just a
subvariety of~$\X$). The dynamical phase space supports the
Hamiltonian and Noether current of the system, which can be used to
obtain conserved quantities of the system and a pre-symplectic
structure on~$\Dyn$. The null directions of the pre-symplectic
structure are seen to correspond to gauge transformations.

Finally, the set~$\Phase$ of gauge orbits on~$\Dyn$ is the
\emph{physical phase space} or, simply, the \emph{phase space}. When
there is no gauge freedom, the dynamical phase space coincides with
the physical phase space. After this reduction from~$\Dyn$
to~$\Phase$, the pre-symplectic structure on~$\Dyn$ becomes a
non-degenerate symplectic structure on~$\Phase$. 

\subsection{Kinematical phase space}

In this section we consider three possible action principles for
electromagnetism on slightly different kinematical phase spaces. The
first is the Lagrangian formulation of
Equations~(\ref{eq:action}--\ref{eq:Lagr}). The second formulation is
the associated Hamiltonian formulation, with the electrostatic
potential~$A_0$ acting as a Lagrange multiplier enforcing the Gauss
law as a constraint. Since the latter is linear, it is possible and
even convenient to impose the Gauss law at the kinematical level
without a Lagrange multiplier. This is the third formulation.

All three kinematical phase spaces are equivalent in that the action
principles defined on them lead to the same space of solutions of the
equations of motion. However, the three kinematical phase spaces are
not isomorphic to each other. The first requires that~$A_0$ be in the
domain of~$\dd$, and that~$E=\partial_t A-\dd A_0$ be
square-integrable. The second alternative allows~$A_0$ to be just
square integrable, but~$E$ must now be in the domain of~$\dd^*$. The
third formulation does without~$A_0$ altogether, but~$E$ must be in the
kernel of~$\dd^*$.

\subsubsection{Lagrangian formulation}

The action of Equation~(\ref{eq:action}) is defined on pairs~$(A,A_0)$
where~$A$ is a~$1$-form and~$A_0$ is a~$0$-form on~$S$, both
time-dependent and such that the quantities $\|A(t)\|$, $\|\partial_t
A(t)\|$, $\|\dd A(t)\|$, $\|A_0(t)\|$ and $\|\dd A_0(t)\|$ are all
square-integrable with respect to~$t$ on any bounded interval. This is
a suitable definition of the kinematical phase space.

The first variation of the Lagrangian of Equation~(\ref{eq:Lagr}) is
the variational~$1$-form on~$\X$
$$
\delta\Lagr[A,A_0]=\partial_t(E,\delta A)-(\dot
E+\dd^*\dd A,\delta A)-(\dd^*E,\delta A_0),
$$ 
where $E=\partial_A\Lagr[A,A_0]=\dot A-\dd A_0$ is the
\emph{electric field}. According to Hamilton's
principle, for physically allowed~$A$ and~$A_0$, the
variation~$\delta\Lagr$ must vanish. The fact that~$\Lagr$ is 
independent of~$\dot A_0$ implies that~$A_0$ is a non-dynamical
Lagrange multiplier field enforcing the constraint
$$
\dd^*E=0.
$$
The dynamical fields are~$A$ and its canonical conjugate,~$E$, which
we combine into a \emph{field configuration}~$X=A\oplus E$.

\subsubsection{Hamiltonian formulation with a Lagrange multiplier}

In terms of the field configuration $X=A\oplus E$ and the Lagrange
multiplier~$A_0$, the original Lagrangian from
Equation~(\ref{eq:Lagr}) can be rewritten as
\begin{equation}
\Lagr[X;A_0]=(E,\dot A)-{1\over 2}\bigl[(E,E)+(\dd A,\dd
  A)\bigr]-(\dd^* E,A_0).
\label{eq:Lagr2}
\end{equation}
This leads to an alternative---and inequivalent---definition of the
kinematical phase space, namely the collection of pairs~$(X,A_0)$
where~$X=A\oplus E$, and such that $\|A(t)\|$, $\|E(t)\|$, $\|\partial_t
A(t)\|$, $\|\dd A(t)\|$, $\|\dd^* E(t)\|$ and~$\|A_0\|$ are all
square-integrable over finite intervals of~$t$.

The Euler--Lagrange equations obtained from the first variation of the
Lagrangian of Equation~(\ref{eq:Lagr2}),
\begin{equation}
\delta\Lagr[A\oplus E;A_0]=\partial_t(E,\delta A)+(\dot A-E-\dd
A_0,\delta E)-(\dot E+\dd^*\dd A,\delta A)-(\dd^* E,\delta A_0)
\label{eq:firstVar}
\end{equation}
yield the Maxwell equations in Hamiltonian form
\begin{eqnarray*}
\dd^* E&=&0\\
\dot A-E&=&\dd A_0\\
\dot E+\dd^*\dd A&=&0
\end{eqnarray*}
Observe that, when the Gauss law is satisfied, the action is
independent of the non-dynamical (and hence arbitrary)~$A_0$, and so
because of the equation~$\dot A=E+\dd A_0$ time evolution is not
uniquely determined by the initial conditions. This is all closely
related to the existence of time-dependent gauge transformations,
which by Equation~(\ref{eq:stgauge}) result in a change of the
Lagrange multiplier field~$A_0$. We can use this gauge freedom to
eliminate the Lagrange multiplier~$A_0$, that is, we perform a time-dependent
gauge transformation to make~$A_0=0$. This is the so-called `temporal
gauge'. Then, the Maxwell equations take the form
\begin{eqnarray}
\label{eq:Max}
\dd^* E&=0&\quad\hbox{(Gauss law constraint)}\\
\dot A-E&=0&\quad\hbox{(Faraday--Lenz law)}\label{eq:FL}\\
\dot E+\dd^*\dd A&=0&\quad\hbox{(Amp\`ere--Maxwell law)}
\label{eq:well}
\end{eqnarray}
on the kinematical phase space. 

\subsubsection{Hamiltonian formulation without Lagrange multipliers}

The partial gauge-fixing of the previous case can be carried out at
the level of the action, leading to a third possible definition of the
kinematical phase space~$\X$, consisting of pairs~$X=A\oplus E$ such
that $\|A(t)\|$, $\|\partial_t A(t)\|$, $\|\dd A(t)\|$, and $\|E(t)\|$
are square-integrable on finite intervals of~$t$, and
that~$\dd^*E(t)=0$ for almost all~$t$. 

We choose this as our preferred kinematical phase space. This means
that, for us,~$\X$ consists of pairs~$X=A\oplus E$ such that
$$
A(t)\oplus E(t)\in\dom\{\dd\from L^2\Omega^1_S\to
L^2\Omega^2_S\}\oplus\ker\{\dd^*\from L^2\Omega^1_S\to L^2\Omega^0_S\}
\qquad\hbox{for almost all}\quad
t
$$
and $\|X(t)\|$ is square-integrable on bounded intervals of~$t$.

On this space,
Equation~(\ref{eq:Max}) is automatically satisfied and the Lagrangian
\begin{equation}
\label{eq:Lagr3}
\Lagr[X]=(E,\dot A)-{1\over 2}\bigl[(E,E)+(\dd A,\dd
  A)\bigr]
\end{equation}
leads to the additional Maxwell Equations~(\ref{eq:FL})
and~(\ref{eq:well}).

\subsection{Dynamical phase space}

The space of solution of the Maxwell equations in the temporal
gauge~(\ref{eq:Max})--(\ref{eq:well}) is the \emph{dynamical phase
space} of the theory.  Because the Maxwell equations are linear, the
space of their solutions is a linear subspace of the kinematical phase
space~$\X$. The global hyperbolicity of~$M$ implies that, in the
temporal gauge, the Maxwell equations form a hyperbolic system of
partial differential equations. Then, each solution of the equations
of motion is uniquely determined by its restriction to a surface of
constant~$t$ (initial data at time~$t$), so each such surface provides
a coordinatization of the dynamical phase space in terms of a pair
of~$1$-forms on~$S$.

In other words, we adopt the point of view that the dynamical phase
space consists of time-dependent solutions~$A\oplus E$ of the
equations of motion, that data~$X(t)=A(t)\oplus E(t)$ at
time~$t$ are a coordinatization of the phase space, and that time
evolution is a change of coordinates in phase space. Under this
interpretation, it can be argued that it is a bad thing to concentrate
too much on the time evolution of initial data. We proceed to do just
this, however.

From any of the definitions of the kinematical phase space~$\X$ in the
previous section it follows that, for almost all~$t$, initial data
$X(t)=A(t)\oplus E(t)$ are such that~$A(t)\in\dom\{\dd\from
L^2\Omega^1_S\to L^2\Omega^2_S\}$ and~$E(t)\in\dom\{\dd^*\from
L^2\Omega^1_S\to L^2\Omega^0_S\}$.  This means that the space of
solutions of Maxwell's equations is isomorphic to a (dense, at least)
subspace of
$$
\Dyn=\dom\{\dd\from L^2\Omega^1_S\to L^2\Omega^2_S\}\oplus\ker\{\dd^*\from
L^2\Omega^1_S\to L^2\Omega^0_S\}.
$$

The Hamiltonian 
\begin{equation}
\label{eq:Ham}
\Ham_t={1\over 2}\Bigl[\bigl(E(t),E(t)\bigr)+\bigl(\dd A(t),\dd
A(t)\bigr)\Bigr] 
\end{equation}
can be directly read off from the form of the Lagrangian in
Equation~(\ref{eq:Lagr3}) and it is preserved by time evolution. What
this means is that, although the Hamiltonian is defined on a
particular surface of constant~$t$, it is independent of~$t$ as long
as~$A\oplus E$ satisfies the equations of motion. In other words, the
Hamiltonian is time-dependent---and thus ill-defined as a single
functional---on the kinematical phase space, but is
coordinate-independent on the dynamical phase space. Moreover,~$\Dyn$
imposes just the right decay and smoothness conditions on~$A(t)$
and~$E(t)$ so that~$\Dyn$ is exactly the space of initial data~$X(t)$
satisfying the Gauss law~$\dd^*E(t)=0$ and for which~$\Ham$ is finite.

The so-called \emph{Noether current} can also be read off directly, in
this case from the total derivative term in the first variation of the
Lagrangian, Equation~(\ref{eq:firstVar}). The Noether current is a
variational~$1$-form on the dynamical phase space~$\Dyn$ which, for
electromagnetism, takes the form
$$
\theta_t=\bigl(E(t),\delta A(t)\bigr).
$$ 
The Noether current can be used to obtain conserved quantities
associated to continuous transformations of the fields. Indeed, 
If~$X=A\oplus E$ is a solution of the equations of motion,
$$
\theta_t-\theta_0=\delta\Act[X].
$$
This means that, if~$X$ depends on a parameter~$\tau$ such
that~$\partial_\tau\Act[X]=0$, then
$$
\theta_t(\partial_\tau)=\bigl(E(t),\partial_\tau A(t)\bigr)
$$ 
is independent of~$t$ and so is a conserved quantity of the
equations of motion. This means~$\theta$ is well-defined
on~$\Dyn$. Conversely, if~$X=A\oplus E$ were not a solution of the equations of
motion the Noether current would depend on~$t$, and so~$\theta$ really
should not be interpreted as a~$1$-form on~$\X$.

For instance, the one-parameter gauge transformation given by
$\partial_\phi A=\dd\phi$ leaves the action invariant, and in that
case~$\theta(\partial_\phi)=(\dd^*E,\phi)$. The conserved quantity
associated to gauge transformations of this form is,
therefore,~$\dd^*E$. Although Gauss' law makes this seem trivial, this
conservation law is nontrivial when the Maxwell equations are coupled
to matter, in which case~$\dd^*E$ equals the electric charge, and
therefore the conservation law associated to gauge invariance is
electric charge conservation. When the surface~$S$ at constant~$t$ has
nontrivial continuous isometries, $\partial_\tau A=L_\xi A$
where~$\xi$ is the Killing field generating the isometries and~$L_\xi$
denotes the Lie derivative with respect to it. In that case, the
Noether current evaluates to~$\theta(\partial_\tau)=(E,L_\xi A)$,
which is the conserved quantity associated to the isometry. This is
one way to define the momentum and angular momentum of the
electromagnetic field on homogeneous, rotationally invariant or
isotropic spaces, such as Minkowski space. It also shows that, when
space has no continuous isometries, there is no global generalization
of the linear and angular momenta of the electromagnetic field.

The variational exterior derivative of the
Noether current is the pre-symplectic structure
\begin{equation}
\label{eq:symplectic}
\omega_t=\bigl(\delta E(t),\delta A(t)\bigr)-\bigl(\delta A(t),\delta E(t)\bigr),
\end{equation}
which is an variational~$2$-form. Like the Hamiltonian, though
ostensibly defined for data on a surface of constant~$t$ and thus
time-dependent, the pre-symplectic structure~$\omega$ is finite and
conserved by time evolution if it is evaluated at a solution~$X$ on
two variations compatible with the equations of motion (that is, two
tangent vectors to~$\Dyn$ at the same~$X\in\Dyn$).  

Because the dynamical phase space~$\Dyn$ is defined by
Equation~(\ref{eq:Max}),~$\omega$ has null directions consisting
precisely of all variations of the form
$$
\partial_\phi A=\dd\phi,
$$
which are the gauge transformations remaining after choosing the
temporal gauge. This means that~$\omega$ is indeed degenerate, and
that the degeneracy is related to gauge freedom.

\subsection{Physical phase space}
\label{sec:analytical.setting}  

We have seen that the Gauss law constraint generates the gauge transformations
$$ 
A \mapsto A + \dd\phi ,
$$ 
and two sets of initial data~$A\oplus E$ and~$A'\oplus E'$ are
physically equivalent if they differ by a transformation of this
form. Thus, taking the quotient of~$\Dyn$ by this equivalence relation
we should obtain the physical phase space of the Maxwell theory,
\begin{equation}
\Phase={\dom\{\dd\maps L^2\Omega^1_S\to
  L^2\Omega^2_S\}\over\overline{\ran}\{\dd\maps L^2\Omega^0_S\to
  L^2\Omega^1_S\}}\oplus\ker\{\dd^*\from L^2\Omega^1_S\to
L^2\Omega^0_S\}.
\label{eq:phase}
\end{equation}
In words, the physical phase space consists of pairs~$[A]\oplus E$
where:~$[A]$ is an equivalence class of
square-integrable~$1$-forms on~$S$ with square-integrable exterior
derivatives modulo~$L^2$ limits of the exterior derivatives of
square-integrable functions on~$S$; and~$E$ is a
square-integrable~$1$-form on~$S$ with vanishing divergence.

Note that the Hamiltonian of Equation~(\ref{eq:Ham}) is manifestly
independent of any choice of representative in the gauge equivalence
class of~$A$. On the other hand, the (now nondegenerate) symplectic
structure~(\ref{eq:symplectic}) is gauge-independent only
because of Gauss' law, as
$$
(A+\dd\beta,E)=(A,E)+(\beta,\dd^* E)=(A,E).
$$

The first direct summand in Equation~(\ref{eq:phase}),
$$
\A={\dom\{\dd\maps L^2\Omega^1_S\to
  L^2\Omega^2_S\}\over\overline{\ran}\{\dd\maps L^2\Omega^0_S\to
  L^2\Omega^1_S\}},
$$
has a natural Hilbert-space norm
\begin{equation}
\label{eqn:norm_A}
\bigl\|[A]\bigr\|_\A^2=\inf_{\phi
\in\Omega^0}(A+\dd\phi,A+\dd\phi)+(\dd A,\dd A),
\end{equation}
which combines the natural norm on a quotient space with the natural
Sobolev norm on~$\dom\{\dd\from L^2\Omega^1_S\to L^2\Omega^2_S\}$. The
second summand is simply
$$
\E=\ker\{\dd^*\maps L^2\Omega^1\to L^2\Omega^0\}
\qquad\hbox{with}\quad
\|E\|_\E^2=(E,E),
$$
and the natural norm on~$\Phase=\A\oplus\E$ is the sum of the two.
The Hamiltonian and symplectic structure on~$\Phase$ are continous
with respect to these norms.

%

\section{Free and oscillating modes}
\label{physical.interpretation}

Since the definition of the physical phase space~$\Phase$ is rather technical,
let us expound on it a bit. A point in the classical phase space is a
pair~$[A]\oplus E$ where: the vector potential~$[A]$ is an equivalence
class of square-integrable $1$-forms modulo gauge transformations,
with square-integrable exterior derivatives; and the electric
field~$E$ is a square-integrable $1$-form satisfying the Gauss law.
Our definition of the physical phase space ensures that it contains
precisely such pairs for which the Hamiltonian (physically, the
energy)~$\Ham$ is finite and the symplectic structure~$\omega$ is
well-defined. It also makes the gauge equivalence relation precise,
and makes precise the sense in which the Gauss law holds.

Note that the physical phase space~$\Phase$ does not necessarily
contain \emph{all} finite-energy initial data for Maxwell's equations,
since we are imposing the additional condition that $([A],[A]) <
\infty$ to make the symplectic structure well-defined. If we omitted
this condition we could define a real Hilbert space consisting of {\it
all} finite-energy initial data for Maxwell's equations, but the
symplectic structure would only be densely defined on this space. This
is a gauge-independent condition because~$([A],[A])$ smallest $L^2$
norm among all the vector potentials in the same gauge-equivalence
class; one could fix the gauge by choosing the representative~$A$ such
that~$(A,A)=([A],[A])$, but that is not necessary.

Observe now that the phase space~$\Phase$ and the Hamiltonian~$\Ham$
are defined very simply in terms of~$\dd$ and~$\dd^*$, and recall the
Kodaira orthogonal-direct-sum decomposition
$$
L^2\Omega^1_S=
\overline{\ran\dd_0} 
\oplus
\ker \Delta_1 
\oplus
\overline{\ran\dd^*_1} 
$$ 
where $\Delta_1=\dd^*_1\dd_1+\dd_0\dd^*_0\from L^2\Omega^1_S\to
L^2\Omega^1_S$. We prove a general version of the Kodaira
decomposition in Section~\ref{sec:Kodaira}. In the present section we
use the decomposition to write~$\Phase$ as the direct sum of a
part~$\Phase_f$ containing the \emph{Aharonov--Bohm modes} or
\emph{free modes}, and a part~$\Phase_o$ containing the more familiar
\emph{oscillating modes} of the electromagnetic field. We will see
that it is convenient to treat the classical dynamics of Maxwell
theory separately on these two parts, but putting the results together
we shall see that time evolution acts as a strongly continuous
1-parameter group of symplectic transformations on~$\Phase$. Note
that, at least in the classical theory, the separation of the
oscillating and free modes is a matter of convenience.

Before embarking on the mathematical details of the Kodaira
decomposition, let us explore its physical significance for the
classical phase space of electromagnetism.

\subsection{The space of pure-gauge potentials}

Observe that, in our definition of the physical phase space,
Equation~(\ref{eq:phase}), we have taken the space of `pure gauge' vector
potentials to be~$\overline{\ran\dd_0}$. This is subtly different from
the common assumption that pure gauge potentials are derivatives of
arbitrary smooth scalar functions. Instead, we are saying they lie in
the {\em closure\/} of the space of derivatives of {\em
square-integrable\/} functions.  While these nuances may seem merely
pedantic, they have have dramatic consequences in certain situations
which we discuss in Section~\ref{sec:harmonic}. The simplest example,
in~$2+1$ dimensions, is when~$S$ is the hyperbolic plane, which has an
infinite-dimensional space of square-integrable $1$-forms that are
exterior derivatives of smooth functions which are not
square-integrable, so the $1$-forms are not pure gauge by our
definition.

Physically, as we are restricting the class of allowed gauge
transformations (essentially to be compactly supported), in general
there will be vector potentials that would na\"\i{}vely be considered
pure gauge but should not, because they involve a change of gauge on
an effectively infinite volume. However, these additional modes cannot
be detected by any experiment carried out on a finite volume, and so
one could argue that they should be discarded after all. However,
these vector potentials are canonically conjugate to static electric
fields with finite energy, and so are required in the canonical
formulation of electromagnetism. This is even more important if these
electric field modes are to be quantized.

Mathematically, our definition is natural thanks to the Kodaira
decomposition, and it leads to consistent classical and quantum
theories, except possibly (see chapters~\ref{sec:harmonic}
and~\ref{chap:qed}) in the case when the space of harmonic vector
potentials is infinite-dimensional.

As for the electric field~$E$, the space~$\ran\dd_0$ is orthogonal
to~$\ker \dd^*_0$, so square-integrable electric fields satisfying
the Gauss' law constraint~$\dd^*_0 E=0$ belong
to~$\ker\Delta_1\oplus\ran\dd^*_1$.

\subsection{Aharonov--Bohm modes}

The space~$\ker \Delta_1$ consists of square-integrable
harmonic~$1$-forms. For any vector potential~$A$ in this space, the
magnetic field~$\dd A$ vanishes. If the manifold~$S$ is compact, Hodge's
theorem asserts that this space is isomorphic to the first de~Rham
cohomology of~$S$, a topological invariant, and vector potentials in
this space can be detected by their holonomies around noncontractible
loops, as in the Aharonov--Bohm effect. Thus, in the compact case, it
makes perfect sense to call~$\ker \Delta_1$ the configuration space of
`Aharonov--Bohm' or `topological' modes of the electromagnetic field.

The situation is subtler if~$S$ is noncompact.  In this case~$\ker
\Delta_1$ is called the `first~$L^2$ cohomology group' of~$S$.
The~$L^2$ cohomology of a non-compact Riemannian manifold can differ
from the de~Rham cohomology, and it depends on the metric, so it is
not a topological invariant. By analogy with the compact case we still
call harmonic vector potentials `Aharonov--Bohm' modes. As we shall
see, sometimes there are Aharonov--Bohm modes even when~$S$ is
contractible. On the other hand, sometimes there are no Aharonov--Bohm
modes when they would be expected on elementary topological
considerations. Finally, the space of Aharonov--Bohm modes may be
infinite-dimensional. These facts make it a bit trickier to understand
vector potentials in~$\ker \Delta$ as topological Aharonov--Bohm
modes.  However, at least for certain large classes of well-behaved
manifolds, it still seems to be possible. We review some of these
results in Section~\ref{sec:harmonic}.

\subsection{Decomposition into free and oscillating modes}

We now apply the Kodaira decomposition to the physical phase
space~$\Phase$, in order to understand the Aharonov--Bohm modes more
deeply, as well as the meaning of the third
summand~$\overline{\ran\delta_1}$ in the Kodaira decomposition. 

The Kodaira decomposition allows us write $\Phase$ as a direct sum $\Phase_o
\oplus \Phase_f$ of `oscillating' and `free' modes of the electromagnetic
field.  The oscillating modes are familiar from electromagnetism on
Minkowski spacetime.  The free modes are those relevant to the
Aharonov--Bohm effect; we call them `free' because the equations of
motion for these modes are mathematically analogous to those of a free
particle, as we shall see.

To see this in detail, first recall that 
$$
\Phase = \A \oplus \e
$$
where
$$
\begin{array}{ccl}
\A &=& \dom \dd_1 / \, \overline{\ran\dd_0} \\
\e &=& \ker \dd^*_1 .
\end{array}
$$
The Kodaira decomposition lets us split~$\A$ and~$\e$ into
`oscillating' and `free' parts:
$$
\begin{array}{ccl}
\A &\cong& \A_o \oplus \A_f \\
\e &=& \e_o \oplus \e_f ,
\end{array}
$$
where 
$$ 
\begin{array}{lclllcl}
\A_o &=& \dom \dd_1 \cap \overline{\ran \dd^*_1}  
&\quad& 
\A_f &=&\ker \Delta         \\
&&&&&& \\
\e_o &=& \overline{\ran \dd^*_1}  
&\quad&
\e_f &=& \ker \Delta .
\end{array}
$$
Note that the difference between~$\A_o$ and~$\e_o$ is coming from the
different norms:~$\bigl\|[A]\bigr\|^2+\|\dd A\|^2$ versus~$\|E\|^2$. This
decomposition lets us write the classical phase space as a direct
sum of real Hilbert spaces
$$ 
\Phase = \Phase_o \oplus \Phase_f , 
$$
where
$$
\begin{array}{ccl}
\Phase_o &=& \A_o \oplus \e_o   \\
\Phase_f &=& \A_f \oplus \e_f    .
\end{array}
$$
This splitting respects the symplectic structure and also the
Hamiltonian on~$\Phase$, so time evolution acts independently on the
oscillating and free part of any initial data $[A]\oplus E \in \Phase$.

\subsection{The oscillating sector}

For modes $A\oplus E \in \Phase_o$, Maxwell's equations say:
$$
\left\{
\begin{array}{l}
\partial_t A = E \cr 
\partial_t E =- \Delta A, \cr
\end{array}
\right. 
$$
a generalization of the equations of motion for a harmonic oscillator.
This is why we call~$\Phase_o$ the phase space of `oscillating' modes.
The Hamiltonian on~$\Phase_o$ is also of harmonic oscillator type:
$$
H[A\oplus E] = {1\over 2} \bigl[(\dd A | \dd A)+(E | E)\bigr].
$$
If we rewrite the above version of
Maxwell's equations as a single integral equation,
we find it has solutions of the form
\begin{equation}
\left( \! \begin{array}{c} A \\ E \end{array} \! \right)
\mapsto T_o(t) 
\left( \! \begin{array}{c} A \\ E \end{array} \! \right) = 
\left( \begin{array}{cc} 
\cos(t\sqrt{\Delta}) & \sin(t\sqrt{\Delta})\,/\,\sqrt{\Delta}\\ 
-\sqrt\Delta\,\sin(t\sqrt{\Delta}) & \cos(t\sqrt{\Delta})  
\end{array} \right) 
\left( \! \begin{array}{c} A \\ E \end{array} \! \right)
\label{eq:time_evolution}
\end{equation}
where we define functions of~$\Delta$ using the functional calculus
\cite{RS}. 
The time evolution operators $T_o(t)$ form 
a strongly continuous group of bounded operators on
$\Phase_o$.  This follows from three facts:
\begin{itemize}
\item $\|T_o(t)\|$ is finite for all~$t$.
\item $T_o(t)T_o(s)=T_o(t+s)$ for all real $s,t$. This involves simple
formal manipulations (as if $\Delta$ were a positive number) 
allowed by the functional calculus.
\item $\lim_{t\to 0}T_o(t)\phi=\phi$ for
all~$\phi\in\X_o$.   This is a straightforward calculation.
\end{itemize}

\subsection{The free sector}

On the other hand, the space~$\Phase_f$ consists of initial data where the
vector potential and electric field are harmonic; these are the states
relevant to the Aharonov--Bohm effect.  For modes $A\oplus E \in \Phase_f$,
Maxwell's equations become
$$
\left\{
\begin{array}{l}
\partial_t A = E \cr
\partial_t E = 0 \cr
\end{array}
\right.
$$
These are analogous to the equations of motion for a free particle on
the line, with~$A$ playing the role of position and~$E$ playing the
role of momentum.   This is why we call~$\Phase_f$ the phase space of
`free' modes. The Hamiltonian on this space is also analogous to the
kinetic energy of a free particle:
$$
H[A\oplus E]={1\over 2}(E\mid E).
$$
Solving the equations of motion, we see that time evolution 
acts on~$\Phase_f$ as follows:
\begin{equation}
\left( \! \begin{array}{c} A \\ E \end{array} \! \right)
\mapsto 
T_f(t) \left( \! \begin{array}{c} A \\ E \end{array} \! \right) =
\left( \begin{array}{cc} 
1 &  t    \\ 
0 &  1  
\end{array} \right)
\left( \! \begin{array}{c} A \\ E \end{array} \! \right)
\label{eq:time_evolution_free}
\end{equation}
The time evolution operators $T_f(t)$ form a norm-continuous group of
bounded operators on $\Phase_f$. Indeed:
\begin{itemize}
\item $1\le\|T_f(t)\|^2\le 2+t^2$, so $\|T_f(t)\|$ is finite for
all~$t$.
\item $T_f(t)T_f(s)=T_f(t+s)$ for all real $s,t$, trivially.
\item $\lim_{t\to 0}T_f(t)=V(0)$ in the norm topology, since 
it is easily seen that~$\|T_f(t)-T_f(0)\|=|t|$. 
\end{itemize}
A key ingredient in these calculations is that~$(\dd A|\dd A)=0$ identically
on~$\Phase_f$.

\section{Summary}

\label{sec:Kodaira}

In this section we summarize the mathematical and physical content of
the present chapter in two results. The first, Theorem~\ref{thm:3+1},
gathers all the important analysis results concerning the exterior
derivative operator on square-integrable differential forms on a
complete Riemannian manifold. The second, Result~\ref{thm:3+1hys} describes the phase space of
vacuum electromagnetism in~$3+1$ dimensions as a real Hilbert space
with a continuous quadratic and nonnegative Hamiltonian, and a
continuous symplectic structure.

We can combine into a single theorem Gaffney's
Proposition~\ref{prop:gaffney} about the operators~$d$ and~$\delta$ on
a complete Riemannian manifold and the version of the Kodaira
decomposition (Proposition~\ref{prop:kodaira}) which was essential to
the physical interpretation of the phase space of Maxwell's theory in
the preceding section:

\begin{theorem}\label{thm:3+1} 
Let~$S$ be a smooth manifold equipped with a complete Riemannian 
metric~$g$.  Then the formally adjoint operators
$$
\xymatrix{C^\infty_0\Omega^k_S\ar@<.5ex>[r]^{\dd_k} &
C^\infty_0\Omega^{k+1}_S\ar@<.5ex>[l]^{\dd^*_k}}
$$
have mutually adjoint closures
$$
\xymatrix{L^2\Omega^k_S\ar@<.5ex>[r]^{\dd_k} &
L^2\Omega^{k+1}_S\ar@<.5ex>[l]^{\dd^*_k}}.
$$
These closed operators satisfy
$$
\ran\dd_{k-1} \subseteq \ker\dd_k, \qquad
\ran \dd^*_k \subseteq \ker \dd^*_{k-1}
$$
and there is a Hilbert-space direct-sum decomposition
$$
L^2\Omega^k =
\overline{\ran d_{k-1}} 
\oplus
\ker \Delta_k
\oplus
\overline{\ran  \delta_k} .
$$
where the Laplacian on $k$-forms, 
$$
\Delta_k = \delta_k d_k + d_{k-1} \delta_{k-1} , 
$$
is a nonnegative densely defined self-adjoint operator on~$L^2
\Omega^k$.
\end{theorem}

\begin{proof}  

The properties of the operators~$\dd$ and~$\dd^*$ are the subject of
Section~\ref{sec:analysis}. We postpone proving the self-adjointness
of the Laplacian to Corollary~\ref{cor:essential} in the next
chapter. To prove the desired direct sum decomposition, we apply the
general Kodaira decomposition (Proposition~\ref{prop:kodaira} below) to
$$
\xymatrix{L^2\Omega^{k-1}\ar@<.5 ex>[r]^{\dd_{k-1}} & 
L^2\Omega^k\ar@<.5 ex>[l]^{\dd^*_{k-1}}\ar@<.5 ex>[r]^{\dd_k} & 
L^2\Omega^{k+1}\ar@<.5 ex>[l]^{\dd^*_k}} 
$$
and obtain 
$$
L^2\Omega^k =
\overline{\ran \dd_{k-1}} 
\oplus
\ker \Delta_k
\oplus
\overline{\ran\dd^*_k} .
$$
where
$$
\Delta_k = \dd^*_k\dd_k + \dd_{k-1} \dd^*_{k-1}
$$
is the Laplacian on $1$-forms.
\end{proof}

It remains only to prove the following general form of the Kodaira
decomposition, which is itself a generalization of the usual Hodge
decomposition for differential forms on a compact Riemannian manifold.

\begin{proposition}[Kodaira decomposition]\label{prop:kodaira}  
If
$$
\begin{CD}
{H}@>{S}>>{H'}@>{T}>>{H''}
\end{CD} 
$$ 
are densely defined closed operators and $\ran S\subseteq\ker T$, then
$$
H' = \overline{\ran T^*}\oplus\ker (T^*T+SS^*)\oplus\overline{\ran S}.
$$
\end{proposition}

\begin{proof}

We break the proof down into a series of lemmas. In the following
results and proofs, all the spaces we will consider will be Hilbert
spaces.  The proofs work equally well for real or complex Hilbert
spaces, but in our application they will be real.

\begin{lemma}\label{lem:ran_ker} 
If
$$
\begin{CD}
{H}@>{T}>>{H'}
\end{CD} 
$$
is a densely defined operator, then
$$
\ker T^*=(\ran T)^\perp
\qquad\hbox{and}\quad
\ker T=(\ran T^*)^\perp\cap\dom T.
$$
\end{lemma}

\begin{proof}  
Since $(\phi\mid T\psi)'=(T^*\phi\mid\psi)$ for all $\phi\in\dom T^*$
and $\psi\in\dom T$, it follows that $\ker T^*\perp\ran T$ and $\ker
T\perp\ran T^*$. Since~$T$ is densely defined, $(\ran
T)^\perp\subseteq\dom T^*$.
\end{proof}

The following lemma guarantees that the closed operators~$\dd$
and~$\dd^*$ satisfy
$$
\ran \dd_{k-1} \subseteq \ker \dd_k, \qquad
\ran \dd^*_k \subseteq \ker \dd_{k-1} .
$$

\begin{lemma}\label{lem:semiexact} 
If
$$
\begin{CD}
{H}@>{S}>>{H'}@>{T}>>{H''}
\end{CD} 
$$ 
are densely defined operators and $\ran S\subseteq\ker T$, then
$$
\ran T^*\subseteq\ker S^*.
$$
\end{lemma}

\begin{proof}  
Since $\ran S\subseteq\ker T$, for all $\phi\in\dom T^*$ and
$\psi\in\dom S$ we have
$$
(T^*\phi\mid S\psi)'=
(\phi\mid TS\psi)''=
(\phi\mid 0)''=
0=(0\mid\psi),
$$
so $\ran T^*\subseteq\ker S^*$.
\end{proof}

\begin{corollary}  
If 
$$
\begin{CD}
{H}@>{S}>>{H'}@>{T}>>{H''}
\end{CD} 
$$ 
are densely defined closable operators and $\ran S\subseteq\ker T$, then
$$
\ran\overline S\subseteq\ker\overline T.
$$
\end{corollary}

\begin{proof}  
By Lemma~\ref{lem:doubledual}, since~$S,T$ are closable,~$S^*$
and~$T^*$ are densely defined and $\overline S=S^{**}$ and $\overline
T=T^{**}$.  Then, $\ran S\subseteq\ker T$ implies $\ran
T^*\subseteq\ker S^*$, so $\ran S^{**}\subseteq\ker T^{**}$. 
\end{proof}

We are now ready to finish the proof of the Kodaira decomposition. The
hypotheses of Proposition~\ref{prop:kodaira} guarantee that~$S^*$
and~$T^*$ are densely defined closed operators and $T=T^{**}$ and
$S=S^{**}$ (Lemma~\ref{lem:doubledual}), so
$$
\ker T=
\ker T^{**}=
(\ran T^*)^\perp
\qquad\hbox{and}\quad
\ker S^*=(\ran S)^\perp.
$$
Then,
$$
H' =\ker T\oplus\overline{\ran T^*}=\ker S^*\oplus\overline{\ran S}
$$
which, together with the inclusions $\ran S\subseteq\ker T$ and $\ran
T^*\subseteq\ker S^*$ (Lemma~\ref{lem:semiexact}), implies
$$
H' =\overline{\ran T^*}\oplus(\ker T\cap\ker S^*)\oplus\overline{\ran S}.
$$
Finally, we know 
$$
(\ker T\cap\ker S^*)\subseteq
\bigl(\ker(T^* T)\cap\ker(SS^*)\bigr)\subseteq
\ker(T^*T+S^*S).
$$
The result then follows from $\ker(T^*T+S^*S)\subseteq(\ker T\cap\ker
S^*)$.  Assume $\psi\in\ker(T^* T+SS^*)$; then
$$
(\psi\mid 0)'
= (\psi\mid(T^*T+SS^*)\psi)'
= (\psi\mid T^*T\psi)' + (\psi\mid S^{**}S^*\psi)'
$$
so that
$$
0 = (T\psi\mid T\psi)'' + (S^*\psi\mid S^*\psi),
$$
which implies $\psi\in\ker T\cap \ker S^*$. 
\end{proof}

We end this chapter with a `physical theorem' gathering all the results of
physical interest about the phase space of electromagnetism 
that we proved 
in
this chapter.

\begin{result}
\label{thm:3+1phys}
Let~$M$ be a $(3+1)$-dimensional static, globally hyperbolic
spacetime, with metric
$$
g_M=e^{2\Phi}(-\dd t^2+g).
$$
Then, electromagnetism on~$M$ with gauge group~$\R$ has as its phase
space the real Hilbert space
$$
\Phase={\dom\{\dd\maps L^2\Omega^1_S\to
  L^2\Omega^2_S\}\over\overline{\ran}\{\dd\maps L^2\Omega^0_S\to
  L^2\Omega^1_S\}}\oplus\ker\{\dd^*\from L^2\Omega^1_S\to
L^2\Omega^0_S\},
$$
with continuous symplectic structure
$$
\omega(X,X')=(E,A')-(E',A)
$$
where~$X=[A]\oplus E$ and~$X'=[A']\oplus E'$ lie in~$\Phase$, and
$$
(\alpha,\beta)=\int_S g(\alpha,\beta)\vol
$$ 
is the canonical inner product induced on~$\Omega^k_S$ by the
optical metric~$g$ on~$S$. The Hamiltonian is the continuous quadratic
form 
$$
H[X]={1\over 2}\bigl[(E,E)+(\dd A,\dd A)\bigr]. 
$$
There phase space splits naturally into two sectors,
$$
\Phase=\Phase_o\oplus\Phase_f,
$$
and the direct summands
$$
\Phase_f=\Phase\cap\ker\Delta
\qquad\hbox{and}\quad
\Phase_o=\Phase\cap\ran\dd^*_1
$$
are preserved by time evolution.
On~$\Phase_o$, time evolution takes the form
$$
\left( \! \begin{array}{c} A \\ E \end{array} \! \right)
\mapsto T_o(t) 
\left( \! \begin{array}{c} A \\ E \end{array} \! \right) = 
\left( \begin{array}{cc} 
\cos(t\sqrt{\Delta}) & \sin(t\sqrt{\Delta})\,/\,\sqrt{\Delta}\\ 
-\sqrt\Delta\,\sin(t\sqrt{\Delta}) & \cos(t\sqrt{\Delta})  
\end{array} \right) 
\left( \! \begin{array}{c} A \\ E \end{array} \! \right)
$$
while on~$\Phase_f$ it takes the form
$$
\left( \! \begin{array}{c} A \\ E \end{array} \! \right)
\mapsto 
T_f(t) \left( \! \begin{array}{c} A \\ E \end{array} \! \right) =
\left( \begin{array}{cc} 
1 &  t    \\ 
0 &  1  
\end{array} \right)
\left( \! \begin{array}{c} A \\ E \end{array} \! \right).
$$
\end{result}

\chapter{$p$-form electromagnetism in~$n+1$ dimensions} 
\label{sec:N+1}

In this chapter we generalize the results of the last chapter to electromagnetism on spacetimes of arbitrary dimension~$n+1$.
As before, we take our spacetime to be of the form~$\R \times S$,
equipped with the Lorentzian metric
$$
g_M = e^{2\Phi}(-dt^2 + g)
$$
where~$g$ is a complete Riemannian metric on~$S$.  The only difference
is that now $S$ is $n$-dimensional instead of 3-dimensional.  However,
this means that Maxwell's equations are no longer conformally
invariant, so the `gravitational potential'~$\Phi$ plays a more
significant role. To see why, recall that the Maxwell action is still
given by
$$
{\Act}[A_M] = -{1\over 2}\int_M g_M(F_M , F_M )\, \vol_M
$$
where
$$
F_M = \dd t\wedge(\partial_t A - \dd A_0) + \dd A.
$$
By Equation~(\ref{eq:stpart2}), 
$$
g_M(F_M,F_M)= 
e^{-4\Phi} \, \bigl[ -g(\partial_t A - dA_0,\partial_t A - dA_0)
                    + g(dA,dA) \bigr]
$$
and, by Equation~(\ref{eq:vol}),
$$
\vol_M = e^{(n+1)\Phi} \vol \wedge dt 
$$
where~$\vol$ is the volume form on space. Hence, we have
$$
{\Act}[A,A_0] = {1\over 2}\int_M 
\bigl [g(\partial_t A - \dd A_0,\partial_t A - \dd A_0) -
g(\dd A,\dd A) \bigr]\, e^{(n-3)\Phi} \vol \wedge dt .
$$ 
The factors involving~$\Phi$ cancel only if $n = 3$, indicating
conformal invariance. In other dimensions, the most elegant way to
deal with these factors involving~$\Phi$ is to \emph{redefine} the
fields~$A_0$ and~$A$ by setting
$$
A_M = e^{-{1\over 2}(n-3)\Phi} (\dd t \wedge A_0+A), 
$$
and then to `twist' the exterior derivative of differential forms on
space, defining a new operator
$$
D = e^{-{1\over 2}(n-3)\Phi} \, \dd \, e^{{1\over 2}(n-3)\Phi} .
$$
The action is then
$$
{\Act}[A,A_0] = {1\over 2}\int_M 
\bigl [g(\partial_t A - DA_0,\partial_t A - DA_0) -
g(DA,DA) \bigr]  \vol \wedge \dd t 
$$ 
which is formally just like equation~(\ref{eq:action}) was in the
$(3+1)$-dimensional case, but with rescaled fields~$A$ and~$A_0$, and
with the twisted exterior derivative~$D$ replacing the usual~$\dd$.

With the help of this formal equivalence, the whole theory goes
through almost exactly as before.  In particular, if we let $L^2
\Omega^p$ be the Hilbert space consisting of all square-integrable
$p$-forms on $S$, there are mutually adjoint operators
$$
\xymatrix{L^2\Omega^0\ar@<.5 ex>[r]^{D_0} & 
L^2\Omega^1\ar@<.5 ex>[l]^{D_0^*}\ar@<.5 ex>[r]^{D_1} & 
L^2\Omega^2\ar@<.5 ex>[l]^{D_1^*}} 
$$
Using the Kodaira decomposition for this sequence we obtain
$$
L^2\Omega^1=
\overline{\ran D_0} 
\oplus
\ker L_1
\oplus
\overline{\ran D_1^*}
$$
where now the Laplacian is replaced by the `twisted Laplacian' $L_1$,
a nonnegative self-adjoint operator on 1-forms given by 
$$
L_1 = D_1^* D_1 + D_0 D_0^*.
$$

In fact, having come this far, it would be a pity not to treat
`$p$-form electromagnetism', a generalization of Maxwell's equations
in which the electromagnetic vector potential is replaced by a
$p$-form. The general case was treated by Henneaux and
Teitelboim~\cite{henneaux86}. For $p = 0$, this theory is just the
massless neutral scalar field.  For $p = 2$, it is the Kalb--Ramond
field arising naturally in string theory~\cite[Section
3.4.5]{green87}\cite{kalb74}, while for $p = 3$ it plays a part in
11-dimensional supergravity \cite{duff99}. All our formulas generalize
painlessly to these theories in the absence of charges. Starting
with the $p$-form~$A_M$ on spacetime, we define a field strength
tensor $F_M = \dd_M A_M$, and take the action of the theory to be
$$
{\Act}[A_M] = -{1\over 2}\int_M g_M(F_M , F_M )\, \vol_M .
$$
This action gives equations of motion and gauge symmetries having
the same form as in Maxwell theory.  Furthermore, if we set
$$
A_M = e^{-{1\over 2}(n-2p-1)\Phi} (\dd t \wedge A_0+A)
$$
and define the twisted differential~$D$ as follows:
$$
D = e^{{1\over 2}(n-2p-1)\Phi} \dd e^{-{1\over 2}(n-2p-1)\Phi} ,
$$
we obtain
$$
{\Act}[A,A_0] = {1\over 2}\int_{\R} 
\bigl [(\partial_t A - DA_0,\partial_t A - DA_0) -
(DA,DA) \bigr]\dd t 
$$ 
in complete analogy with ordinary Maxwell theory.  This allows us to
generalize all our results on Maxwell's equations to the $p$-form
case.

\section{Spacetime Geometry}
\label{sec:N+1-geometry}

We model spacetime as an $(n+1)$-dimensional smooth manifold~$M$ with
a Lorentzian metric of signature $(-+\cdots +)$. We assume that $M =
\R\times S$ for some smooth manifold~$S$, and that the metric on~$M$
is of the form
$$
g_M=-e^{2\Phi} dt^2 +g_S 
$$ 
where~$g_S$ is a Riemannian metric on~$S$ and~$\Phi$ is a smooth
real-valued function on~$S$. As in Equation~(\ref{eq:optical}) we
write
$$
g_M=e^{2\Phi}(-dt^2 + g)
$$
where the `optical metric'~$g$ is given by $g = e^{-2\Phi} g_S$.  We
assume that~$g$ makes~$S$ into a complete Riemannian manifold, since
this is a necessary and sufficient condition for~$M$ to be a globally
hyperbolic spacetime with the surfaces $\{t = c\}$ as Cauchy surfaces.
For a more complete discussion, refer back to
Section~\ref{sec:3+1-geometry}.  As before, all fields on spacetime
carry the subscript~`$M$'; fields on space are written without subscript
or with the subscript~$0$.  To study $p$-form electromagnetism we need
to fix an integer~$p$ with $0 \le p \le n$. Then, any
$k$-form~$\alpha_M$ on~$M$ can be uniquely decomposed as
$$
\alpha_M = e^{-{1\over 2}(n-2p-1)\Phi}  (\dd t \wedge \alpha_0+ \alpha)
$$
where~$\alpha$ is a time-dependent $k$-form on~$S$ and~$\alpha_0$ a
time-dependent $(k-1)$-form on~$S$.  As explained in the previous
section, the strange-looking factor involving~$\Phi$ is chosen to
simplify things later.

The metric~$g_M$ induces a metric on the $k$-forms on spacetime, which
we also call~$g_M$, and similarly for the metric~$g$ on space.  In terms
of spatial and temporal parts, these are related by:
\begin{equation}
g_M(\alpha_M,\alpha_M')
=e^{-(n+2k-2p-1) \Phi} \left[ -g(\alpha_0,\beta_0)+ g(\alpha,\beta) \right].
\label{eq:newinner}
\end{equation}

Assuming that~$S$ is oriented, the metrics~$g_M$ and~$g$ determine
volume forms~$\vol_M$ on~$M$ and~$\vol$ on~$S$, which are related by
\begin{equation}
\vol_M = e^{(n+1) \Phi} \vol \wedge \dd t .
\label{eq:newvol}
\end{equation}
Again, it would be possible to deal with the nonorientable case by
working with densities instead of forms.  As before, we define an
inner product $(\cdot \mid \cdot)$ on $k$-forms on space by
Equation~(\ref{eq:inner}), namely
$$
(\alpha,\beta) = \int_S g(\alpha,\beta)\,\vol,
$$
and define~$L^2 \Omega^k$ to be the space of measurable
$k$-forms~$\alpha$ on~$S$ such that $(\alpha \mid \alpha) < \infty$.

We define the twisted exterior derivative $D_k \maps C_0^\infty \Omega^k_S
\to C_0^\infty \Omega^{k+1}_S$ by
\begin{equation}
\label{Dk}
D_k = e^{{1\over 2}(n-2p-1)\Phi} \dd_k e^{-{1\over 2}(n-2p-1)\Phi} .
\end{equation}
This operator has a formal adjoint
\begin{equation}
\label{Dkdagger}
D_k^\dagger = e^{-{1\over 2}(n-2p-1)\Phi} \delta_{k+1} e^{{1\over
2}(n-2p-1)\Phi}  
\end{equation}
meaning that 
\begin{equation}
(D_k^\dagger \alpha,\beta) = (\alpha, D_k \beta)
\label{eq:formal_adjoint_n+1} 
\end{equation}
whenever $\alpha \in C_0^\infty \Omega^{k+1}$ and
$\beta \in C_0^\infty \Omega^k$.  In what follows we shall
omit the subscript~`$k$' from the operators~$D_k$ and~$D_k^\dagger$
when it is clear from context.

In Section~(\ref{sec:N+1-theorem}) we shall show that these operators have
mutually adjoint closures~$\overline{D_k}\from L^2\Omega^k_S\to
L^2\Omega^{k+1}_S$ and~$D_k^*\from L^2\Omega^{k+1}_S\to\Omega^k_S$,
and that the operators~$D_kD_k^\dagger$ and~$D_k^\dagger D_k$ are both
essentially self-adjoint, meaning that their respective
closures,~$\overline{D_k}D_k^*$ and~$D_k^*\overline{D_k}$, are their
unique self-adjoint extensions~\cite[\S VIII.2]{RS}.

\section{$p$-Form electromagnetism}
\label{sec:N+1-maxwell}

In $p$-form electromagnetism we take the vector potential as a $p$-form
on spacetime,~$A_M$, and take the action to be
$$
{\Act}[A_M] = -{1\over 2}\int_M g_M(F_M,F_M)\, \vol_M
$$
where the field strength tensor~$F_M$ is given by
$$
F_M=\dd_MA_M
$$
In terms of the twisted exterior derivative defined in
Equation~(\ref{Dk}), the field strength tensor equals
$$
\begin{array}{ccl}
F_M &=& (dt \wedge \partial_t + d) A_M   \\
    &=& (dt \wedge \partial_t + d) e^{-{1\over 2}(n-2p-1)\Phi} 
        (dt \wedge A_0 + A)  \\
    &=&  e^{-{1\over 2}(n-2p-1)\Phi}(dt \wedge \partial_t + D)
        (dt \wedge A_0 + A)  \\
    &=&  e^{-{1\over 2}(n-2p-1)\Phi}
        \bigl[dt \wedge (\partial_t A - DA_0) + DA\bigr] .
\end{array}
$$
With the help of equations~(\ref{eq:newinner})--(\ref{eq:newvol}), 
this means that the action can be written as
\begin{eqnarray}
\label{eq:action_n+1}
{\Act} &=& {1\over 2}\int_\R \int_M 
 \bigl[ g(\partial_t A - DA_0, \partial_t A - DA_0) - 
 g(DA,DA) \bigr] \vol_M \nonumber\\
&& \nonumber\\
&=&\displaystyle
{1\over 2}\int_\R
\bigl[(\partial_t A - DA_0, \partial_t A - DA_0) -
(DA, DA)\bigr] \, dt .
\end{eqnarray}
Note the complete analogy with Equation~(\ref{eq:action}). This
action gives the following equations of motion:
$$
\left\{
\begin{array}{rcl}
\partial_t DA & = & D A_0 \\
\partial_t^2A &=& - D^\dagger D A + \partial_t DA_0. \\
\end{array}
\right.
$$

The equations of $p$-form electromagnetism admit gauge symmetries of
the form 
$$
A_M \mapsto A_M + \dd_M \beta_M
$$ 
where~$\beta_M$ is a $(p-1)$-form on spacetime. Thus, to obtain
evolution equations, we work in temporal gauge, which amounts to
setting $A_0 = 0$.  The above equations can then be written as
$$
\left\{
\begin{array}{rcc}
D^\dagger E & = & 0 \\
\partial_t A & = & E \\
\partial_t E & = & - D^\dagger D A. \\
\end{array}
\right.
$$
The Gauss law constraint $D^\dagger E = 0$ generates gauge
transformations of the form
$$ 
A \mapsto A + D \beta 
$$ 
where~$\beta$ is a $(p-1)$-form on spacetime.  Two pairs~$A\oplus E$
are physically equivalent if they differ by such a transformation.
Thus, ignoring analytical subtleties, the phase space of $p$-form
electromagnetism consists of pairs~$[A]\oplus E$ where~$[A]$ is an
equivalence class of $p$-forms on~$S$ modulo those of the
form~$D\beta$ (twisted-exact), and~$E$ is a $p$-form on~$S$
satisfying~$D^\dagger E = 0$ (twisted-divergenceless).  The
Hamiltonian on this phase space is easily seen to be
$$
H\bigl[[A]\oplus E\bigr]={1\over 2}\bigl[(DA, DA) +(E, E) \bigr] .
$$
and the symplectic structure is
$$
\omega\bigl[[A]\oplus E,[A']\oplus E'\bigr]= (A, E') -(E, A') .
$$
Again as in the case of~$3+1$ dimensions, $(D\beta, E)=(\beta,
D^\dagger E)=0$ implies that the symplectic structure is
gauge-invariant.

All these formulas have analogues in Section \ref{sec:3+1}, so to
generalize all the results of that section we only need to generalize
Theorem \ref{thm:3+1} to the present context.  In other words, first
we must show that the operators
$$
\xymatrix{C^\infty_0\Omega^k\ar@<.5ex>[r]^{D} &
C^\infty_0\Omega^{k+1}\ar@<.5ex>[l]^{D^\dagger}}
$$
have mutually adjoint closures, which we write as
$$
\xymatrix{L^2 \Omega^k\ar@<.5ex>[r]^{D} &
L^2 \Omega^{k+1}\ar@<.5ex>[l]^{D^*}}.  
$$
Then we must prove a version of the Kodaira decomposition
saying that
$$
L^2\Omega^p =
\overline{\ran D_{p-1}} 
\oplus
\ker L_p 
\oplus
\overline{\ran D_p^*} 
$$
where the twisted Laplacian on $k$-forms,
$$
L_p = D_p^* D_p + D_{p-1} D_{p-1}^* , 
$$
is a nonnegative self-adjoint operator on~$L^2 \Omega^p$.  We do all
this in Section~\ref{sec:N+1-theorem} below.

Using these facts, we define the classical phase space for
$p$-form electromagnetism to be
$$
\Phase  = \A \oplus \e
$$
where
$$
\begin{array}{ccl}
\A &=& \dom D_p / \, \overline{\ran D_{p-1}}   \\
\e &=& \ker D_p .
\end{array}
$$
As with the Maxwell theory in~$3+1$ dimensions,~$\Phase$ becomes a real
Hilbert space space if we define
$$
\|[A]\oplus E\|^2=
( [A], [A] ) + ( d A, d A) + (E \mid E), 
$$
where~$([A],[A]')$ can be defined on gauge equivalence classes using
the fact that, by the Kodaira decomposition,~$L^2\Omega^p /
\overline{\ran D_{p-1}}$ is canonically isomorphic to~$\ran
D_p^\perp$, which inherits an inner product by virtue of being a
subspace of~$L^2\Omega^p$.

As before, we can split the spaces~$\A$ and~$\e$ 
into `oscillating' and `free' parts:
$$
\begin{array}{ccl}
\A &=& \A_o \oplus \A_f \\
\e &=& \e_o \oplus \e_f ,
\end{array}
$$
where 
$$ 
\begin{array}{lclllcl}
\A_o &=& \dom D_p \cap \overline{\ran D_p^* }  
&\quad& 
\A_f &=&\ker L_p         \\
&&&&&& \\
\e_o &=& \overline{\ran D_p^*}  
&\quad&
\e_f &=& \ker L_p .
\end{array}
$$

These decompositions let us write the classical phase space
as a direct sum of real Hilbert spaces:
$$ 
\Phase = \Phase_o \oplus \Phase_f , 
$$
where
$$
\begin{array}{ccl}
\Phase_o &=& \A_o \oplus \e_o   \\
\Phase_f &=& \A_f \oplus \e_f    .
\end{array}
$$
This is also a direct sum of symplectic vector spaces, and the
Hamiltonian is a sum of separate Hamiltonians on~$\Phase_o$ and~$\Phase_f$.
As a result, time evolution acts by symplectic transformations,
independently on the oscillating and free parts of any initial
data~$[A]\oplus E \in \Phase$.

For modes~$[A]\oplus E \in \Phase_o$, the Hamiltonian resembles that
of a harmonic oscillator:
$$
H[A\oplus E] = {1\over 2} \bigl[(DA, DA)+(E, E)\bigr]
$$
and the equations of motion are
$$
\left\{
\begin{array}{l}
\partial_t A = E \cr 
\partial_t E =- LA, \cr
\end{array}
\right. 
$$
where we write the twisted Laplacian~$L_p$ simply as~$L$.
The solutions of the corresponding integral equation are given by
\begin{equation}
\left( \begin{array}{c} A \\ E \end{array} \right)
\mapsto 
\left( \begin{array}{cc} 
\cos(t \sqrt{L} ) &   
\sin(t \sqrt{L}) \, / \,\sqrt{L}    \\ 
-\sqrt{L} \,\sin(t\sqrt{L}) &
\cos(t \sqrt{L})  
\end{array} \right) 
\left( \begin{array}{c} A \\ E \end{array} \right)
\label{eq:time_evolution_twisted}
\end{equation}
where we use the functional calculus to define functions of~$L$. The
proof that this is a strongly continuous~$1$-parameter group of
bounded operators is essentially the same as the one sketched after
Equation~(\ref{eq:time_evolution}).

For modes~$[A]\oplus E \in \Phase_f$, the Hamiltonian resembles that of a
free particle:
$$
H\bigl[[A]\oplus E\bigr] = {1\over 2}(E,E)
$$
and the equations of motion are 
$$
\left\{
\begin{array}{l}
\partial_t A = E \cr
\partial_t E = 0. \cr
\end{array}
\right.
$$
The solutions of the equations of motion are given by
$$
\left( \begin{array}{c} A \\ E \end{array} \right)
\mapsto 
\left( \begin{array}{cc} 
1 &  t    \\ 
0 &  1  
\end{array} \right)
\left( \begin{array}{c} A \\ E \end{array} \right)
$$
Note that, in the case of free modes, nothing besides the
definition of the Laplacian has changed from the case of~$1$-forms in
$3+1$ dimensions. In particular, time evolution is given by the very same
Equation~(\ref{eq:time_evolution_free}).

\section{Mathematical details}
\label{sec:N+1-theorem}

The results we need to make our work in the previous section 
rigorous are all contained in this theorem:

\begin{theorem}\label{thm:N+1}  
Let $S$ be a smooth $n$-dimensional manifold equipped with a complete
Riemannian  metric $g$, and let $\Phi$ be a smooth real-valued function
on $S$.  Fix an integer $0 \le p \le n$.  Then for any integer $k$, the
operators
$$
\xymatrix{C^\infty_0\Omega^k_S\ar@<.5ex>[r]^{D_k} &
C^\infty_0\Omega_S^{k+1}\ar@<.5ex>[l]^{D_k^\dagger}}
$$
defined in equations (\ref{Dk}) and (\ref{Dkdagger})
have mutually adjoint closures, which we write as
$$
\xymatrix{L^2\Omega^k_S\ar@<.5ex>[r]^{D_k} &
L^2\Omega^{k+1}_S\ar@<.5ex>[l]^{D_k^*}}
$$
These closures satisfy 
$$
\ran D_{k-1} \subseteq \ker D_k, \qquad 
\ran D_k^* \subseteq \ker D_{k-1}^* , 
$$
and we obtain a direct sum decomposition
$$
L^2\Omega^k =
\overline{\ran D_{k-1}} 
\oplus
\ker L_k 
\oplus
\overline{\ran D_k^*} .
$$
where the twisted Laplacian on $k$-forms, 
$$
L_k = D_k^* D_k + D_{k-1} D_{k-1}^* , 
$$
is a nonnegative densely defined self-adjoint operator on~$L^2 \Omega^k$.
\end{theorem}

\begin{proof} Because of the twisting of the exterior derivative
operator in Equation~(\ref{Dk}), one cannot simply apply the proof of
Theorem~\ref{thm:3+1}. The reason is that Gaffney's
Proposition~\ref{prop:gaffney} depends on the specific properties of
the `untwisted'~$\dd$ and~$\delta$.  However, the generalization is in fact
true, essentially because~$e^f \dd e^{-f}$ and~$\dd$ have the
same first-order part whenever $f$ is a smooth function.
This is made precise by an argument 
due to Chernoff, which uses the concept of the `symbol' of a differential
operator.  This argument implies both a generalization of
Proposition~\ref{prop:gaffney} and the self-adjointness of the twisted
Laplacian.

We begin by recalling Chernoff's formalism~\cite{chernoff73}, which is
the key to proving this theorem. Let~$S$ be a Riemannian manifold with
metric~$g$, and let~$E$ be any vector bundle on~$S$ whose fiber at
each point $x \in S$ is equipped with an inner
product~$\langle\cdot,\cdot\rangle_x$ depending smoothly on~$x$.  The
space of smooth compactly supported sections of this vector bundle,
denoted~$C^\infty_0 E$, is given an inner product
$$
(\alpha\mid\beta) = \int_S \langle\alpha(x),\beta(x)\rangle_x \; \vol_S,
$$
where~$\vol_S$ is the canonical volume form on~$S$.  The Hilbert space
completion of~$C^\infty_0 E$ with respect to this inner product is
denoted~$L^2 E$.

Assume that $T \maps C^\infty_0 E \to C^\infty_0 E$ is a first-order
linear differential operator on~$E$.  Its formal adjoint~$T^\dagger$ is
again a first-order linear differential operator, defined by requiring
that
$$
(\alpha\mid T\beta)=(T^\dagger\alpha\mid\beta)
\qquad\hbox{for all}\quad
\alpha,\beta \in C^\infty_0 E.
$$  
The `symbol' of~$T$ is defined by
$$
\sigma(\dd f,\alpha)= T(f \alpha)- fT\alpha
$$
for any~$C^\infty_0$ function~$f$ and any~$\alpha \in C^\infty_0 E$. 
Note that~$\sigma(\dd f,\alpha)$ is a function on~$S$ whose value at any 
point depends only on the values of~$\dd f$ and~$\alpha$ at that point.

If~$T+T^\dagger$ is equal to multiplication by a smooth function, we say
the differential equation $\partial_t \alpha = T\alpha$ is a `symmetric
hyperbolic system'.  At any point~$x \in S$, solutions of this
equation propagate at the speed
$$
c(x) =
\sup\bigl\{\|\sigma(\dd f,\alpha)\|_x \colon \; \|\dd f\|_x=\|\alpha\|_x=1\bigr\}
$$
where~$\|\dd f\|_x$ is the norm of~$\dd f$ at the point~$x$, defined using
the Riemannian metric~$g$, and~$\|\alpha\|_x$ is the norm of~$\alpha$
at the point~$x$, defined using the inner product on the fiber of~$E$
at~$x$.

Chernoff then essentially proves the following theorem. Note that the
Hilbert spaces appearing in this theorem are complex, so to apply it
to our real Hilbert spaces we need to complexify them.

\begin{lemma}[Chernoff] \label{lem:chernoff} 
If the metric~$c^{-2}g$ makes~$S$ into a complete Riemannian manifold,
the symmetric hyperbolic system $\partial_t \alpha =T \alpha$ with
initial data in~$C^\infty_0 E$ has a unique solution on~$\R\times S$
which is in~$C^\infty_0 E$ for all $t \in \R$.  Moreover, if~$T$ is
formally skew-adjoint ($T+T^\dagger=0$), then~$-iT$ and all its powers
are essentially self-adjoint on~$C_0^\infty E$.
\end{lemma}

\begin{proof}[Sketch of proof] The basic idea is that when we solve the
differential equation $\partial_t \alpha = T\alpha$, perturbations
propagate at speed~$1$ with respect to the metric~$c^{-2}g$.  If this
metric is complete, information can never reach spacelike infinity in
a finite amount of time.  Thus, given compactly supported smooth
initial data, the equation $\partial_t \alpha = T\alpha$ has a
solution~$\alpha(t,x)$ such that~$\alpha(t,\cdot)$ is compactly
supported for all~$t$---and smooth, by general results on hyperbolic
systems.

If~$T$ is formally skew-adjoint, one can show that the inner product
of two solutions is constant as a function of time:
$$
\begin{array}{ccl}
{d\over dt} (\alpha(t,\cdot) \mid \beta(t,\cdot)) &=&
(T \alpha(t,\cdot) \mid \beta(t,\cdot)) +
(\alpha(t,\cdot) \mid T \beta(t,\cdot)) \\
&=& (\alpha(t,\cdot) \mid T^\dagger \beta(t,\cdot)) +
(\alpha(t,\cdot) \mid T \beta(t,\cdot)) \\
&=& 0 .
\end{array}
$$
The crucial point here is that~$\alpha(t,\cdot)$ and~$\beta(t,\cdot)$
are compactly supported for all~$t$, so there are no boundary terms:
we only need the fact that~$T$ and~$T^\dagger$ are formal adjoints.

It follows that time evolution defines a one-parameter group of
inner-product-preserving transformations of~$C_0^\infty E$, which by
density extends uniquely to a one-parameter unitary group~$U(t)$
on~$L^2 E$.  One can show that~$C_0^\infty E$ forms a `dense invariant
subspace of $C^\infty$ vectors' for~$U(t)$; in other words,
that~$C_0^\infty E$ is a dense subspace of~$L^2 E$, and that given
initial data~$\alpha$ in this subspace, the solution~$U(t) \alpha$
remains in this subspace for all times, defining an infinitely
differentiable function from~$\R$ to~$L^2 E$. By a theorem of
Nelson~\cite[Lemma 10.1]{nelson59}, this implies that~$-iT$ and all
its powers are essentially self-adjoint on the domain~$C_0^\infty E$,
and that the closure of~$-iT$ generates the one-parameter
group~$U(t)$.  The only new thing to check here is the existence of
the derivatives~${d^n \over dt^n} U(t) \alpha$, which one can show by
repeatedly using the differential equation ${d\over dt} U(t) \alpha =
-iT U(t)\alpha$.
\end{proof}

This result applies without modification to first-order differential
equations like the Dirac equation.  To apply it to our problem, we
resort to a well-known trick, taking $-iT$ to be the operator
$$
\left( \begin{array}{cc} 
0 & D_k^\dagger    \\ 
D_k & 0  
\end{array} \right) .
$$
The essential self-adjointness of this operator will imply that
$D_k$ and $D_k^\dagger$ have mutually adjoint closures:

\begin{lemma}\label{lem:direct_sum}  
Let $H_1$ and $H_2$ be Hilbert spaces and let 
$$
\xymatrix{H_1 \ar@<.5ex>[r]^{A} &
H_2 \ar@<.5ex>[l]^{B}}
$$
be densely defined operators that are formal adjoints of one another:
$$
\langle A \phi, \psi \rangle_1 = 
\langle \phi , B \psi \rangle_2  
\qquad\hbox{for all}\quad
\phi \in \dom A , \psi \in \dom B.
$$
Let $H = H_1 \oplus H_2$ and let $S$ be the densely
defined operator
$$
\left( \begin{array}{cc} 
0 & B    \\ 
A & 0  
\end{array} \right)
$$
on $H$.  If $S$ is essentially self-adjoint, then $A$ and
$B$ have mutually adjoint closures. 
\end{lemma}

\begin{proof} 
It is easy to verify that the closure of $S$ is
$$
\left( \begin{array}{cc} 
0 &  \overline{B}    \\ 
\overline{A} &  0  
\end{array} \right)
$$
while the adjoint of the closure of $S$ is
$$
\left( \begin{array}{cc} 
0 &  (\overline{A})^*    \\ 
(\overline{B})^* &  0  
\end{array} \right)  .
$$
If $S$ is essentially self-adjoint, these two
operators are equal.  This implies that 
$$  
(\overline{A})^* = \overline{B}  
$$
and 
$$ 
(\overline{B})^* = \overline{A}  
$$
so the closures of~$A$ and~$B$ are mutually adjoint.  
\end{proof}

\begin{lemma}\label{lem:essential}  
Suppose $S$ is a complete Riemannian manifold and $\Phi$ 
a smooth real-valued function on $S$.   Let
$$
T \maps L^2 \Omega^k_S \oplus L^2 \Omega^{k+1}_S \to
        L^2 \Omega^k_S \oplus L^2 \Omega^{k+1}_S 
$$
be the densely defined operator 
$$
\left( \begin{array}{cc} 
0 & iD_k^\dagger    \\ 
iD_k & 0  
\end{array} \right) .
$$
Then $-iT$ and all its powers are essentially self-adjoint on
$C^\infty_0\Omega^k\oplus C^\infty_0\Omega^{k+1}$.
\end{lemma}

\begin{proof}  
We show that the hypotheses of Lemma \ref{lem:chernoff} apply to the
operator $T$.  Clearly $T$ is formally skew-adjoint, so it suffices to
check that the equation $\partial_t \alpha = T\alpha$ has propagation
speed $c = 1$. 

First we consider the case where $\Phi = 0$, so $D = \dd$ and
$D^\dagger = \delta$.  The symbol of the operator $\dd$ is
$$
\sigma_\dd(\dd f,\alpha)=
\bigl(\dd(f\alpha)- f \dd\alpha\bigr)= \dd f \wedge\alpha
$$ 
for any $\alpha \in C_0^\infty \Omega^p_S$.  The symbol of $\delta$ is 
$$
\sigma_\delta(\dd f,\beta)= -i_{\dd f} \beta
\qquad\hbox{for any}\quad
\beta \in C_0^\infty \Omega^{p+1}_S,
$$
since
$$
\bigl( \sigma_\delta(\dd f,\beta), \gamma \bigr) =
\bigl( \delta(f\beta) - f \delta \beta, \gamma \bigr) =
-\bigl( \beta, \dd(f\gamma) - f \dd\gamma \bigr) =
-\bigl( \beta, \dd f \wedge\gamma \bigr) =
-\bigl( i_{\dd f} \beta, \gamma \bigr)
$$
for any $\gamma \in C_0^\infty \Omega^p_S$.
It follows that the symbol of $T$ is
$$
\sigma_T(\dd f,\alpha \oplus\beta)= i(i_{\dd h}\beta \oplus \dd f\wedge\alpha).
$$
To compute the propagation speed, note first that
$$
\begin{array}{ccl}
\|\sigma(\dd f,\alpha\oplus\beta)\|_x^2 
&=& \|\dd f\wedge\alpha\|_x^2 + \|i_{\dd f}\beta \|_x^2 \\
&\le& \|\dd f\|^2 \left(\|\alpha\|_x^2 + \|\beta \|_x^2 \right) \\
&=& \|\dd f\|^2 \left(\|\alpha \oplus \beta \|_x^2 \right) 
\end{array}
$$
so the propagation speed is $\le 1$.  In fact the propagation speed is 
exactly $1$, since equality is achieved by letting $\dd f=\dd x_1$, 
$\alpha = \dd x_2\wedge\cdots \wedge \dd x_{k+1}$, and $\beta = 0$
near $x$, where $\dd x_1,\ldots,\dd x_n$ is a coordinate frame orthogonal 
at $x$.  

To deal with the general case where $\Phi$ is nonzero, note that for
any first-order linear differential operator $X$ and any smooth
real-valued function~$h$, the operator $e^hXe^{-h}$ has the same
symbol as~$X$.  In particular, the operators~$\dd$ and~$D$ have the same
symbol, as do~$\delta$ and~$D^\dagger$.  It follows that~$T$ always
has the same symbol as it does in the special case where~$\Phi = 0$,
so the propagation speed is always~$1$. 
\end{proof}

\begin{corollary} \label{cor:essential}  
Under the same hypothesis as Lemma \ref{lem:essential},
the operators
$$
\xymatrix{C^\infty_0\Omega^k_S\ar@<.5ex>[r]^{D_k} &
C^\infty_0\Omega^{k+1}_S\ar@<.5ex>[l]^{D_k^\dagger}} 
$$
have mutually adjoint closures, and the operators~$D_k^\dagger D_k$
and~$D_k D_{k-1}^\dagger$ are essentially self-adjoint
on~$C_0^\infty \Omega^k$.
\end{corollary}

\begin{proof}  The first part follows immediately from Lemmas
\ref{lem:direct_sum} and \ref{lem:essential}. For the second part, note
by Lemma~\ref{lem:essential} that $T^2 = D_k^\dagger D_k \oplus D_k
D_k^\dagger$ is essentially self-adjoint on~$C^\infty_0\Omega^k \oplus
C^\infty_0\Omega^{k+1}$.   This implies that~$D_k^\dagger D_k$ and~$D_k
D_k^\dagger$ are essentially self-adjoint.    
\end{proof}

We can now complete the proof of Theorem~\ref{thm:N+1}.

If we now use~$D_k$ and~$D_k^*$ to stand for the mutually adjoint
closures of the operators~$D_k$ and~$D_k^\dagger$,
Lemma~(\ref{lem:semiexact}) implies that 
$$
\ran D_{k-1} \subseteq \ker D_k, \qquad 
\ran D_k^* \subseteq \ker D_{k-1}^* , 
$$
so we can apply the Kodaira decomposition (Proposition~\ref{prop:kodaira})
to see that
$$
L^2\Omega^k =
\overline{\ran D_{k-1}} 
\oplus
\ker L_k 
\oplus
\overline{\ran D_k^*} .
$$
where 
$$
L_k = D_k^* D_k + D_{k-1} D_{k-1}^* .
$$

To conclude we only need to show that~$L_k$ is a non-negative
self-adjoint operator.  With respect to the Kodaira decomposition this
operator takes the block diagonal form
$$
\left( \begin{array}{ccc} 
D_{k-1} D_{k-1}^* & 0 & 0   \\ 
0 & 0 & 0  \\
0 & 0 & D_k^* D_k
\end{array} \right) .
$$
It thus suffices to show that that $D_k^* D_k$ and $D_{k-1}
D_{k-1}^*$ are nonnegative and self-adjoint.  By Lemma
\ref{lem:essential} we know these operators are essentially
self-adjoint when restricted to $C_0^\infty \Omega^k$.  So all that
remains is to show that they are nonnegative. But $(x\mid D_k^* D_k
x)=(D_k x\mid D_k x)\ge 0$ for all $x\in\dom D_k^* D_k$. 
\end{proof}

We end this section with a `physical theorem' entirely analogous to
the Result~\ref{thm:3+1phys} stated at the end of last chapter.

\begin{result}
\label{thm:n+1phys}
Let~$M$ be a $(n+1)$-dimensional static globally hyperbolic
spacetime, with metric
$$
g_M=e^{2\Phi}(-\dd t^2+g).
$$
Then, $p$-form electromagnetism on~$M$ with gauge group~$\R$ has as
its phase space the real Hilbert space
$$
\Phase={\dom\{D_p\maps L^2\Omega^p_S\to
  L^2\Omega^{p+1}_S\}\over\overline{\ran}\{D_{p-1}\maps L^2\Omega^{p-1}_S\to
  L^2\Omega^p_S\}}\oplus\ker\{D_{p-1}^*\from L^2\Omega^{p}_S\to
L^2\Omega^{p-1}_S\},
$$
where
$$
D_p=e^{{1\over 2}(n-2p-1)\Phi}\dd_p e^{-{1\over 2}(n-2p-1)\Phi}
$$
is the \emph{twisted exterior derivative}. The phase space 
admits a continuous symplectic structure
$$
\omega(X,X')=(E,A')-(E',A)
$$
where~$X=[A]\oplus E$ and~$X'=[A']\oplus E'$ lie in~$\Phase$ and
$$
(\alpha,\beta)=\int_S g(\alpha,\beta)\vol
$$ 
is the canonical inner product induced on~$\Omega^k_S$ by the
optical metric~$g$ on~$S$. The Hamiltonian is the continuous quadratic
form 
$$
H[X]={1\over 2}\bigl[(E,E)+(D_p A,D_p A)\bigr]. 
$$
The phase space splits naturally into two sectors,
$$
\Phase=\Phase_o\oplus\Phase_f,
$$
and the direct summands
$$
\Phase_f=\Phase\cap\ker L
\qquad\hbox{and}\quad
\Phase_o=\Phase\cap\ran D^*_{p}
$$
are preserved by time evolution.
On~$\Phase_o$, time evolution takes the form
$$
\left( \! \begin{array}{c} A \\ E \end{array} \! \right)
\mapsto T_o(t) 
\left( \! \begin{array}{c} A \\ E \end{array} \! \right) = 
\left( \begin{array}{cc} 
\cos(t\sqrt{L_p}) & \sin(t\sqrt{L_p})\,/\,\sqrt{L_p}\\ 
-\sqrt L_p\,\sin(t\sqrt{L_p}) & \cos(t\sqrt{L_p})  
\end{array} \right) 
\left( \! \begin{array}{c} A \\ E \end{array} \! \right)
$$
while on~$\Phase_f$ it takes the form
$$
\left( \! \begin{array}{c} A \\ E \end{array} \! \right)
\mapsto 
T_f(t) \left( \! \begin{array}{c} A \\ E \end{array} \! \right) =
\left( \begin{array}{cc} 
1 &  t    \\ 
0 &  1  
\end{array} \right)
\left( \! \begin{array}{c} A \\ E \end{array} \! \right).
$$
\end{result}

\chapter{Hodge--de~Rham theory on noncompact manifolds} 
\label{sec:harmonic}

As we have seen, the space of harmonic differential forms, consisting
of closed and coclosed differential forms, plays a special role in
the analysis of the phase space of Maxwell's equations: it corresponds
to the space of physical vector potentials with vanishing magnetic
field (Aharonov--Bohm effect), and also to static electric fields with
no finite sources (charge without charge).

When space is compact, it is well known that the Hodge-de~Rham theorem
identifies the square-integrable, smooth and real cohomologies of a
space, and that the space of square-integrable harmonic forms
coincides with the kernel of the Hodge Laplacian~$\Delta=\dd\delta
+\delta\dd$.

When space is noncompact everything becomes more complicated. To begin
with, the definition of the codifferential~$\delta$ involves
integration by parts. As a result, unless space is complete in the
optical metric it may be
impossible to define the codifferential (and hence the Laplacian)
without specifying boundary conditions at infinity. When the optical
metric on space is
complete, not only is there an unambiguous definition of the
codifferential and Laplacian, but the space of $L^2$~harmonic forms is
identified with the kernel of the Hodge Laplacian, and it has a
square-integrable cohomology interpretation. However, the
square-integrable cohomology is not a topological invariant, as it
depends crucially on the geometry at infinity.

These are the main questions one can ask about the Laplacian~$\Delta$
on a complete Riemannian manifold~\cite{lott97,carron01,carron02}:
\begin{enumerate}
\item Is the dimension of~$\ker\Delta_p$ finite or infinite? In
      physical terms, this is
      the dimension of the space of $p$-form Aharonov--Bohm modes. 
\item What are sufficient conditions for~$\ker\Delta_p$ to be trivial
      or finite-dimensional?
\item If $\ker\Delta_p$ is finite-dimensional, does it have a
      topological interpretation?
\item Is~$0$ in the essential spectrum of~$\Delta_p$? Physically, this
      signals the presence of \emph{infrared divergences} for massless
      $p$-form fields. Conversely, if the essential spectrum is
      bounded away from~$0$, we have a mass gap
      for a free massless field induced by the spatial geometry at infinity! Note that it is possible for~$0$ to
      be in the essential spectrum of the Laplacian even
      if~$\ker\Delta_p$ is trivial, and that the most familiar example
      of this is Euclidean~$\R^n$.
\end{enumerate}
\noindent The answer to all of these questions depends on the
behaviour of the curvature of the optical metric at infinity, so even a massless field may
acquire an `effective mass'.  In this chapter we collect some known facts and open
issues about the space~$\ker\Delta_p$ of harmonic~$p$-forms and the
spectrum of the Laplacian on a complete Riemannian manifold~$S$, and
give physical interpretations of them. Although this chapter is a
review, it points out how rich the subject is compared to the
amount of attention it has received from physicists.

This chapter is based in part on the excellent review of harmonic
forms on noncompact manifolds by Carron~\cite{carron01} (in French),
which includes his finite-dimensionality results~\cite{carron99}
obtained from Sobolev-type inequalities involving the
curvature. Another paper of his~\cite{carron02} (in English) contains a shorter overview, and a
geometrical interpretation of the~$L^2$ cohomology of manifolds with
flat ends (which are known to have finite cohomologies). The~$L^2$
cohomology of hyperbolic manifolds is described by Lott~\cite{lott97}. The case of geometrically finite hyperbolic
manifolds was obtained by Mazzeo and Phillips~\cite{mazzeo90},
including a calculation of the essential spectrum of the
Laplacian. Mazzeo also calculated the cohomology and essential
spectrum of the Laplacian for conformally compact metrics~\cite{mazzeo88}. The~$L^2$ cohomology for rotationally symmetric
manifolds was obtained by Dodziuk~\cite{dodziuk79}.

An additional complication is the `twisting' of the cohomology
complex:
$$
\xymatrix{L^2\Omega^{k-1}_S\ar[r]^{\dd}\ar[d]^{e^{{1\over
2}(n-2p-1)\Phi}}&L^2\Omega^k_S\ar[r]^{\dd}\ar[d]^{e^{{1\over
2}(n-2p-1)\Phi}}&L^2\Omega^{k+1}_S\ar[d]^{e^{{1\over 2}(n-2p-1)\Phi}}\\
L^2\Omega^{k-1}_S\ar[r]^{D}&L^2\Omega^k_S\ar[r]^{D}&L^2\Omega^{k+1}_S}
$$ 
For compact $S$ the (smooth) function $\Phi$ is bounded,
multiplication by $e^{{1\over 2}(n-1-2p)\Phi}$ is bi-continuous on each
$L^2\Omega^k$, the twisted $L^2$ cohomology coincides with the
ordinary $L^2$ cohomology, and the latter with the de~Rham cohomology
by Hodge's theorem. For non-compact $S$, however, $\Phi$ might be
unbounded, in which case the twisted $L^2$ cohomology complex need not
be isomorphic to the ordinary $L^2$ cohomology complex, which we know
already can be very much unlike the de~Rham cohomology complex for
which we have some intuition. Note that if~$n+1=2(p+1)$ (when $p$-form
electromagnetism is conformally invariant) there is no twisting of the
cohomology complex, so the only subtleties are the differences between
the $L^2$ and de~Rham cohomologies.

Since the twisted Laplacian~$DD^*+D^*D$ has not been studied in nearly
as much detail as the ordinary Hodge Laplacian, we know little about
its behaviour. Therefore, when~$\Phi$ is unbounded, most of what we will
say in this chapter is directly applicable only to the cases
where~$p$-form electromagnetism is conformally invariant---\ie,
$p$-form electromagnetism in~$2(p+1)$-dimensional spacetime, which
includes the classical case of~$1$-forms in~$3+1$ dimensions.
 
%

\section{Cohomologies galore}

Let~$C^\infty\Omega^k_S$ denote the space of smooth~$k$-forms on
space. The exterior derivative
$$
\dd\colon C^\infty\Omega^k_S \to C^\infty\Omega^{k+1}_S
$$
gives rise to the smooth (de~Rham) complex
$$
\xymatrix{C^\infty\Omega^{k-1}_S\ar[r]^{\dd}&C^\infty\Omega^k_S\ar[r]^{\dd}&C^\infty\Omega^{k+1}_S}
$$
and the smooth de~Rham cohomology is
$$
H^k(S) \colon = {Z^k(S)\over B^k(S)} \colon = {\ker\{\dd\colon C^\infty\Omega^k_S\to
C^\infty\Omega^{k+1}_S\}\over \dd C^\infty\Omega^{k-1}_S}.
$$
It is the content of de~Rham's theorem that~$H^k(S)$ is isomorphic to
the real cohomology of the manifold,~$H^k(S;\R)$.

Recall that we used compactly-supported smooth differential forms to
derive the Maxwell equations. Denoting the space of smooth,
compactly-supported $k$-forms by~$C^\infty_0\Omega^k_S$, we have the
complex
$$
\xymatrix{C^\infty_0\Omega^{k-1}_S\ar[r]^{\dd}&C^\infty_0\Omega^k_S\ar[r]^{\dd}&C^\infty_0\Omega^{k+1}_S}
$$
and the compactly-supported smooth cohomology is defined by
$$
H^k_0(S) \colon = {Z^k_0(S)\over B^k_0(S)} \colon = {\ker\{{\dd}\colon C^\infty_0\Omega^k_S\to
C^\infty_0\Omega^{k+1}_S\}\over \dd C^\infty_0\Omega^{k-1}_S}.
$$
If~$S$ is the interior of a compact manifold~$M$ with
boundary~$\partial M$, then~$H^k_0(S)$ is isomorphic to the real
relative cohomology of~$M$, denoted~$H^k(M,\partial M;\R)$.

In fact, the derivation of the Maxwell equations and the definition of
the codifferential~$\delta$ required an inner product on the space of
differential forms. If~$L^2\Omega^k_S$ denotes the space of
square-integrable $k$-forms on~$S$, then we have the complex
$$
\xymatrix{L^2\Omega^{k-1}_S\ar[r]^{\dd}&L^2\Omega^k_S\ar[r]^{\dd}&L^2\Omega^{k+1}_S} 
$$
where~$\dd$ is the densely-defined operator obtained by closing the
exterior differential defined on compactly-supported, smooth
differential forms.  The reduced~$L^2$ cohomology is
$$
H^k_2(S) \colon = {Z^k_2(S)\over B^k_2(S)} \colon = {\ker\{{\dd}\colon L^2\Omega^k_S\to
L^2\Omega^{k+1}_S\}\over \overline{\dd L^2\Omega^{k-1}_S}}.
$$
and has a natural Hilbert-space topology. Note also that
$\overline{\dd L^2\Omega^{k-1}_S}=\overline{\dd
C^\infty_0\Omega^{k-1}_S}$. This cohomology space is not a topological
invariant, but it is quasi-isometrically invariant, even bi-Lipschitz
homotopy invariant~\cite{lott97}. We also know that, when the metric
on~$S$ is complete, $H^k_2(S)\simeq\ker\Delta_k$.

Finally, the absolute~$L^2$ cohomology is
$$
H^k_{2,a}(S) \colon = {Z^k_{2,a}(S)\over B^k_{2,a}(S)} \colon = {\ker\{\dd\colon L^2\Omega^k(S)\to
L^2\Omega^{k+1}(S)\}\over \dd L^2\Omega^{k-1}(S)}.
$$ 
This coincides with the reduced cohomology when~$0$ is not in the
essential spectrum of~$\Delta$ (in particular, when~$M$ is compact),
but otherwise it is infinite-dimensional. The absolute cohomology has
nicer algebraic properties than the relative cohomology, such as the
Mayer--Vietoris sequence, but it is not a Hilbert space because it
involves a quotient by a non-closed subspace~\cite{mazzeo90}. It is common
usage to refer to the reduced~$L^2$ cohomology as simply the~$L^2$
cohomology.

In short, the problem is that our intuition about cohomology is based
on compact spaces, and that there the (compactly-supported) smooth and
(absolute/reduced) square-integrable cohomologies all coincide, and
moreover are isomorphic to the cohomologies obtained by combinatorial
methods. Since in the non-compact case all of these cohomologies may
be different, the question arises of which cohomology to use. This
choice has physical implications for electromagnetism: both
classically, through Wheeler's concept of ``charge without charge''
arising through ``field lines trapped by the topology of spacetime'';
and quantumly, through the Aharonov--Bohm effect and mass gaps induced
by the metric when the spectrum of the Laplacian is bounded away
from zero. If we were using~$U(1)$ as the gauge group instead
of~$\R$, topology would manifest itself also through topological terms
in the action (``topological mass'') and topologically stable
solutions (solitons and monopoles).

As we have pointed out,~$p$-form electromagnetism on
an~$(n+1)$-dimensional spacetime is conformally invariant if the
relation~$n+1=2(p+1)$ is satisfied. In all other cases we have seen
that the phase space of classical electromagnetism can be described
most conveniently in terms of the twisted differential operator
$$
D = e^{{1\over 2}(n-2p-1)\Phi} \dd e^{-{1\over 2}(n-2p-1)\Phi}.
$$
The following commutative diagram
$$
\xymatrix{C^\infty_0\Omega^{k-1}_S\ar[r]^{\dd}\ar[d]^{e^{{1\over
2}(n-2p-1)\Phi}}&C^\infty_0\Omega^k_S\ar[r]^{\dd}\ar[d]^{e^{{1\over
2}(n-2p-1)\Phi}}&C^\infty_0\Omega^{k+1}_S\ar[d]^{e^{{1\over 2}(n-2p-1)\Phi}}\\
C^\infty_0\Omega^{k-1}_S\ar[r]^{D}&C^\infty_0\Omega^k_S\ar[r]^{D}&C^\infty_0\Omega^{k+1}_S}
$$ 
where the downward arrows represent multiplication operators, is an
isomorphism of cohomology complexes as long as~$\Phi$ is smooth. If we
complete all the spaces in the~$L^2$ norm and close all operators we
still obtain two cohomology complexes, but the diagram
$$
\xymatrix{L^2\Omega^{k-1}_S\ar[r]^{\dd}\ar[d]^{e^{{1\over
2}(n-2p-1)\Phi}}&L^2\Omega^k_S\ar[r]^{\dd}\ar[d]^{e^{{1\over
2}(n-2p-1)\Phi}}&L^2\Omega^{k+1}_S\ar[d]^{e^{{1\over 2}(n-2p-1)\Phi}}\\
L^2\Omega^{k-1}_S\ar[r]^{D}&L^2\Omega^k_S\ar[r]^{D}&L^2\Omega^{k+1}_S}
$$ 
now has vertical arrows which, depending on the behaviour of~$\Phi$
and~$\dd\Phi$ at spatial infinity, may be only densely defined, and so
definitely not isomorphisms. Therefore, the ordinary (top) and twisted
(bottom) chain complexes may not be isomorphic, and so the
`twisted' $L^2$ cohomology based on~$D$ may not be isomorphic to the
ordinary one based on~$\dd$, even though the smooth cohomologies are
in fact isomorphic.

This may come about in several ways. On the one hand, the closure of
the multiplication operator depends on the behaviour of the
function~$\Phi$ at infinity. Indeed, for a~$k$-form~$\alpha$ to be in
the domain of~$\overline{e^{{1\over 2}(n-1-2p)\Phi}}$ it is necessary
that both~$\alpha$ and~$e^{{1\over 2}(n-1-2p)\Phi}\alpha$ be
square-integrable. Also, although multiplication by~$e^{{1\over
2}(n-1-2p)\Phi}$ is an isomorphism between spaces of smooth forms, its
closure need not be invertible if~$\Phi$ is unbounded. On the other
hand, the closure of~$D=\dd+\bigl[p+1-{n+1\over
2}\bigr](\dd\Phi)\wedge$ depends on the behaviour of~$\dd\Phi$ at
infinity, which can be wild even if~$\Phi$ is bounded.

\section{Known results}

In this section we present a summary of known results on the
(reduced)~$L^2$ cohomology of a Riemannian manifold. Since Poincar\'e duality still holds in the form
$H^k_2(S)\simeq H^{n-k}_2(S)$, so only the cases~$0\le 2k\le n$
need be considered. 

The zeroth cohomology~$H^0_2(S)$ is $1$-dimensional if~$S$ has finite
volume, and trivial otherwise. This is easy to understand since,
essentially, the question is whether constants are square-integrable
or not. This is the first difference with the compact case. 

Other than in the extreme dimensions~$0$ and~$n$, very little can be
said in general. For instance, Anderson~\cite{anderson85} proves that,
if~$n>1$,~$a>|n-2p|$ and~$a\ge 1$, there are complete Riemannian
manifolds diffeomorphic to~$\R^n$, with curvature bounded by~$-a^2\le
K\le 1$ and such that their $p$th square-integrable cohomology~$H^p_2$
is infinite-dimensional.

For rotationally symmetric $n$-dimensional manifolds with metric
$$
ds^2=dr^2+f(r)^2d\theta^2,
$$ 
where~$d\theta$ is the standard metric on~$S^{n-1}$, the
square-integrable cohomology is~$H^k_2=\{0\}$ if~$k\neq 0,n/2,n$. As
we know, when $k=0,n$ the cohomology depends on the volume of
spacetime. Finally, when~$k=n/2$, $H^k_2=\{0\}$ if
$\int^\infty{ds\over f(s)}=\infty$, and infinite-dimensional
otherwise. This is because of conformal invariance of the cohomology
in the middle dimension, and the fact that convergence of the integral
correlates with conformal compactness~\cite{dodziuk79}. A remarkable
consequence of this is that, when space a two-dimensional cylinder,
the square-integrable cohomology~$H^1_2$ never matches the smooth
cohomology, which is one-dimensional and is generated by~$\dd\theta$.

A complete Riemannian manifold is conformally compact if it is
diffeomorphic to the interior of a compact manifold~$M$ with boundary,
and the metrics of the two manifolds at corresponding points are
proportional by a function called the conformal factor:
$$
g_M=\rho^2 g.
$$ 
The conformal factor~$\rho$ has the effect of ``pushing the
boundary of~$M$ to infinity''. For this it is necessary that
$\int_\gamma \rho^{-1}\dd s_M$ diverge whenever~$\gamma$ is a
(finite-length) curve in~$M$ with at least one endpoint on~$\partial
M$, so the conformal factor must vanish precisely on~$\partial
M$. Conformal compactification makes precise the idea of ``ideal
boundary at infinity'' of a noncompact manifold, and it was introduced
into general relativity as an important tool by Penrose.

Mazzeo~\cite{mazzeo88} studies the case where the conformal factor
satisfies the additional regularity condition that~$\dd\rho$ does not
vanish on~$\partial M$, in which case the manifold is asymptotically
hyperbolic. He then proves that a complete conformally
compact~$n$-dimensional Riemannian manifold has finite-dimensional
cohomology groups except possibly for the middle dimensions, and gives a
topological interpretation of them:
$$
H^k_2\simeq\left\{
\begin{array}{ll}
H^k(M,\partial M,\R) & k<(n-1)/2 \\
H^k(M,\R) & k>(n+1)/2
\end{array}
\right.
$$ 
Moreover, if~$-a^2$ is the most negative limiting curvature at
infinity, then the essential spectrum of the Laplacian~$\Delta_k$ is
$$
\sigma_{\mathrm{ess}}(\Delta_k)=\left\{
\begin{array}{ll}
[a^2(n-2k-1)^2,\infty) & k<n/2 \\
\{0\}\cup[a^2/4,\infty) & k=n/2 \\
{[a^2(n-2k+1)^2,\infty)} & k>n/2
\end{array}
\right.
$$ 
In particular, if~$n=2k$, the $k$th cohomology group is
infinite-dimensional and, if $|n-2k|\le 1$, the essential spectrum
extends all the way to~$0$. For hyperbolic manifolds which are
geometrically finite (\ie, having no tubular ends), Mazzeo and
Phillips~\cite{mazzeo90} prove that the cohomology of the middle
dimensions~$k=(n\pm 1)/2$ is finite-dimensional and has a topological
interpretation. These results are extended by Lott~\cite{lott97} to
the case of hyperbolic $3$-manifolds which are diffeomorphic to the
interior of a compact manifold with boundary and geometrically
infinite. In particular, Lott proves that, if such a space is `nice'
(has incompressible ends and its injectivity radius does not go to
zero at infinity), the kernel of the Laplacian on~$1$-forms is
finite-dimensional. He also provides a variety of results on the
spectrum of the Laplacian on~$1$-forms.

These results have a direct physical interpretation when~$p$-form
electromagnetism is conformally invariant, as otherwise one has to
consider an appropriately twisted~$L^2$ cohomology complex for which
there are no known general results. We have pointed out that,
when~$\Phi$ and~$\dd\Phi$ are both bounded, the twisted $L^2$
cohomology complex is isomorphic to the untwisted one, and so the
above-mentioned results can be applied directly. In the general case,
it is reasonable to assume that the behaviour of the twisted
cohomology will be at least as rich as that of the ordinary~$L^2$
cohomology. In the conformally invariant cases, we have the following
possible physical interpretations:
\begin{itemize}
\item the massless scalar field ($0$-form electromagnetism)
in~$1+1$ dimensions. In this case, since the space manifold~$S$ is
assumed to be noncompact, it is
diffeomorphic to~$\R$. Global hyperbolicity then requires that the optical metric give~$S$ infinite
length, and so~$H^0_2=\{0\}$ because the constant field is not square
integrable. In other words, square-integrable fields must go to zero
at infinity. 
\item ordinary ($1$-form) electromagnetism in~$3+1$ dimensions. If
space is spherically symmetric there are no harmonic,
square-integrable~$1$-forms according to~\cite{dodziuk79}. This is not
a surprise since the first de~Rham cohomology is also trivial. In more
general cases, if the space manifold~$S$ is conformally compact the
spectrum of the Laplacian reaches all the way to~$0$ (physically, the
photon does not acquire a mass), but the
dimension of the kernel of the Laplacian is not known in
general. Anderson's example~\cite{anderson85} shows that it is
possible for this space of non-standard Aharonov-Bohm modes to be
infinite-dimensional.  
\item when~$p$-form electromagnetism is conformally invariant the
dimension of space is~$p=(n-1)/2$, and we are always in one of the ``middle
dimension'' cases where the dimension of the space of harmonic vector
potentials remains unresolved, although for a large class of manifolds
it is known that the essential spectrum of the Laplacian is all
of~$[0,\infty)$ and so there is no mass gap.
\end{itemize}
\noindent 
In case~$\Phi$ and~$\dd\Phi$ are bounded, the dimension of the space
of ``twisted'' harmonic~$p$-forms is independent of~$\Phi$, and so we
can draw valid physical conclusions about non-standard Aharonov--Bohm
modes even in the absence of conformal invariance. The lower bounds to
the spectrum of the Laplacian may be critically dependent on~$\Phi$,
so any inferences we make from the~$\Phi=0$ case are probably
unwarranted, but still enticingly point to situations where the
phenomenon of mass gaps might occur. The physical interpretation of
the~$L^2$ cohomology results in the cases when electromagnetism is not
conformaly invariant follows.
\begin{itemize}
\item the massless scalar field in~$n+1$ dimensions has at most a
one-dimensional space of harmonic solutions. This depends on whether
the function
$$
f=e^{{1\over 2}(n-1)\Phi}
$$
is square-integrable with respect to the optical metric.  Also, if
space is conformally compact and the curvature at infinity is bounded
below by~$-a^2$, then the essential spectrum of the Laplacian
is~$[a^2(n-1),\infty)$. This means that, if the dimension of space is
$n>1$, the free massless scalar field can have a 
mass gap in the~$\Phi=0$ case.
\item ordinary electromagnetism in~$2+1$ dimensions can have an
infinite-dimensional space of harmonic vector potentials even in the
rotationally symmetric case, including when the optical metric on
space is that of the hyperbolic plane. In addition, if space is
conformally compact the~$\Phi=0$ mass gap is~$a^2/4$, where~$-a^2$ is the
lower bound to the curvature at infinity. When the optical metric on
space is conformally compact and of dimension~$4+1$ or higher, the
space of harmonic vector potentials is isomorphic to the first
cohomology of~$M$ relative to its boundary, and so there are no
non-standard Aharonov--Bohm modes. When the curvature at infinity is
bounded below by~$-a^2$, the essential spectrum of the ordinary
Laplacian is~$[a^2(n-3),\infty)$ if~$n\ge 3$, signaling the
possibility of topological mass gaps in~$4+1$ dimensions or higher, at
least when~$\Phi=0$.
\item for~$p$-form electromagnetism, there is an infinite-dimensional
space of harmonic vector potentials if space is a~$2p$-dimensional and
rotationally symmetric or conformally compact. In the latter case,
there is a~$\Phi=0$ mass gap of~$a^2/4$. If~$|n-2p|>1$ there are no
non-standard Aharonov-Bohm modes, but the~$\Phi=0$ mass gap is zero only
if~$n=2p\pm 1$. 
\end{itemize}

\part{Quantum electromagnetism}

The apparent truism that a quantum mechanical theory\index{quantum
mechanical theory} needs to be cast in classical
language\index{classical language} in order to correlate its
predictions with our experience, a point that Niels Bohr\index{Niels
Bohr} made into a cornerstone of his philosophy of quantum
mechanics\index{philosophy of quantum mechanics}, has practical
consequences for the development of quantum descriptions of physical
systems\index{physical system}. This is because a physical system will
be described operationally or geometrically in inevitably classical
terms, and this information needs to be fashioned into a quantum
theory\index{quantum theory} whose predictions need to be, again,
reexpressed in classical terms. In addition, the process of
constructing a classical theory\index{classical theory} from
operational or geometric data is so well-understood that it is
convenient to construct the quantum theory by first constructing a
classical theory from the data and then `quantizing' it.

Quantization\index{quantization} is a catch-all term for any process
taking as input a classical mechanical system\index{classical
mechanical system}, and producing as output a quantum mechanical
system\index{quantum mechanical system} reducing to the original
classical system\index{classical system} in an appropriate
limit. Quantization would ideally be algorithmic or functorial, but it
turns out to be neither, although formulating quantization in
algebraic language\index{algebraic language} seems to bring it closest
to the goal of functoriality. 

In algebraic terms, a classical mechanical system\index{classical
mechanical system} is defined by specifying a Poisson
algebra\index{Poisson algebra} of observables, while any associative
algebra can play the role of algebra of observables\index{algebra of
observables} for a quantum system\index{quantum system}. The Dirac
quantization prescription~\cite[Chapter IV]{dirac57}\index{Dirac
quantization prescription} `promotes' the commuting classical
observables\index{classical observables} to operators satisfying the
Heisenberg commutation relations\index{Heisenberg commutation
relations}
$$
[\hat f,\hat g]=i\hbar\{\widehat{f,g}\},
\index{$\hat f$!quantized observable}
\index{[~,~]!commutator}
\index{$\hbar$!Planck's constant}
\index{$f$!observable}
\index{$\{~,~\}$!Poisson bracket}
$$ 
where~$\{f,g\}$ is the Poisson bracket\index{Poisson bracket} of
the classical observables~$f$ and~$g$, $[\hat f,\hat g]$ is the
commutator\index{commutator} of their quantum counterparts, and
Planck's constant~$\hbar$\index{Planck's constant} measures the
departure from classical behaviour\index{classical behaviour} (where
observables commute). It is not hard to convince oneself that, because
the algebra of quantum observables\index{quantum observables} is
nonabelian, the operation~$f\mapsto\hat f$ cannot be an algebra
homomorphism\index{algebra homomorphism}. That
is,~$\widehat{fg}\neq\hat f\hat g$ in general. Physicists call this
fact `operator ordering\index{operator ordering} ambiguities'.

An operator algebra of quantum observables\index{quantum observables}
realizing the canonical commutation relations achieves
quantization\index{quantization} in a kinematical sense, but the
physical and dynamical content of the theory comes about by means of a
specific representation of the quantum observables as an algebra of
(unbounded) linear operators on a Hilbert space\index{Hilbert space}
of quantum states\index{quantum states}. Each representation is
associated to a choice of `vacuum expectation'\index{vacuum
expectation} on the algebra of observables\index{algebra of
observables} and it is known that, for systems with infinitely many
degrees of freedom, different states may lead to unitarily
inequivalent representations. The choice of representation can be
narrowed down by the need to recover an appropriate classical limit,
and by requiring that physical symmetries be implemented unitarily.

The classical limit is encoded in the correspondence
principle\index{correspondence principle}, by which we mean the
following. The Poisson algebra\index{Poisson algebra} of classical
observables\index{classical observables} consists of smooth functions
on a symplectic manifold (phase space\index{phase space}) playing the
role of state space\index{state space} for the classical
theory\index{classical theory}. The correspondence
principle\index{correspondence principle} requires that, for any phase
space point~$x\in\Phase$\index{$x$!phase space
point}\index{$\Phase$!phase space} and any
observable~$f$\index{$f$!observable}, there should be a quantum
state~$\ket{x}$\index{$\ket{x}$!quantized phase space point} such that
the expected value of~$\hat f$\index{$\hat f$!quantized observable} in
the state~$\ket{x}$ equals the classical value~$f(x)$, if not exactly,
at least in the limit~$\hbar\to 0$. That is,
$$
\matElem{x}{\hat f}{x}=f(x)+O(\hbar).
\index{$\matElem{x}{\hat f}{y}$!matrix element of~$\hat f$}
$$

There is one last requirement that a sensible quantization must
satisfy, and that is that physical symmetries be represented by
unitary operators on the Hilbert space of quantum states of the
system.

In the case where the classical phase space\index{classical phase
space} is a vector space, the linear observables can be identified
with the points of the phase space\index{phase space} itself, and so
the Heisenberg commutation relations\index{Heisenberg commutation
relations} can be implemented on the phase space. In
Chapter~\ref{chap:linear} we develop the
quantization\index{quantization} of an abstract linear system and
develop the concept of a quasioperator on Fock space, and in
Chapter~\ref{chap:qed} we apply this to Maxwell's
equations\index{Maxwell's equations} for the electromagnetic
field\index{electromagnetic field} and express the dynamics of the
quantized electromagnetic field in terms of Wilson loops
quasioperators.


The work most closely akin to ours is that of
Dimock~\cite{dimock92}. Like us, Dimock constructs a $C^*$-algebra of
observables for the electromagnetic field, but he does not exhibit any
states or Hilbert-space representations. He notes in passing that ``in
any case such [Hilbert-space] representations exist, say by a Fock
space construction''. We discuss below some ways in which a Fock space
representations may fail to exist. 

Because Dimock describes the classical theory in the covariant
canonical formalism, he is forced to focus on ``the algebraic
structure of the theory, not in the specification of particular
states''. In our terms, Dimock quantizes the electromagnetic field as
a `general boson field'. He also constructs a classical Poisson
bracket, and his quantization procedure is equivalent to our general
linear quantization. Dimock does show that different Hilbert-space
representations lead to~$*$-isomorphic~$C^*$-algebras of
observables. This form of equivalence, however, obviates the possible
physical consequences of unitary inequivalence of Hilbert-space
representations, and for this reason Dimock's paper suffers from what
Earman and coauthors critically term ``algebraic imperialism''
in~\cite{earman}.

Dimock does not show that the classical canonical transformations
associated to changes in the choice of Cauchy surface are implemented
unitarily on the $C^*$-algebras of quantum observables, because that
is simply not true. In fact, Torre and Varadarajan~\cite{torre} show
that, even in the case of free scalar fields on a flat spacetime of
dimension higher than two, there is no unitary transformation between
the Fock representations associated to arbitrary initial and final
Cauchy surfaces. They point out that unitary implementability is
easily obtained if the Cauchy surfaces are related by a spacetime
isometry, though. They also mention related results of Helfer (no
unitary implementation of the $S$-matrix if the `in' and `out'
states are Hadamard states)~\cite{helfer96}, and of van~Hove (only a
small subgroup of the classical canonical transformations is unitarily
implementable)~\cite{hove51}.

Another paper addressing specifically the quantization of the
electromagnetic field is the one by Corichi~\cite{corichi}. Corichi
stresses that Fock quantization depends crucially on the linear
structure of phase space, and characterizes the Fock quantization
procedure as ``completely elementary''.

Here we perform Fock quantization of Maxwell's equations on a static,
globally hyperbolic spacetime with a trivial~$\R$ bundle on
it. Presumably this can be extended to stationary spacetimes, but not
beyond that because of the need for a nontrivial group of
isometries. The treatment of nontrivial or~$\U(1)$ bundles should
require only straightforward modifications, but one of the lessons of
our work is that sometimes there are surprises in store even for
topics as well-understood as electromagnetism.

In chapter~\ref{chap:qed}, because of the appearance of negative
powers of the Laplacian~$\Delta$ (or the twisted Laplacian~$L_p$ in
the general case) in the process, we will be forced to restrict Fock
quantization to the space~$\Phase_o$ of oscillating modes of the
electromagnetic field. Also, for mathematical convenience one often
assumes that~$\Delta\ge\epsilon>0$ for some~$\epsilon$, which is true
when space is compact but not necessarily when it is
noncompact. However, we do not do this as one cannot exclude the
possibility that the spectrum of~$L_p$ or~$\Delta$ reach all the way
to~$0$ because that is the case in physically interesting situations
such as Minkowski space.

\chapter{Coherent-state quantization of linear
systems}\label{chap:linear}   

In this chapter we present a rigorous framework for quantization of
linear dynamics based on the ideas of Irving Segal. 

Segal pioneered the idea of of formalizing quantum mechanics in
terms of algebras of observables\index{algebras of observables},
making Hilbert spaces play the subordinate role of supporting linear
representations of them. These Hilbert spaces can, in fact, be
constructed from the abstract algebra of observables by means of the
Gel'fand--Na\u{\i}mark--Segal
construction\index{Gel'fand--Na\u{\i}mark--Segal construction} using a
single \emph{state}\index{GNS state} or, in physics parlance,
\emph{vacuum expectation}\index{vacuum expectation value}.

Implicit in the work of Segal is a concept of \emph{general boson
field}\index{general boson field} associated to any linear phase
space\index{linear phase space}, which formalizes the Heisenberg
commutation relations\index{Heisenberg commutation relations} among
field operators in terms of exponentiated field operators, using the
so-called \emph{Weyl relations}\index{Weyl relations}. This has the
advantage of avoiding the technicalities of unbounded
operators\index{unbounded operators}. In addition, physical symmetries
are readily implemented as automorphisms of the Weyl algebra.

Segal introduced the related concept of \emph{free boson
field}\index{free boson field}, which can be constructed from a phase
space equipped with a compatible complex structure\index{compatible
complex structure}. Segal's free boson field axiomatizes the
properties of the usual of Fock space\index{Fock space}, and the
axiomatic approach makes it transparent that the Fock\index{Fock
representation}, Schr\"odinger\index{Schr\"odinger representation} and
Bargmann--Segal\index{Bargmann--Segal representation} representations
of linear quantum fields are all unitarily equivalent. Within this
framework, Segal also studied the problem of representing time
evolution unitarily on Fock space\index{Fock space}, and the stability
of the generator of unitary time evolution\index{unitary time
evolution}, namely whether the quantum Hamiltonian\index{Hamiltonian}
is bounded below.

Here we put together both ideas, and the result is a new construction
of the free boson field\index{free boson field} based on
\emph{coherent states}\index{coherent states}. In this construction we
not only associate to each linear functional\index{linear functional}
on phase space\index{phase space} a field operator\index{field
operator} but, given a choice of vacuum state\index{vacuum state}, we
can associate to each point in phase space\index{phase space point} a
coherent state\index{coherent state}. The collection of all coherent
states indexed by points of phase space spans the Hilbert space of
quantum states\index{quantum state space} of the theory, and the
result is what Segal called the \emph{general boson field}. The free
boson field, which as we have mentioned is unitarily equivalent to the
Fock representation, is obtained by means of a GNS state with Gaussian
statistics.

We find that the mathematical process of quantization can be
understood with reference to three physical guiding principles: the
canonical commutation relations, the correspondence principle, and the
unitary implementation of physical symmetries.

We proceed as follows: we first construct the Weyl algebra\index{Weyl
algebra} of observables associated to a linear phase
space\index{linear phase space}, and then choose a compatible complex
structure\index{compatible complex structure} on the Phase space,
which amounts to
selecting a vacuum expectation\index{vacuum
expectation} on the Weyl algebra, with the help of the correspondence
principle and the requirement that time evolution\index{time
evolution} be unitarily and stably implemented.

Coherent states\index{coherent states} are most useful because many
classical equations hold exactly between expectation
values\index{expectation value} on coherent states. Thus, by using
coherent states, our quantization\index{quantization} procedure never
loses sight of the correspondence principle\index{correspondence
principle}. In addition, the vacuum expectation value\index{vacuum
expectation value} acts as a generating function\index{generating
function} of the matrix elements\index{matrix element} of field
operators\index{field operator} between coherent states\index{coherent
state}, not only for ordinary field operators\index{field operator}
but also for their Wick powers\index{Wick power} (called
normal-ordered operators\index{normal-ordered operator} in
physics). As an unexpected bonus, using matrix elements\index{matrix
elements} between coherent states\index{coherent states} one can
define normal-ordered Wilson loops\index{Wilson loops} as
quasioperators\index{quasioperators} without the need for
regularization\index{regularization}.

Segal's treatment of the free boson field\index{free boson field} is
presented in~\cite{BSZ}. A comprehensive physical treatment of the
coherent states\index{coherent states} of the electromagnetic
field\index{electromagnetic field} can be found in~\cite[Chapter
11]{mandel95}.

\section{The general boson field}

The development that follows may seem idiosyncratic to those familiar
with the traditional quantization methods and the notations used in
physics. In particular, we insist on distinguishing the phase
space~$\Phase$ from its dual~$\Phase^*$. There are some good reasons
for this. At the present stage of development of mathematical physics,
the most compelling reason for studying the quantization of a linear
systems is as a springboard for quantization of nonlinear systems, or
as a testing ground for ideas suggested by the study of nonlinear
systems. Our approach is motivated by the fact that the ordinary
quantization of linear systems makes use of several identifications
that can only be made for a linear system. Adopting the view that a
classical mechanical system is characterized by its Poisson algebra of
observables, the cotangent space at each point of phase space acquires
a symplectic structure. When the phase space~$\Phase$ is linear, the
following identifications can be made: the dual~$\Phase^*$ can be
identified with the linear observables, and the restriction of the
Poisson bracket to~$\Phase^*$ is a symplectic structure. Also, the
cotangent spaces to each point of phase space are canonically
isomorphic to each other and to~$\Phase^*$, and the globally-defined
symplectic structure on~$\Phase^*$ makes~$\Phase$ isomorphic
to~$\Phase^*$ and also endows it with a symplectic structure. All of
these identifications, and even the possibility of considering
itself~$\Phase$ to be a symplectic vector space, are accidents of
linearity. Accordingly, we will avoid making use of these features as
much as possible. Every time we are forced to make use of one of these
identifications, it will be a sign that the procedure cannot be
readily generalized to nonlinear situations.

\subsection{Linear phase spaces}

We start by formalizing the notion of \emph{linear phase space}, which
is the necessary classical input of our quantization procedure. 

\begin{definition}[linear phase space]\label{def:linearPhaseSpace}
A \emph{linear phase space} is a reflexive real topological vector
space~$\Phase$ whose dual~$\Phase^*$ is a \emph{symplectic vector
space\/}. That is,~$\Phase^*$ is a topological vector space equipped
with a \emph{symplectic structure}: a continuous, skew-symmetric
bilinear form~$\omega$ which is \emph{weakly nondegenerate} in the
sense that the duality map~$*\from\Phase^*\to\Phase$ given by
$$
\omega(f,g)=f(g^*)
\qquad\hbox{for all}\quad
f,g\in\Phase^*
$$ 
is injective.
\end{definition}

\begin{proof}[Note]
Without the assumption that~$\Phase$ is reflexive, the duality map
would be~$*\from\Phase^*\to\Phase^{**}$. This would have a bearing on
the definition of the Hilbert space of quantum states below.

A finite-dimensional vector space has a unique Hausdorff topology, and
any infinite-dimensional vector space can be topologized
algebraically~\cite[\S 1.2]{BSZ}; in either case the continuity
of~$\omega$ is vacuously true. In general, the dual~$\Phase^*$ of a
topological vector space is itself naturally a topological vector
space, with the $\hbox{weak-}*$ topology making every element
of~$\Phase$ a continuous linear functional on~$\Phase^*$. If~$\Phase$
has a normed topology,~$\Phase^*$ can also be given the (normed)
strong operator topology. In either case,~$\Phase\subseteq\Phase^{**}$
is a continuous inclusion.
\end{proof}

The right notion of automorphism of a linear phase space is the
following. Recall that, if~$T\from\Phase\to\Phase$ is linear, there is
a unique linear map ~$T^*\from\Phase^*\to\Phase^*$ called its
\emph{dual} such that
$$
(T^*f)(x)=f(Tx)
\qquad\hbox{for all}\quad
x\in\Phase,f\in\Phase^*.
$$ 

\begin{definition}[automorphism of a linear phase space]\label{def:automorph}
An automorphism of the linear phase space~$\Phase$ is a continuous
invertible linear map~$T\from\Phase\to\Phase$ whose dual
map~$T^*\from\Phase^*\to\Phase^*$ preserves the symplectic structure on~$\Phase^*$.
\end{definition}

The \emph{space of states} of a classical system is its physical phase
space~$P$, namely the space of gauge equivalence classes of
solutions of its equations of motion. Similarly, its \emph{algebra of
observables} consists of smooth gauge-invariant functions of solutions
to the equations of motion,~$C^\infty(P)$. The classical algebra
of observables is naturally a Poisson algebra, but the physical phase
space~$P$ need not be a Poisson manifold, let alone a symplectic
vector space. For instance, in Yang--Mills theory~$P$ is some
sort of `singular infinite-dimensional variety', a concept without a
precise definition. Continuous non-gauge symmetries of the physical
system are are represented by automorphisms of the Poisson algebra of
observables generated through Poisson brackets with appropriate
observables: the conserved quantities associated to the symmetries via
Noether's theorem.

Suppose, then, that not only~$C^\infty(P)$ is a Poisson algebra with
Poisson bracket~$\{~,~\}$ but that~$P$ is a manifold. The Poisson
bracket defines a 
bivector~$\omega\from\Omega^2(P)\to\R$ given by
$$
\omega(\dd f,\dd g)=\{f,g\}
\qquad\hbox{for all}\quad
f,g\in C^\infty(\Phase).
$$ 
If~$\omega$ is non-degenerate at~$x\in P$, the space~$T^*_xP$
becomes a symplectic vector space. In physical terms,~$x$ is a field
configuration and~$T^*_xP$ is the space of linear observables in the
vicinity of this field configuration. This is the only symplectic
vector space that can be constructed in a natural way from the phase
space~$P$, and Definition~\ref{def:linearPhaseSpace} applies
with~$\Phase=T_xP$ and~$\omega=\omega_x$. In these favourable cases,
symmetries of field configurations~$x\in P$ are Poisson maps
leaving~$x$ fixed, which induce linear symplectic transformations
of~$T_x^*P$.

Identifying all the~$T_x^*P$ amounts to choosing a trivialization
of~$T^*P$, and this is natural only if~$P$ is a linear space admitting
a canonical flat connection. In that case, each of the~$T_x^*P$ is
canonically isomorphic to~$P^*$ itself. When the equations of motion
are linear, one can take~$\Phase=P$ in
Definition~\ref{def:linearPhaseSpace}, and restrict one's attention to
linear observables and symmetry transformations.

\subsection{Quantizing a linear phase space}

Linear quantization\index{quantization} is a process ``promoting''
each~$x\in\Phase$\index{$x$!point of phase space} to a unit
vector~$\ket{x}$\index{$\ket{x}$!quantized phase space point} in a
suitable Hilbert space~$\K$\index{$\K$!quantum state space}, and
each~$f\in\Phase^*$ to a self-adjoint operator~$\hat f$\index{$\hat
f$!quantized observable} on~$\K$, in such a way that the Heisenberg
commutation relations\index{Heisenberg commutation relations}
\begin{equation}\label{HeisCommRel}
[\hat f,\hat g]=i\omega(f,g)\id_\K
\qquad\hbox{for all}\quad
f,g\in\Phase^*
\end{equation}
hold. Equation~\ref{HeisCommRel} is a restricted form of the Dirac
quantization prescription\index{Dirac quantization prescription},
since it is applied only to linear observables on~$\Phase$, and not to
arbitrary ones as it was originally formulated. In addition, the
correspondence principle\index{correspondence principle} is required
to hold in the form
\begin{equation}
\label{eq:corr}
\matElem{x}{\hat f}{x}=f(x)
\qquad\hbox{for all}\quad
x\in\Phase,f\in\Phase^*,
\end{equation}
without allowing for corrections of order~$\hbar$.
Finally, one would hope to represent every physical
symmetry~$T\from\Phase\to\Phase$ as a unitary
operator~$U_T\from\K\to\K$ in such a way that~$U_S U_T=U_{ST}$ for all
symplectic maps~$S,T\from\Phase\to\Phase$. As we shall see, in 
general this is only possible for a subgroup of linear symplectic
transformations of~$\Phase$ and, in fact, choosing a small subgroup of
physical symmetries that must be unitarily implemented can be enough
to determine~$\K$, sometimes uniquely. Time evolution is always
required to be a physical symmetry and, in this sense, the dynamics
determine the quantization.

\subsubsection{Canonical commutation relations}

The Heisenberg relations cannot be implemented on an algebra of
bounded operators~\cite[\S 13.6]{rudin91}, and so
Equation~\ref{HeisCommRel} must be understood as holding on the
(hopefully) dense domain of~$[\hat f,\hat g]$ in~$\K$. This is only
the first of a long list of nuisances that arise from necessarily
dealing with unbounded operators, but all the same we encode it as a
definition.

\begin{definition}[Heisenberg system]
A \emph{Heisenberg system}\index{Heisenberg system} on a symplectic
vector space~$(\Phase^*,\omega)$ is a real-linear map~$\Phi\maps
f\mapsto\Phi(f)$ from~$\Phase^*$ to the self-adjoint operators on some
complex Hilbert space~$\K$, satisfying the \emph{Heisenberg
commutation relations}\index{Heisenberg commutation relations}
$$
\bigl[\Phi(f),\Phi(g)\bigr]=i\omega(f,g)\id_\K.
\qquad\hbox{for all}\quad
f,g\in\Phase^*
$$ 
as an operator equation holding on the common domain
of~$\Phi(f)\Phi(g)$ and~$\Phi(f)\Phi(g)$, which is assumed to be
dense. The operator~$\Phi(f)$ is called the \emph{Heisenberg
operator}\index{Heisenberg operator} associated to~$f\in\Phase^*$. 
\end{definition}

In other words, linear quantization is partially achieved by
constructing a Heisenberg system on the space of linear
observables~$(\Phase^*,\omega)$. However, there are lots of Heisenberg
systems that have nothing to do with physics, examples of which can be
found in~\cite{R&S,BSZ}, so for honest quantum physics one needs to
impose some additional regularity on the Heisenberg systems. This is
achieved in an somewhat circuitous way by considering the unitary
groups supposedly generated by the Heisenberg
operators. Heuristically, if~$\Phi(f)$ is a Heisenberg operator
on~$\K$, the operator~$W(f)=e^{-i\Phi(f)}$ is unitary and,
since~$\bigl[\Phi(f),\Phi(g)\bigr]$ commutes with both~$\Phi(f)$
and~$\Phi(g)$, the Baker--Campbell--Hausdorff
formula\index{Baker--Campbell--Hausdorff formula} applies, giving
$$
e^{-i\Phi(f)}e^{-i\Phi(g)}=e^{-i\Phi(f+g)}e^{-{1\over 2}[\Phi(f),\Phi(g)]}.
$$ 
We take this heuristic calculation as the motivation of our next
definition. 

\begin{definition}[Weyl algebra]
The \emph{Weyl algebra}\index{Weyl algebra} on a symplectic vector
space space~$(\Phase^*,\omega)$, is the
complex~$*$-algebra~$\W(\Phase^*,\omega)$\index{$\W(\Phase^*,\omega)$!Weyl
algebra} generated by the set
$\W(\Phase^*)=\bigl\{\W(f)\bigr\}_{f\in\Phase^*}$,\index{$\W(\Phase^*)$!Weyl
operators of~$\Phase^*$}\index{$\W(f)$!Weyl operator
of~$f\in\Phase^*$} of \emph{Weyl operators}, modulo the
\emph{unitarity relations}
$$
\W(f)^*=\W(-f)
\qquad\hbox{for all}\quad
f\in\Phase^*
$$ 
and the \emph{Weyl relations}\index{Weyl relations}
$$
\W(f)\W(g)=e^{\omega(f,g)/2i}\W(f+g)
\qquad\hbox{for all}\quad
f,g\in\Phase^*.
$$
\end{definition}

\begin{proof}[Note]
Because the Weyl relations\index{Weyl relations} reduce products of
Weyl operators\index{Weyl operator} to single Weyl operators, the Weyl
algebra~$\W(\Phase,\omega)$ coincides with the linear span
of~$\W(\Phase)$. In fact,~$\W(\Phase)$ is a basis
of~$\W(\Phase,\omega)$.
\end{proof}

Heuristically, because of the Baker--Campbell--Hausdorff
formula\index{Baker--Campbell--Hausdorff formula} above, one would expect
that a Heisenberg system can be constructed from a representation of
the Weyl algebra as an algebra of operators on a suitable Hilbert
space. Such a representation is called a Weyl system\index{Weyl
system}. We will consistently use the fonts~$\W$ and~$W$ to
distinguish the \emph{abstract} Weyl algebra~$\W(\Phase,\omega)$, and
its generators~$\W(x)$, from Weyl systems~$W$ associated to
\emph{concrete} Hilbert-space representations of the Weyl algebra.

\begin{definition}[Weyl system]
A \emph{Weyl system}\index{Weyl system} on the symplectic vector
space~$(\Phase^*,\omega)$ is a continuous mapping $W\from\Phase^*\to
U(\K)$, where~$U(\K)$ is the group of unitary operators on the complex
Hilbert space~$\K$ with the strong operator topology,
and~$W$ satisfies the Weyl relations\index{Weyl relations}
$$
W(f)W(g)=e^{\omega(f,g)/2i}W(f+g)
\qquad\hbox{for all}\quad
f,g\in\Phase^*.
$$
\end{definition}

\begin{proof}[Note]
Since a Weyl system is required to be continuous in the strong
operator topology on~$U(\K)$, the map~$t\mapsto W(tf)$ is a
strongly-continuous one-parameter subgroup of~$U(\K)$. By Stone's
theorem~\cite[\S VIII.4]{RS}, this one-parameter subgroup has a
self-adjoint generator~$\Phi(f)$ such that $W(f)=e^{-i\Phi(f)}$.
\end{proof}

\begin{lemma}
\label{lem:WeylHeis}
If~$W\from\Phase^*\to U(\K)$ is a Weyl system on the symplectic vector
space~$(\Phase^*,\omega)$ then~$\Phi\from\Phase^*\to L(\K)$ is a
Heisenberg system on~$(\Phi^*,\omega)$. In addition, for
all~$x,y\in\Phase$, the operator~$\Phi(f)+i\Phi(g)$ is closed
and~$\Phi(f+g)$ is the closure of~$\Phi(f)+\Phi(g)$.
\end{lemma}

\begin{proof}[Sketch of proof]
Differentiating the Weyl relation\index{Weyl relations}
$$
W(tf)W(tg)=e^{t^2\omega(f,g)/2i}W\bigl(t(f+g)\bigr)
$$
twice and setting~$t=0$, one obtains that~$f\mapsto\Phi(f)$ is additive
and satisfies the Heisenberg commutation relations
$$
\bigl[\Phi(f),\Phi(g)\bigr]=i\omega(f,g)\id_\K.
$$ 
The proof of the closure properties of the Heisenberg operators is
in~\cite[\S 1.2]{BSZ}. 
\end{proof}

At this point, a theorem of von~Neumann~\cite[\S VIII.5]{R&S}
guarantees that all Weyl systems\index{Weyl system} on a
finite-dimensional phase space are unitarily equivalent. At any rate,
we see that Weyl systems are the right formalization of
Equation~(\ref{HeisCommRel}), the Heisenberg commutation
relations. The following lemma shows one reason why it is convenient
to insist that physical symmetries be represented by \emph{linear}
symplectic maps on~$\Phase$.

\begin{lemma}
Suppose that~$\gamma\from\W(\Phase^*,\omega)\to\W(\Phase^*,\omega)$ is
a~$*$-algebra endomorphism such that
$$ 
\hbox{for every}\quad 
f\in\Phase^*,
\qquad
\gamma\bigl(\W(f)\bigr)=\W(g)
\qquad\hbox{for some}\quad
g\in\Phase^*,
$$ 
and suppose furthermore that the
map~$T^*\from(\Phase^*,\omega)\to(\Phase^*,\omega)$ given by~$T^*f=g$ is
continuous. Then,~$T^*$ is in fact linear and preserves the symplectic
structure~$\omega$. If, in addition,~$\gamma$ is an automorphism,
then~$T^*$ is invertible, that is,~$T$ is an automorphism of the
linear phase space~$\Phase$. 
\end{lemma}

What this means is that the formalization of quantization using Weyl
systems is best suited to the case when physical symmetries---in
particular, time evolution---are linear.

\begin{proof}
Assuming~$\gamma$ is a $*$-algebra endomorphism, 
$$
\gamma\bigl(\W(f)\bigr)\gamma\bigl(\W(h)\bigr)=\gamma\bigl(\W(f)\W(h)\bigr)
$$
so, applying the definition of~$T^*$ on the left-hand side and the Weyl
relations on the right-hand side,
$$
\W(T^*f)\W(T^*h)=\gamma\bigl(e^{\omega(f,h)/2i}\W(f+h)\bigr).
$$
Now, the Weyl relations on the left-hand side and the properties
of~$\gamma$ on the right-hand side imply
$$
e^{\omega(T^*f,T^*h)/2i}\W(T^*f+T^*h)=e^{\omega(f,h)/2i}\W\bigl(T^*(f+h)\bigr).
$$ 
Since all the~$\{\W(f)\}_{f\in\Phase^*}$ are linearly independent by
construction, it follows that~$T^*$ is additive and
preserves~$\omega$. Finally, continuous additive functions are linear.
\end{proof}

The converse of this result is also true. 

\begin{lemma}\label{lem:symmetry}
If~$T\from\Phase\to\Phase$ is an automorphism of the linear phase
space~$\Phase$, then there exists a unique~$*$-algebra
automorphism~$\gamma(T)\from\W(\Phase^*,\omega)\to\W(\Phase^*,\omega)$
determined by
$$
\gamma(T)\suchthat\W(T^*f)\mapsto\W(f)
\qquad\hbox{for all}\quad
f\in\Phase^*
$$ 
and such that~$\gamma(ST)=\gamma(S)\gamma(T)$.
\end{lemma}

In other words,~$\gamma$ is the unique representation of the group of
symplectic automorphisms of~$(\Phase^*,\omega)$ as~$*$-algebra
automorphisms of~$\W(\Phase^*,\omega)$ mapping the set of
generators~$\{\W(f)\suchthat f\in\Phase^*\}$ to itself. This result is
related to~\cite[Corollary 5.1.1]{BSZ}.

\begin{proof}
Applying~$\gamma(T)$ to both sides of the Weyl relation
$$
W(T^*f)W(T^*g)=e^{\omega(T^*f,T^*g)/2i}W\bigl(T^*(f+g)\bigr)
\qquad\hbox{for all}\quad
f,g\in\Phase^*
$$
we obtain
$$
W(f)W(g)=e^{\omega(T^*f,T^*g)/2i}W\bigl(f+g\bigr)
\qquad\hbox{for all}\quad
f,g\in\Phase^*,
$$
so~$\gamma(T)$ is an automorphism because~$T^*$ is symplectic. Also,
if~$S,T\from\Phase\to\Phase$ are two automorphisms of~$\Phase$,
\begin{eqnarray*}
\gamma(S)\gamma(T)\W\bigl((ST)^*f\bigr)
&=&
\gamma(S)\gamma(T)\W(T^*S^*f)\\
&=&
\gamma(S)\W(S^*f)\\
&=&
\W(f)\\
&=&
\gamma(ST)\W\bigl((ST)^*f\bigr)
\end{eqnarray*}
for all~$f\in\Phase^*$.
\end{proof}

\begin{definition}[general boson field]
If~$(\Phase^*,\omega)$ is a symplectic vector space, the \emph{general boson
field} over it is the pair~$(\W,\gamma)$ where~$\W\suchthat
f\mapsto\W(f)$ is the map from~$\Phase^*$ to~$\W(\Phase^*,\omega)$,
and~$\gamma$ is the representation of automorphisms of~$\Phase$
by~$*$-automorphisms of~$\W(\Phase^*,\omega)$ mentioned in
Lemma~\ref{lem:symmetry}.
\end{definition}

\begin{proof}[Note]
This definition is implicit in \cite[\S 5.3]{BSZ}.
\end{proof}

In sum, given any linear phase space space~$\Phase$ with
dual~$(\Phase^*,\omega)$ one can construct the associated Weyl
algebra~$\W(\Phase^*,\omega)$, which supports a
representation~$\gamma$ of the automorphisms of~$\Phase$
as~$*$-algebra automorphisms of~$\W(\Phase^*,\omega)$. In addition,
any Weyl system on~$(\Phase^*,\omega)$, that is, any strongly
continuous representation of~$\W(\Phase^*,\omega)$ as unitary
operators on a complex Hilbert space~$\K$ provides a realization of
the Heisenberg commutation relations. This is the general boson field
on~$\Phase$.

\subsubsection{Correspondence principle}

The general boson field realizes the canonical commutation relations
and the physical symmetries of a linear system, but it does not
provide a complete quantization of a linear phase space, as there are
a few lingering issues. The first is how to actually construct Weyl
systems. The second is whether the correspondence principle is
satisfied. The third is whether physical symmetries are implemented
unitarily on the supporting Hilbert space of the Weyl system. It turns
out that all three are related. In this section we will first use the
Gel'fand--Na\u{\i}mark--Segal construction to produce Weyl systems,
and then use the correspondence principle and unitary implementability
of physical symmetries to select the Weyl systems that produce
physically sensible quantizations.

The following example constructs the so-called \emph{Schr\"odinger
representation}\index{Schr\"odinger representation} of the Heisenberg
commutation relations\index{Heisenberg commutation relations} in one
dimension.

\begin{proof}[Example]
We choose units such that~$\hbar=1$. Let~$\K=L^2(\R)$ and, for
each~$f=(a,k)\in\R^2$, define
$$
\bigl[W(f)\psi\bigr](x)=e^{-ik(x-a/2)}\psi(x-a)
\qquad\hbox{for all}\quad
\psi\in\K,
$$
which clearly makes~$W(f)$ a unitary operator on~$\K$. Also, 
$$
W(f)W(f')=e^{(ka'-k'a)/2i}W(f+f')
$$
so~$W$ is a Weyl system\index{Weyl system} on the linear phase
space~$\Phase=\R^2$ with
$$
\Phase^*=\bigl\{f=(a,k)\in\R^2\bigr\}
\qquad\hbox{and}\quad
\omega(f,f')=ka'-k'a.
$$
The Heisenberg operators are given by
$$
\Phi(f)\psi(x)=(kx-ia\partial_x)\psi(x).
$$
This Heisenberg system is called the Schr\"odinger
representation\index{Schr\"odinger representation}.

Given that~$\Phi$ is linear, it might seem odd that the momentum
coordinate~$k$ appears as the coefficient of the operator of
multiplication by~$x$, which we would usually with the position
operator. In addition, the symplectic structure~$\omega(f,f')=ka'-k'a$
seems backwards. We now proceed to explain these features of the
representation.

The configuration space is~$\R$ with coordinate function~$q\from\R
\to\R$ satisfying~$q(x)=x$, and the phase space is~$\Phase=\R^2$ with
coordinate functions~$q,p\from\R^2\to\R$ ($p$ being the momentum
coordinate function). Then,~$\dd p$ and~$\dd q$ are a basis
of~$\Phase^*$, and~$(a,k)$ are coordinates on~$\Phase^*$ with respect
to that basis. That is, we identify~$f=(a,k)$ with~$f=a\dd p+k\dd
q$. This is the correct pairing despite what our intuition might
suggest, namely pairing~$a$ with~$q$ since they both refer to the same
quantity (position), because~$ps+qk$ has homogeneous units of action
while~$qs+pk$ is not a homogeneous quantity. We are, in fact, omitting
factors of Planck's constant~$\hbar$ as we have chosen `natural
units' in which~$\hbar=1$ according to custom.

The linear observables~$q$ (position) and~$p$ (momentum) on~$\Phase$
have Poisson bracket
$$
\{q,p\}=1.
$$
Accordingly, the dual~$\Phase^*$ is generated by~$\dd q,\dd p$ with
symplectic structure
$$
\omega(\dd q,\dd p)=\{q,p\}=1.
$$
In other words,
$$
\Phase^*=\{f=k\dd q+a\dd p\suchthat a,k\in\R\}
$$
and the symplectic structure on~$\Phase^*$ is
$$
\omega(f,f')=\omega(k\dd q+a\dd p,k'\dd q+a'\dd p)=ka'-k'a.
$$
So, the apparently contradictory
$$
\{q,p\}=1
\qquad\hbox{and}\quad
\omega\bigl((a,k),(a',k')\bigr)=ka'-k'a
$$
are entirely consistent. Then, we have
$$
\Phi(\dd q)\phi(x)=x\phi(x)
\qquad\hbox{and}\quad
\Phi(\dd p)\phi(x)=-i\partial_x\phi(x)
$$
as expected, and if~$f=(a,k)$,
$$
\Phi(f)=a\Phi(\dd p)+k\Phi(\dd q).
$$
\end{proof}

It is clear how this representation can be extended to any finite
number of dimensions, and by the theorem of von~Neumann alluded to
after Lemma~(\ref{lem:WeylHeis}), these representations are unique up
to unitary equivalence. For the infinite-dimensional case relevant to
field theories, though, one needs to use the
Gel'fand--Na\u\i{}mark--Segal
construction\index{Gel'fand--Na\u\i{}mark--Segal construction}, which
is based on the concept of a \emph{state}\index{state!GNS state} and
leads to possibly unitarily inequivalent representations.

\begin{definition}[GNS state]\label{def:GNS}
A \emph{state}\index{state!GNS state} on a~$*$-algebra~$A$ is a linear
functional
$$
\langle~\rangle\from A\to\C
$$
which is \emph{nonnegative}
$$
\langle a^*a\rangle\ge 0
\qquad\hbox{for all}\quad
a\in A,
$$
and \emph{normalized}
$$
\langle 1\rangle=1.
$$ 
\end{definition}

\begin{proof}[Note]
The usage here is completely analogous to that for linear functionals
on vector spaces. A purely algebraic definition of linear functional
on a vector space requires that it be defined everywhere, but when a
topology is introduced one finds it useful to consider discontinuous,
densely-defined linear functionals. In the same vein, as long as the
algebra~$A$ is not assumed to have a topology, one must require that
states be defined on all of~$A$. However, if~$A$ has a topology making
addition and multiplication continuous, then one can talk about
continuous or bounded states, and also about discontinuous,
densely-defined states. At this point,~$\W(\Phase^*,\omega)$ does not
have a topology defined on it so states on it should be defined
everywhere. On the other hand, the Weyl system~$W(\Phase^*,\omega)$
on~$\K$ is given the strong operator topology, and so densely-defined
states make sense on it. In fact, we will use a state
on~$\W(\Phase^*,\omega)$ to construct~$\K$, and it is not guaranteed
that the state will be everywhere defined on it.
\end{proof}

A state\index{state!GNS state} on~$\W(\Phase^*,\omega)$ defines a
nonnegative-definite sesquilinear form~$\langle~\mid~\rangle$
on~$\W(\Phase^*,\omega)$ by means of
$$
\langle \W\mid\W'\rangle\colon=\langle \W^*\W'\rangle
\qquad\hbox{for all}\quad
\W,\W'\in\W(\Phase^*,\omega).
$$
Note that,
since~$\langle\W(f)\mid\W(g)\rangle=\langle\W(-f)\W(g)\rangle$, 
\begin{equation}
\label{eq:InnProd}
\langle\W(f)\mid\W(g)\rangle=e^{i\omega(f,g)/2}\langle\W(g-f)\rangle
\qquad\hbox{for all}\quad
f,g\in\Phase^*.
\end{equation}
The associated nonnegative quadratic form
$$
|\W|^2=\langle\W|\W\rangle
$$ 
is finite on all of~$\W(\Phase,\omega)$, since
$$
\bigl|\W(f)\bigr|=1
\qquad\hbox{for all}\quad
f\in\Phase^*.
$$ 
However, it can only be guaranteed to be a seminorm, because it is
possible that~$\langle~\rangle$ has a kernel. However, this kernel is
necessarily invariant under multiplication by elements
of~$\W(\Phase^*,\omega)$. Indeed, that~$|\W|=0$ is equivalent
to~$\langle\W(f)\mid\W\rangle=0$ for all~$f\in\Phase^*$. But then
$$
\langle\W(f)\mid\W(g)\W\rangle=\langle e^{\omega(f,g)/2i}\W(f-g)\mid\W\rangle
\qquad\hbox{for all}\quad
f\in\Phase^*
$$
implies that~$|\W(g)\W|=0$ for all~$g\in\Phase^*$.

By the standard
procedure---namely, taking the quotient of~$\W(\Phase^*,\omega)$ by
the null subspace of~$|~|$ and completing the result with respect
to~$|~|$ (which is a norm after quotienting by the null
subspace)---one can construct a complex Hilbert space~$\K$ with inner
product~$\langle~\mid~\rangle$. The invariance of the null space
of~$|~|$ under the multiplicative action of~$\W(\Phase^*,\omega)$
implies that~$\W(\Phase^*,\omega)$ acts on~$\K$.

This is a version of the Gel'fand--Na\u\i{}mark--Segal
construction\index{Gel'fand--Na\u\i{}mark--Segal construction}. We now
show that we can give a description of~$\K$ in terms of the phase
space~$\Phase$. For this, we draw the following definition
from~\cite[\S 5.3]{BSZ}.

\begin{definition}[characteristic functional]\label{def:genFunct}
If~$\langle~\rangle$ is a state on the Weyl
algebra~$\W(\Phase^*,\omega)$, its \emph{characteristic
functional}~$\mu\maps\Phase^*\to\C$ is given by 
\begin{equation}
\label{eq:genFunct}
\mu(f)\colon=\langle\W(f)\rangle
\qquad\hbox{for all}\quad
f\in\Phase^*.
\end{equation}
We say the state~$\langle~\rangle$ is~\emph{regular} if, for
every~$f\in\Phase^*$, the function 
$$
t\mapsto\mu(tf)
\qquad(t\in\R)
$$
is twice differentiable at~$t=0$. 
\end{definition}

\begin{proof}[Note]
We will find it convenient to introduce the following notation:
$$
\partial_f\mu(g)=\left.{\partial\over\partial t}\right|_{t=0}\mu(g+tf).
$$
\end{proof}

\begin{theorem}
\label{thm:genBosField}
Let~$(\Phase^*,\omega)$ be a symplectic vector space. Then, given a
regular state~$\langle~\rangle$ on~$\W(\Phase^*,\omega)$ with
characteristic function~$\mu$, there is an~$x\in\Phase$ such that
$$
i\partial_f\mu(0)=f(x)
\qquad\hbox{for all}\quad
f\in\Phase^*.
$$
Then, the collection of formal
symbols~$\Psi=\bigl\{\ket{x+f^*}\suchthat f\in\Phase^*\bigr\}$
generates a complex vector space with the following properties: 
\begin{enumerate}
\item the sesquilinear form
\begin{equation}\label{eqn:innProd}
\langle x+f^*\mid x+g^*\rangle=e^{\omega(g,f)/2i}\mu(g-f)
\end{equation}
makes the span of~$\Psi$ into a complex pre-Hilbert space whose
Hilbert space completion is denoted~$\K$
\item there is a Weyl system~$W\from\Phase^*\to U(\K)$
on~$(\Phase^*,\omega)$, given by 
\begin{equation}\label{eq:WeylSystem}
W(f)\ket{x+g^*}=e^{\omega(f,g)/2i}\ket{x+f^*+g^*}
\qquad\hbox{for all}\quad
f,g\in\Phase^*
\end{equation}
\item the unit vector~$\ket{x}\in\K$ is a cyclic vector of the Weyl
system~$W(\Phase^*,\omega)$
\item the associated Heisenberg system~$\Phi\from\Phase^*\to L(\K)$
satisfies
$$
\matElem{x+g^*}{\Phi(f)}{x+g^*}=f(x+g^*)
\qquad\hbox{for all}\quad
f,g\in\Phase^*.
$$
\end{enumerate}  
\end{theorem}

The last property states that the Heisenberg system obtained from the
regular state~$\langle~\rangle$ satisfies the correspondence
principle. Namely, the expected value of the quantum
observable~$\Phi(g)$ in the quantum state~$\ket{x+f^*}$ equals the
value of the classical observable~$g\in\Phase^*$ in the classical
state~$x+f^*\in\Phase$. 

Because we have not assumed that the symplectic structure~$\omega$
makes the duality map~$*\from\Phase^*\to\Phase$ onto, it is possible
that~$x\neq f^*$ for any~$f\in\Phase^*$, in which case the collection
of indices~$\{x+f^*\suchthat f\in\Phase^*\}$ is an affine subspace
of~$\Phase$.
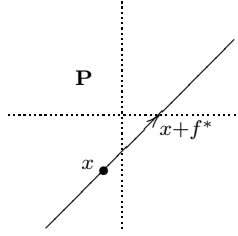
\begin{figure}
$$
\begin{xy}
(-5,5)*{\scriptstyle\Phase};
(-15,0)*{};(15,0)**\dir{.};
(0,-15)*{};(0,15)**\dir{.};
(-10,-15)*{};(15,10)**\dir{-}?(.3)*\dir{*}+(-2,1)*{\scriptstyle
    x}?(.6)*\dir{>}+(0,-2)*\rlap{$\scriptstyle x+f^*$};
\end{xy}
$$
\caption{schematic representation of the relative coherent states as
  an affine subspace of phase space}
\end{figure}
In other words, if~$\omega$ is only
weakly and not strongly nondegenerate, not every classical state
in~$\Phase$ has a counterpart in~$\K$.

The physical interpretation of the vector~$x\in\Phase$ is that of a
classical `background' field configuration,
since~$\matElem{x}{\Phi(f)}{x}=f(x)$ for every classical linear
observable~$f\in\Phase^*$. Clearly any other density operator in~$\K$
can be used to define a state leading to a unitarily equivalent Weyl
system, possibly with a different background field
configuration. If~$x$ is not of the form~$g^*$ for any~$g\in\Phase^*$,
it will actually be impossible to eliminate the background altogether
by a unitary change of representation.

Finally, the fact that the span of~$\Psi$ is dense in the Hilbert
space~$\K$ will be used consistently in the sequel to 
characterize densely defined linear operators and sesquilinear forms
on~$\K$.

\begin{proof}
This proof has a curious way of pulling itself up by its own
bootstraps: the main conceptual difficulty is that, in order to show
that~$f\mapsto i\partial_f\mu(0)$ is a continuous
linear functional on~$\Phase^*$ one needs to have the Weyl system~$W$
in place. We proceed by constructing~$\Psi$ and~$\K$ before
the names~$\ket{x+f^*}$ are available, and then renaming the vectors
after~$x$ is shown to have the advertised properties.

We will temporarily denote by~$\psi_f\in\K$ the image of~$\W(f)$ under
the GNS construction described immediately before
Definition~\ref{def:genFunct}. We denote~$\Psi=\{\psi_f\suchthat
f\in\Phase^*\}$. It follows immediately from Equation~(\ref{eq:InnProd})
that
\begin{equation}\label{InnProd}
\langle\psi_f\mid\psi_g\rangle=e^{i\omega(f,g)/2}\mu(g-f)
\qquad\hbox{for all}\quad
f,g\in\Phase^*.
\end{equation}
The span of~$\Psi$, which consists of unit vectors, is dense
in~$\K$ with respect to this inner product. Recall also that, in the
lead-up to Definition~\ref{def:genFunct}, we showed that the action
of~$\W(\Phase^*,\omega)$ on itself by left multiplication projects
to~$\K$ by virtue of the invariance of the null space
of~$\langle~\mid~\rangle$. This action, namely the Weyl relations,
passes to the quotient as
\begin{equation}\label{eq:preWeylSystem}
W(f)\psi_g=e^{\omega(f,g)/2i}\psi_{f+g}
\qquad\hbox{for all}\quad
f,g\in\Phase^*.
\end{equation}

We are now ready to construct a Weyl system~$W\from\Phase^*\to
U(\K)$. The following lemma shows that the Hilbert space~$\K$
automatically supports a Weyl sistem on~$\W(\Phase^*,\omega)$.

\begin{lemma}\label{lem:Weyl}
Suppose that a regular state~$\langle~\rangle$ is given on the Weyl
algebra~$\W(\Phase,\omega)$ and the GNS construction is performed
resulting in the Hilbert space~$\K$, as just described. Then, Equation~(\ref{eq:preWeylSystem})
defines a map~$W\from\Phase^*\to U(\K)$ which is a Weyl
system\index{Weyl system} on~$(\Phase^*,\omega)$. In addition,
the unit vector~$\psi_0\in\K$ is a cyclic vector of the Weyl
system~$W\from\Phase^*\to U(\K)$.
\end{lemma}

\begin{proof}
First, we need to show that~$W(f)\in U(\K)$ for
all~$f\in\Phase^*$. Indeed, observe that~$W(f)$ maps~$\Psi$ to itself
and that, for all~$f,g,h\in\Phase^*$,
\begin{eqnarray*}
\langle W(f)\psi_g\mid W(f)\psi_h\rangle
&=&
e^{\omega(f,h-g)/2i}\langle\psi_{f+g}\mid\psi_{f+h}\rangle\\
&=&
e^{\omega(f,h-g)/2i}e^{i\omega(f+g,f+h)/2}\mu(h-g)\\
&=&
e^{i\omega(g,h)/2}\mu(h-g)=\langle\psi_g\mid\psi_h\rangle
\end{eqnarray*}
This implies that $W(f)$ is an invertible isometry on the span
of~$\Psi$. Then, by density of the span of~$\Psi$ in~$\K$ and
linearity, it follows that~$W(f)$ is unitary on~$\K$. 

Now, we need to show that, for all~$f,g,h\in\Phase^*$,
$$
W(f)W(g)\,\psi_h=e^{\omega(f,g)/2i\,}W(f+g)\,\psi_h.
$$
The left-hand side is equal to
$$
W(f)e^{\omega(g,h)/2i}\psi_{g+h}=
e^{\omega(g,h)/2i}\,e^{\omega(f,g+h)/2i}\,\psi_{f+g+h},
$$
and the right-hand side is equal to
$$
e^{\omega(f,g)/2i}W(f+g)\psi_h=e^{\omega(f,g)/2i}\,e^{\omega(f+g,h)/2i}\,\psi_{f+g+h}.
$$

To show strong continuity of the Weyl system~$W$ we need to show that,
if~$f_n\to f$ in~$\Phase^*$, then~$W(f_n)\to W(f)$ in the strong
operator topology on~$U(\K)$. To this end, we consider
$$
\bigl[W(f)-W(g)\bigr]\psi_h=e^{\omega(f,h)/2i}\psi_{f+h}-e^{\omega(g,h)/2i}\psi_{g+h}.
$$
Then, 
$$
\bigl\|\bigl[W(f)-W(g)\bigr]\psi_h\bigr\|^2=2\Real\bigl[1-
e^{i\omega(f-g,h)+i\omega(f,g)/2}\mu(g-f)\bigr],
$$ 
which indeed vanishes as~$f-g\to 0$ because of the continuity
of~$\omega$ and~$\mu$ and the antisymmetry of~$\omega$.

Finally, the unit vector~$\psi_0\in\K$ is a cyclic vector of the Weyl
system~$W\from\Phase^*\to U(\K)$ because~$W(f)\psi_0=\psi_f$ for
all~$f\in\Phase^*$, and the collection of all~$\psi_f$ is dense
in~$\K$.
\end{proof}

We now study the Heisenberg system associated to the Weyl system
defined in Lemma~\ref{lem:Weyl}.

\begin{lemma}\label{lem:Heis}
In the hypotheses of
Lemma~\ref{lem:Weyl},~$\langle\psi_g\mid\Phi(f)\psi_g\rangle$
and~$\|\Phi(f)\psi_g\|$ are both finite for
all~$f,g\in\Phase^*$. Moreover, 
$$
\langle\psi_g\mid\Phi(f)\psi_g\rangle=\omega(f,g)+\langle\psi_0\mid\Phi(f)\psi_0\rangle
$$
and
$$
\|\Phi(f)\psi_g\|^2-\|\Phi(f)\psi_0\|^2=\langle\psi_g\mid\Phi(f)\psi_g\rangle^2-\langle\psi_0\mid\Phi(f)\psi_0\rangle^2.
$$
\end{lemma}

\begin{proof}
Observe that, if~$\psi_g$ is in the domain of~$\Phi(f)$, then
$$
\langle\psi_g\mid\Phi(f)\psi_g\rangle=i\left.{\partial\over\partial t}\right|_{t=0}
\langle\psi_g\mid W(tf)\psi_g\rangle
$$
and
$$
\|\Phi(f)\psi_g\|^2=\langle\psi_g\mid\Phi(f)^2\psi_g\rangle=-\left.{\partial^2\over\partial t^2}\right|_{t=0} 
\langle\psi_g\mid W(tf)\psi_g\rangle.
$$ 
Conversely, since~$\Phi(f)$ is a closed operator, the finiteness
of~$-\left.(\partial^2/\partial t^2)\right|_{t=0} \langle\psi_g\mid
W(tf)\psi_g\rangle$ would imply that~$\psi_g$ is in the domain
of~$\Phi(f)$. We now show this.

First we use the definition of the Weyl system given in
Lemma~\ref{lem:Weyl} to compute the matrix elements of the unitary
operator~$W(f)$ between arbitrary elements of~$\Psi$:
$$
\langle\psi_g\mid
W(f)\psi_h\rangle=e^{\omega(f,g+h)/2i+i\omega(g,h)/2}\mu(f-g+h)
\qquad\hbox{for all}\quad f,g,h\in\Phase^*.
$$ 
When~$f=0$, this matrix element reduces to Equation~(\ref{InnProd})
for~$\langle\psi_g\mid\psi_h\rangle$

Differentiating the matrix element~$\langle\psi_g\mid
W(tf)\psi_g\rangle$ twice with respect to~$t$ and setting~$t=0$ one obtains
$$
\|\Phi(f)\psi_g\|^2=\bigl[\omega(f,g)\bigr]^2+2i\omega(f,g)\partial_f\mu(0)-\partial^2_f\mu(0),
$$ 
which is finite by the assumption that~$\mu(tf)$ is
twice-differentiable. Particularizing to~$g=0$ we obtain 
$$
\partial^2_f\mu(0)=-\|\Phi(f)\psi_0\|^2.
$$

One obtains the matrix elements of the Heisenberg
operator~$\Phi(f)$ by differentiating the matrix
element~$\langle\psi_g\mid W(tf)\psi_h\rangle$ with respect to~$t$ and
setting~$t=0$, namely:
$$
\langle\psi_g\mid
\Phi(f)\psi_h\rangle=\Bigl[{1\over 2}\omega(f,g+h)\mu(h-g)+i\partial_f\mu(h-g)\Bigr]e^{i\omega(g,h)/2}.
$$
If, in particular, $h=g$,
$$
\langle\psi_g\mid
\Phi(f)\psi_g\rangle=\omega(f,g)+i\partial_f\mu(0).
$$
The case~$g=0$ shows that
$$
i\partial_f\mu(0)=\langle\psi_0\mid\Phi(f)\psi_0\rangle,
$$ 
and the result follows by elementary algebraic manipulations.
\end{proof}

At this point, we can assert that
$$
i\partial_f\mu(0)=\langle\psi_0\mid\Phi(f)\psi_0\rangle=f(x)
$$ 
for some~$x\in\Phase$ (had we not assumed that~$\Phase$ is
reflexive, we could only deduce that~$x\in\Phase^{**}$). This takes
care of the first conclusion of the theorem. If we now make the
identification~$\psi_f\sim\ket{x+f^*}$, it follows that
$$
\matElem{x+g^*}{\Phi(f)}{x+g^*}=f(x+g^*)
$$
because~$\omega(f,g)=f(g^*)$.
\end{proof}

\begin{definition}[relative coherent states]
\label{def:coh}
Given a regular state~$\langle~\rangle$ on~$\W(\Phase,\omega)$, the
element~$x\in\Phase$ such
that
$$
i\partial_f\mu(0)=f(x)
\qquad\hbox{for all}\quad
f\in\Phase^*
$$ 
is called the \emph{background} for~$\langle~\rangle$.
The image of~$\W(f)$ inside~$\K$ by the GNS construction, denoted
by~$\ket{x+f^*}$, is called a \emph{coherent state
relative}\index{relative coherent state} to the
state~$\langle~\rangle$. We denote the set of 
relative coherent states by~$\Psi=\{\ket{x+f^*}\colon f\in\Phase^*\}$.
\end{definition}

\begin{proof}[Note]
One of the conclusions of Lemma~\ref{lem:Heis} is that the variance
(mean-square deviation from the mean) of
the observable~$\Phi(g)$ in state~$\ket{x+f^*}$ is
$$
\var_{x+f^*}(g)=\matElem{x+f^*}{\Phi(g)^2}{x+f^*}-\matElem{x+f^*}{\Phi(g)}{x+f^*}^2,
$$ 
which is independent of~$f\in\Phase^*$. In other words, the standard
deviation of each observable~$\Phi(g)$ is the same on all relative
coherent states. Note that we are not claiming that the relative
coherent states are minimal-uncertainty states in the sense that they
saturate the inequality in Heisenberg's
uncertainty principle, but it is true that if any one relative
coherent state is a minimal-uncertainty state, all of them will
be. Our definition of relative coherent state includes as
special cases the ordinary coherent states of the harmonic oscillator
and quantum optics, but also the so-called `squeezed states' and many
others, which may or may not be pure states.
\end{proof}

The problem of quantizing a linear phase
space~$(\Phase,\omega)$\index{linear phase space} can thus be partly
solved by finding a state\index{GNS state}~$\langle~\rangle$ on the
Weyl algebra~$\W(\Phase,\omega)$\index{Weyl algebra}. This leads to a
Weyl system on~$(\Phase^*,\tilde\omega)$ and so to Heisenberg
operators~$\Phi(f)$ satisfying the canonical commutation relations and
the correspondence principle, albeit possibly with a nontrivial
background.

Still, the canonical commutation relations and the correspondence
principle together are far from sufficient to uniquely determine the
quantization and, unless~$\Phase$ is finite-dimensional, different
states may lead to unitarily inequivalent Weyl systems\index{Weyl
system}. The problem remains how to construct or identify
representations suitable for particular physical applications. In the
next section we investigate the implications of requiring that
physical symmetries, in particular time evolution, be implemented
unitarily.

\subsubsection{Unitary representation of physical symmetries}

Having quantized the phase space itself, we now consider the
quantization of dynamics and, more generally, physical symmetries. The
ultimate goal is to represent physical symmetries as unitary operators
on the quantum state space~$\K$. The linear phase spaces we are
considering have been defined as topological vector spaces whose duals
are symplectic vector spaces, and are associated to a natural concept
of automorphism. Here we limit our attention to those physical symmetries
which can be represented by automorphisms of the physical phase space
in the sense of~Definition~\ref{def:automorph}.

Putting together Lemma~\ref{lem:symmetry}, Equation~(\ref{eq:WeylSystem}) and
Definition~\ref{def:coh}, we obtain the following result.

\begin{lemma}\label{lem:Gamma}
Assume that~$\langle~\rangle$ is a regular state
on~$\W(\Phase^*,\omega)$, with background~$x\in\Phase$. Given any automorphism~$T\from\Phase\to\Phase$ of the linear phase
space~$\Phase$, there is a densely defined linear
map~$\Gamma(T)\from\K\to\K$ such that
\begin{equation}
\label{eq:symAct}
\Gamma(T)\ket{x+Tf^*}=\ket{x+f^*}.
\end{equation}
This map intertwines the unitary
operators~$W(f)$, that is,
\begin{equation}
\label{eq:intertwine}
\Gamma(T)W(T^*f)=W(f)\Gamma(T)
\qquad\hbox{for all}\quad
f\in\Phase^*,
\end{equation}
and satisfies~$\Gamma(ST)=\Gamma(S)\Gamma(T)$. 
\end{lemma}

It is worth remarking that, when the background~$x\in\Phase$ is
a fixed point of the automorphism~$T\from\Phase\to\Phase$, we have the
nicer formula
$$
\Gamma(T)\ket{Ty}=\ket{y}
\qquad\hbox{for all}\quad
y=x+f^*
\qquad\hbox{with}\quad
f\in\Phase^*.
$$

\begin{proof}
By Lemma~\ref{lem:symmetry}, there is a unique
automorphism~$\gamma(T)$ of the $*$-algebra~$\W(\Phase,\omega)$ such
that
$$
\gamma(T)\W(T^*f)=\W(f)
\qquad\hbox{for all}\quad
f\in\Phase^*
$$ 
and satisfying~$\gamma(ST)=\gamma(S)\gamma(T)$. The GNS construction
preceding Definition~\ref{def:coh} produces a unique densely-defined
linear operator~$\Gamma(T)\from\K\to\K$ defined on the dense span
of~$\Psi$ by Equation~(\ref{eq:symAct}). In
addition, since
$$
\Gamma(S)\Gamma(T)\ket{x+STf^*}=\Gamma(S)\Gamma(T)\ket{x+{T^*S^*f}^*}=\Gamma(S)\ket{x+Sf^*},
$$
$\Gamma(ST)=\Gamma(S)\Gamma(T)$ holds on the span of the relative
coherent states, which is a common dense domain of all three operators
and is left invariant by them.

As for the intertwining of the Weyl operators,
Equation~(\ref{eq:WeylSystem}) implies
$$
\Gamma(T)W(T^*f)\ket{x+Tg^*}=e^{\omega(T^*f,T^*g)/2i}\Gamma(T)\ket{x+T(f^*+g^*)}
$$
which, because~$T^*$ preserves~$\omega$ and by definition
of~$\Gamma(T)$, equals 
$$
e^{\omega(f,g)/2i}\ket{x+f^*+g^*}.
$$
Similarly,
$$
W(f)\Gamma(T)\ket{x+Tg^*}=W(f)\ket{x+g^*},
$$
and the result follows, again by Equation~(\ref{eq:WeylSystem}).
\end{proof}

Perhaps surprisingly,~$\Gamma(T)$ is not necessarily an isometry
of~$\K$ despite the fact that it preserves the norm of all the
relative coherent states~$\ket{x+f^*}$. However, it should not be
surprising that unitarity is obtained when the characteristic
functional~$\mu$ is preserved by~$T^*$.

\begin{lemma}\label{lem:unitary}
In the hypotheses of Lemma~\ref{lem:Gamma}, the operator~$\Gamma(T)$
extends uniquely to a unitary operator on~$\K$ if, and only if,~$T$
preserves~$\langle~\rangle$ in the sense that
$$
\mu(T^*h)=\mu(h)
\qquad\hbox{for all}\quad
h\in\Phase^*.
$$
\end{lemma}

In other words, the invertible operator~$\Gamma(T)$ is unitary on~$\K$
if, and only if, the characteristic functional~$\mu$ is constant on orbits
of~$T^*$.

\begin{proof}
From Equation~(\ref{eqn:innProd}) it follows that
$$
\langle x+Tf^*\mid
x+Tg^*\rangle=e^{\omega(T^*g,T^*f)/2i}\mu\bigl(T^*(g-f)\bigr)
$$ 
so~$\Gamma(T)$ is an isometry on the span of the relative coherent
states if, and 
only if,~$\mu(T^*h)=\mu(h)$ for
all~$h\in\Phase^*$. An isometry is unitary if and only if it is
invertible.
\end{proof}

It follows that, for a whole subgroup~$G$ of automorphisms of~$\Phase$
to be unitarily implemented on~$\K$ by~$\Gamma$, it is necessary and
sufficient that~$\mu$ be constant on the orbits of the whole
subgroup. It is possible that unitary representations other
than~$\Gamma$ exist, and in fact that is guaranteed when~$\Phase^*$ is
finite-dimensional. Now, if~$G$ is a continuous group generated by a
Poisson algebra~$\g$ of classical observables on~$\Phase$, this is
equivalent to the characteristic functional~$\mu$ having vanishing
Poisson brackets with all the elements of~$\g$. In particular, if time
evolution is to be implemented unitarily, the characteristic
functional of the state must be a constant of the motion. It is in
this precise sense that the dynamics can be said to determine the
quantization.

\subsection{Summary}

Putting Theorem~\ref{thm:genBosField}
and~Lemmas~\ref{lem:Gamma}--\ref{lem:unitary} together we obtain the
following theorem listing the properties of representations of the
general boson field.

\begin{theorem}
\label{thm:GenBosField}
Let~$(\Phase,\omega)$ be a linear phase space, let~$\langle~\rangle$
be a regular GNS state on the Weyl algebra~$\W(\Phase,\omega)$ with
characteristic function~$\mu$. Let the background~$x\in\Phase$
associated to~$\mu$ be defined by
$$
i\partial_f\mu(0)=f(x)
\qquad\hbox{for all}\quad
f\in\Phase^*,
$$
and let~$\Psi=\{\ket{x+f^*}\mid
f\in\Phase^*\}$. Then, 
\begin{enumerate}
\item the sesquilinear form
$$
\bracket{x+f^*}{x+g^*}=e^{i\omega(f,g)/2}\mu(g-f)
$$
makes the span of~$\Psi$ into a complex pre-Hilbert space whose
Hilbert-space completion is denoted~$\K$
\item there is a Weyl system~$W\from\Phase^*\to U(\K)$
on~$(\Phase^*,\omega)$, given by 
$$
W(f)\ket{x+g^*}=e^{\omega(f,g)/2i}\ket{x+g^*+f^*}
\qquad\hbox{for all}\quad
f\in\Phase,g\in\Phase^*
$$
\item the associated Heisenberg system~$\Phi\from\Phase^*\to L(\K)$
satisfies
$$
\matElem{x+g^*}{\Phi(f)}{x+g^*}=f(x+g^*)
\qquad\hbox{for all}\quad
f,g\in\Phase^*
$$
\item there is a group homomorphism~$\Gamma$
mapping automorphisms~$T\from\Phase\to\Phase$ to invertible linear
operators on~$\K$, given by 
$$
\Gamma(T)\ket{x+Tf^*}=\ket{x+f^*}
\qquad\hbox{for all}\quad
f\in\Phase^*
$$
and satisfying
$$
\Gamma(T)W(T^*f)=W(f)\Gamma(T)
\qquad\hbox{for all}\quad
f\in\Phase^*
$$
\item the unit vector~$\ket{x}\in\K$ is a
cyclic vector of the Weyl system~$W(\Phase^*,\tilde\omega)$
\item $\Gamma(T)$ is unitary if, and only if,~$\mu$ is constant on
orbits of~$T$.
\end{enumerate}
\end{theorem}

Since~$\Gamma$ is defined on symplectic and not unitary
transformations the generators are not self-adjoint and it is not
clear that there is a meaningful notion of positivity of~$\Gamma$, in
contrast with the free boson field below. In other words, there seems
to be no way to define what a stable representation of the general
boson field is. Also, we have not specified a topology on the
automorphisms of~$\Phase$, so we cannot prove continuity of~$\Gamma$.

\begin{proof}
\begin{enumerate}
\item This is Equation~(\ref{eqn:innProd}) from
      Theorem~\ref{thm:genBosField}. 
\item This is the content of Lemma~\ref{lem:Weyl}, which was part of
  the proof of Theorem~\ref{thm:genBosField}. 
\item This is by Lemma~\ref{lem:Heis}, also part of the proof of Theorem~\ref{thm:genBosField}. 
\item This is Lemma~\ref{lem:Gamma}.
\item This is part of the conclusions of Lemmas~\ref{lem:Weyl}
      and~\ref{lem:Gamma}. 
\item This is the content of Lemma~\ref{lem:unitary}.
\end{enumerate}
\end{proof}

\section{The free boson field}

\label{sec:freeBosonField}

In certain cases, the space~$\Phase^*$ of linear observables on the
physical phase-space a classical theory is not only a real symplectic
space, but also admits a complex Hilbert space~$\Hilb$ such that and
the symplectic structure~$\omega$ is the imaginary part of the complex
inner product. This is the algebraic setting in which Segal~\cite{BSZ}
defined his concept of a free boson field, which is an axiomatic
definition of the usual Fock representation of free quantum fields. In
the present section we develop tools and techniques specific to Fock
quantization and that will be needed later on.

\begin{definition}[free boson field]
The \emph{free boson field} over a complex Hilbert space~$\Hilb$
consists of
\begin{enumerate}
\item a complex Hilbert space~$\K$
\item a Weyl system~$W\from\Hilb\to U(\K)$
\item a continuous representation~$\Gamma\from U(\Hilb^\dagger)\to U(\K)$
  satisfying
$$
\Gamma(U)W(z)\Gamma(U)^{-1}=W(Uz)
\qquad\hbox{for all}\quad
z\in\Hilb
$$
\item a unit vector~$\nu\in\K$ which is invariant under~$\Gamma(U)$
  for all~$U\in U(\Hilb^\dagger)$ and a cyclic vector of~$W(\Hilb)$
\end{enumerate}
such that $\Gamma$ is positive in the sense that, if the one-parameter
group~$U(t)\subset U(\Hilb^\dagger)$ has a nonnegative self-adjoint
generator~$A$, then~$\partial\Gamma(A)$, which denotes the
self-adjoint generator of the
group~$\Gamma\bigl(U(t)\bigr)\from\K\to\K$, is a nonnegative
self-adjoint operator on~$\K$.
\end{definition}

\begin{proof}[Note] 
The positivity condition can be weakened to apply only to a single
operator~$A$, and the free boson field is unique up to unitary
equivalence~\cite[\S 1.10]{BSZ}.
\end{proof}

Now, if~$\Hilb$ is a complex Hilbert space with inner
product~$\langle~,~\rangle$ and norm~$\|~\|$, and one
defines~$h(f,g)=\Real{\langle f,g\rangle}$
and~$\omega(f,g)=\Imag{\langle f,g\rangle}$, then~$\Hilb$ becomes a
real Hilbert space with inner product~$h$ and norm~$\|~\|$,
and~$\omega$ is a continuous symplectic structure on~$\Hilb$. If we
denote by~$\Phase$ the real dual of~$\Hilb$, then~$\Phase$ is a linear
phase space in the sense of Definition~\ref{def:linearPhaseSpace},
with~$(\Phase^*,\omega)=(\Hilb,\omega)$. In addition, we have a
map~$*\from\Hilb\to\Phase$ defined by
$$
g^*(f)=\omega(f,g)
\qquad\hbox{for all}\quad
f,g\in\Hilb.
$$

With this notation, the following is a consequence of
Theorem~\ref{thm:GenBosField}.

\begin{theorem}\label{thm:FreeBosField}
Let~$\Hilb$ be a complex Hilbert space with inner product~$\langle
~,~\rangle$ and norm~$\|~\|$. Define~$h$, and~$\omega$
on~$\Hilb\cong\Phase^*$ and~$*\from\Hilb\to\Phase$ as above. Then, the
representation of the general boson field on~$\W(\Phase^*,\omega)$
given by the regular state with characteristic functional
$$
\mu(f)=e^{-\|f\|^2/4}
\qquad\hbox{for all}\quad
f\in\Hilb
$$
is the free boson field on~$\Hilb$, with 
\begin{enumerate}
\item $\K$ being the completion of the span
of~$\Psi=\{\ket{f^*}\suchthat f\in\Hilb\}$ with respect to the complex
inner product
$$
\bracket{f^*}{g^*}=e^{\omega(g,f)/2i}e^{-\|g-f\|^2/4}
$$
\item $W$ being the Weyl system on~$\W(\Hilb,\omega)$ given by
$$
W(f)\ket{g^*}=e^{ig^*(f)/2}\ket{g^*+f^*}
\qquad\hbox{for all}\quad
f,g\in\Hilb
$$
\item $\Gamma$ being defined by
$$
\Gamma(U)\ket{f^*}=\ket{(Uf)^*}
\qquad\hbox{for all}\quad
f\in\Hilb
$$
\item $\nu=\ket{0}$
\end{enumerate}
In addition, the mean and variance of~$\Phi(g)$ in the
state~$\ket{x}$ are
$$
\matElem{f^*}{\Phi(g)}{f^*}=\omega(g,f)
\qquad\hbox{and}\quad
\var_{f^*}(g)={1\over 2}\|g\|^2
\qquad\hbox{for all}\quad
x,f\in\Hilb.
$$
\end{theorem}

\begin{proof}
All the numbered properties of the free boson field are immediate
consequences of Theorem~\ref{thm:GenBosField}. It only remains to show
positivity of the representation~$\Gamma$. 

Assume that~$U=e^{-itA}\in U(\Hilb)$ with~$\langle Af,f\rangle=\langle
f,Af\rangle\ge 0$ for all~$f\in\Hilb$. Then,
\begin{eqnarray*}
\matElem{f^*}{\partial\Gamma(A)}{f^*}
&=&
\left.i{\partial\over\partial t}\right|_{t=0}\matElem{f^*}{\Gamma(e^{-itA})}{f^*}\\
&=&
\left.i{\partial\over\partial t}\right|_{t=0}\bracket{f^*}{(e^{-itA}f)^*}\\
&=&
\left.i{\partial\over\partial t}\right|_{t=0}e^{\omega(e^{-itA}f,f)/2i}e^{-\|(e^{-itA}-1)f\|^2/4}\\
&=&
{1\over 2}\omega(-iAf,f)={1\over 2}\langle f,Af\rangle\ge 0\\
\end{eqnarray*}
for all~$f\in\Hilb$. 

As for the mean and variance of the Heisenberg observables, note that
\begin{equation}\label{eq:matElem}
\matElem{f^*}{W(g)}{h^*}=e^{\omega(g,h+f)/2i}e^{\omega(h,f)/2i}e^{-\|h+g-f\|^2/4}.
\end{equation}
In particular, if~$f=h$, 
$$
\matElem{f^*}{W(g)}{f^*}=e^{-i\omega(g,f)}e^{-\|g\|^2/4}.
$$
But this is precisely the characteristic functional of a Gaussian
random variable with mean~$\omega(g,f)$ and variance~${1\over 2}\|g\|^2$.
\end{proof}

\subsection{Normal-ordered functions}

The Weyl system~$W\from\Hilb\to U(\K)$ has an associated real-linear
Heisenberg system~$\Phi\from\Hilb\to L(\K)$. From this real-linear map
we can construct complex-linear and complex-antilinear
maps~$a,a^\dagger\from\Hilb\to L(\K)$ with the help of the complex
structure of~$\Hilb$. The \emph{creation operator}
$$
a^\dagger(f)={\Phi(f)-i\Phi(if)\over\sqrt 2}
\qquad\hbox{for all}\quad
f\in\Hilb
$$
is complex-linear, and its adjoint the \emph{annihilation operator}
$$
a(f)={\Phi(f)+i\Phi(if)\over\sqrt 2}
\qquad\hbox{for all}\quad
f\in\Hilb
$$
is complex-antilinear. The creation and annihilation operators satisfy
the commutation relations
$$
\bigl[a(f),a(g)\bigr]=0,
\qquad
\bigl[a(f),a^\dagger(g)\bigr]=\langle f,g\rangle,
\qquad\hbox{and}\quad
\bigl[a^\dagger(f),a^\dagger(g)\bigr]=0
$$ 
for all $f,g\in\Hilb$.

It is now easy to prove that coherent states are joint eigenstates of every~$a(f)$.

\begin{lemma}\label{lem:annihilation}
If~$f,g\in\Hilb$ then
$$
a(g)\ket{h^*}={\langle g,h\rangle\over i\sqrt 2}\ket{h^*}.
$$
\end{lemma}

\begin{proof}
Equation~(\ref{eq:matElem}) implies that
$$
{\matElem{f^*}{W(g)}{h^*}\over\bracket{f^*}{h^*}}=e^{(\langle
  f,g\rangle-\langle g,h\rangle)/2}e^{-\|g\|^2/4}
$$
so the matrix elements of the Heisenberg operators satisfy
\begin{equation}\label{eq:Heisenberg}
{\matElem{f^*}{\Phi(g)}{h^*}\over\bracket{f^*}{h^*}}={i\over 2}\bigl[\langle
  f,g\rangle-\langle g,h\rangle\bigr]
\end{equation}
which implies
$$
{\matElem{f^*}{\Phi(ig)}{h^*}\over\bracket{f^*}{h^*}}=-{1\over 2}\bigl[\langle
  f,g\rangle+\langle g,h\rangle\bigr]
$$
and so
$$
\matElem{f^*}{a(g)}{h^*}={\langle g,h\rangle\over i\sqrt 2}\bracket{f^*}{h^*}
$$
By the density of the span of the coherent states in~$\K$, the result
follows. 
\end{proof}

We now use this property to prove a remarkable formula for the matrix
elements of `normal-ordered' functions of Heisenberg operators. We
first introduce the definition of normal-ordered powers of Heisenberg
operators, or Wick powers. The Wick powers are obtained by expressing
the Heisenberg operator in terms of creation and annihilation
operators, expanding the product and rearranging each monomial to have
all creation operators to the left of all the annihilation operators,
discarding all commutators. 

\begin{definition}[Wick power]
If~$f\in\Hilb$, the~$n$th \emph{Wick power} or \emph{normal-ordered power} of the Heisenberg operator~$\Phi(f)$ is
the operator on~$\K$ given by
$$
\Wick{\Phi(f)^n}={1\over
    2^{n/2}}\sum_{m=0}^n{n\choose m}a^\dagger(f)^m a(f)^{n-m}.
$$
\end{definition}

We first show that Wick powers are densely defined on~$\K$; what is
more, their domain always contains the coherent states.

\begin{lemma}
For all~$n\in\N$ and all~$f\in\Hilb$, the Wick
power~$\Wick{\Phi(f)^n}$ is densely defined on~$\K$.
\end{lemma}

In other words, for all~$f\in\Hilb$, the coherent states
are~\emph{$C^\infty$ vectors for~$\Phi(f)$} \cite[\S X.6]{RS}.

\begin{proof}
That the domain of~$\Wick{\Phi(f)^n}$ is dense in~$\K$ will follow from
the fact that it contains the coherent states, whose span is dense
in~$\K$. Since the coherent states are eigenstates of the annihilation
operators, it is clear that any power of annihilation operators is
densely defined on~$\K$. Also, the creation operators are defined
on the coherent states because they are linear combinations of the
Heisenberg operators, to which Lemma~\ref{lem:Heis} applies. However,
the question is whether higher powers of the creation operators
are defined on coherent states. Since powers of annihilation operators
are polynomials in the Heisenberg operators, the result will follow if
we can show that arbitrary powers of Heisenberg operators are defined
on coherent states. The techniques used to prove Lemma~\ref{lem:Heis}
generalize to this situation.

Indeed, observe that~$W(f)=e^{-i\Phi(f)}$ implies that
$$
\|\Phi(f_1)\cdots\Phi(f_n)\ket{g^*}\|^2=\matElem{g^*}{\Phi(f_n)\cdots\Phi_{f_1}^2\cdots\Phi(f_n)}{g^*}
$$
equals
$$
\Bigl.i^{2n}{\partial^{2n}\over\partial
t_1\cdots\partial t_{2n}}\Bigr|_{t_i=0}\matElem{g^*}{W(t_1f_n)\cdots
W(t_nf_1)W(t_{n+1}f_1)\cdots W(t_{2n}f_n)}{g^*}.
$$
Since the matrix element is proportional
to~$\mu(t_1f_n+\cdots+t_{2n}f_n)$, it follows
that the squared norm $\|\Phi(f_1)\cdots\Phi(f_n)\ket{g^*}\|^2$ is a linear combination
of derivatives of~$\mu(0)$ of order up 
to~$2n$. 
It is easily checked
that the characteristic functional of the free boson
field,
$$
\mu(f)=e^{-\|f\|^2/4}
$$
is infinitely differentiable, and the
result follows.
\end{proof}

Just how well coherent states and Wick powers get along is made evident by
the following result.

\begin{lemma}\label{lem:Wick}
The matrix elements of Wick powers on coherent states satisfy
$$
{\matElem{f^*}{\Wick{\Phi(g)^n}}{h^*}\over\bracket{f^*}{h^*}}=\Bigl({\matElem{f^*}{\Phi(g)}{h^*}\over\bracket{f^*}{h^*}}\Bigr)^n
$$
whenever~$f,g,h\in\Hilb$.
\end{lemma}

\begin{proof}
By repeated application of Lemma~\ref{lem:annihilation},
\begin{eqnarray*}
\matElem{f^*}{\Wick{\Phi(g)^n}}{h^*}
&=&
{1\over
    2^{n/2}}\sum_{m=0}^n{n\choose m}\matElem{f^*}{a^\dagger(g)^m a(g)^{n-m}}{h^*}\\
&=&
{1\over
    2^{n/2}}\sum_{m=0}^n{n\choose m}\Bigl({\langle f,g\rangle\over -i\sqrt 2}\Bigr)^m\Bigl({\langle
g,h\rangle\over i\sqrt 2}\Bigr)^{n-m}\bracket{f^*}{h^*}\\
&=&\bracket{f^*}{h^*}\Bigl({i\over 2}\bigl[\langle f,g\rangle-\langle g,h\rangle\bigr]\Bigr)^n\\
\end{eqnarray*}
and the result follows by Equation~(\ref{eq:Heisenberg}).
\end{proof}

\subsection{Quasioperators}

\label{sec:quasioperators}

Let us look again at Equation~(\ref{eq:Heisenberg}):
$$
{\matElem{f^*}{\Phi(g)}{h^*}\over\bracket{f^*}{h^*}}={i\over 2}\bigl[\langle
  f,g\rangle-\langle g,h\rangle\bigr]
\qquad\hbox{for all}\quad f,g,h\in\Hilb.
$$ 
In this equation the right-hand side, being multilinear, is much
better behaved as a function of~$f,g,h\in\Hilb$ than one would expect
from the object on the left-hand side: recall that~$f\mapsto\ket{f^*}$
is not a linear map from~$\Hilb$ to~$\K$, and also that~$\Phi(g)$ is
an unbounded operator on~$\K$. This is extremely useful, as
it allows one to make sense of the expression on the left-hand side in
cases where~$f$ is so singular that~$\Phi(f)$ does not exist as an
operator on~$\K$. 

Specifically, suppose that we are given a classical linear
observable~$f$ which is too singular to be an element of~$\Phase^*=\Hilb$. Often-used examples of this come readily to mind, since
typically~$\Phase$ is a space of square-integrable tensor-valued
differential forms on a manifold and these have no pointwise values
nor can they be integrated on submanifolds. Thus, classical
observables such as~$A\mapsto A(x)$ or~$A\mapsto\oint_\gamma A$ do
not, in general, admit quantum analogues defined by the techniques
introduced so far. In the case of the free boson field we can see
explicitly that, if~$\|g\|=\infty$, then any attempt at constructing
the unitary operator~$W(g)$ will fail,
as~$\lim_{\|g\|\to\infty}\matElem{f^*}{W(g)}{h^*}=0$ because it
contains a leading factor of~$e^{-\|g\|^2/4}$. Accordingly, there is
no coherent state~$\ket{g}$ nor is a nonzero Heisenberg
operator~$\Phi(g)$ obtainable by taking derivatives of~$W(g)$.

However, if there is a scale of
spaces~$\Hilb_0\subseteq\Hilb\subseteq\Hilb_0^\dagger$,
Equation~(\ref{eq:Heisenberg}) makes sense
for~$g\in\Hilb_0^\dagger$ as long as~$f,h\in\Hilb_0$. If the
span of the coherent states~$\{\ket{f}\suchthat f\in\Hilb_0\}$ is
dense in~$\K$, then~$\Phi(g)$ is well-behaved enough for most
practical purposes. We now make this idea precise by means of the
concept of \emph{quasioperator}, and prove that things are in fact as
we suggest.

\begin{definition}[quasioperator]
Let~$\K_0$ be a topological vector space with a dense continuous
inclusion into the Hilbert space~$\K$. A \emph{quasioperator} on~$\K$ with
domain~$\K_0$ 
is a continuous
sesquilinear form~$Q\from\K_0\times\K_0\to\C$, antilinear in the first argument and
linear in the second. 
\end{definition}

\begin{proof}[Note]
Whenever there is a scale of
spaces~$\K_0\subseteq\K\cong\K^\dagger\subseteq\K_0^\dagger$, we will refer to
elements of~$\K_0$ as the space of \emph{regular} elements of~$\K$,
and~$\K_0^\dagger$ as the space of \emph{singular} ones. In other
words, a quasioperator on~$\K$ maps regular elements of~$\K$ to
`singular elements of~$\K$'. While possibly hair-raising to the
mathematician, this manner of speaking is actually very useful in
physical reasoning. For instance, we call the Dirac delta a `singular
function' even though it is not, strictly speaking, a function.
%
\end{proof}

We now assume that the $\Hilb_0\subseteq\Hilb$ is a complex
topological vector space and that the inclusion map is continuous,
with dense range. We call the elements of~$\Hilb_0$ \emph{regular
observables}. The map~$*\from\Hilb\to\Phase$ restricts to a
map~$*\from\Hilb_0\to\Phase$ whose image~$\Phase_0$ is the space of
\emph{regular field configurations}. The dual~$\Hilb_0^\dagger$ is the
space of~\emph{singular observables}. Our goal is to extend the
Heisenberg system~$\Phi$ from~$\Hilb$ to~$\Hilb_0^\dagger$. If~$g$ is
a singular observable~$\Phi(g)$ will be defined as a quasioperator.

Recall now that the collection of coherent
states~$\Psi=\{\ket{f^*}\suchthat f\in\Hilb\}$ spans a dense subspace
of the Fock space~$\K$. We will call the coherent states
in~$\Psi_0=\{\ket{f^*}\suchthat f\in\Hilb_0\}$ \emph{regular coherent
states}. We now show that the span of the regular coherent states is
also dense in~$\K$.  

\begin{lemma}
Let~$\Hilb_0\subseteq\Hilb$ be a topological vector space with a dense
continuous inclusion into~$\Hilb$. Then, if~$f_n\in\Hilb_0$ for all~$n$
and~$\lim_{n\to\infty}f_n=f$ in the topology of~$\Hilb$, then
$$
\lim_{n\to\infty}\ket{f_n^*}=\ket{f^*}
$$  
in the topology of~$\K$.
\end{lemma}

\begin{proof}
For all~$g\in\Hilb$,
$$
\bracket{g^*}{f_n^*}-\bracket{g^*}{f^*}=e^{\omega(f_n,g)/2i}\mu(g-f_n)-e^{\omega(f,g)/2i}\mu(g-f).
$$ 
By the continuity of~$\omega$ and~$\mu$ on~$\Hilb$ and the density
of the~$\ket{g^*}$ in~$\K$, the result follows.
\end{proof}

We now let~$\K_0$ be the span of~$\Psi_0$, consisting of finite linear
combinations of regular coherent states, topologized algebraically.
We are then ready to define~$\Phi(g)$ as a quasioperator on~$\K_0$.

\begin{lemma}\label{lem:quasioperator1}
For every~$g\in\Hilb_0^\dagger$ there is a unique quasioperator~$\Phi(g)$
on~$\K$ with domain~$\K_0$ such that
$$
{\matElem{f^*}{\Phi(g)}{h^*}\over\bracket{f^*}{h^*}}
={i\over 2}\bigl[\langle
  f,g\rangle-\langle g,h\rangle\bigr]
\qquad\hbox{for all}\quad f,h\in\Hilb_0.
$$
\end{lemma}

Note that, when~$g\in\Hilb$, the matrix elements of the ordinary
Heisenberg operator~$\Phi(g)$ provide a quasioperator of this form. In
this sense, this construction extends the definition of the Heisenberg
operator~$\Phi(g)$ from regular~$g$ to singular~$g$.

\begin{proof}
Consider the function from~$\Psi_0\times\Psi_0$
$$
\ket{f^*}\times\ket{h^*}\mapsto{i\over 2}\bigl[\langle
  f,g\rangle-\langle g,h\rangle\bigr]\bracket{f^*}{h^*}
\qquad\hbox{for all}\quad
f,h\in\Hilb_0,
$$
which is clearly jointly continuous in the topology of~$\K_0$. This function
extends by linearity to a continous sesquilinear form
on~$\K_0$, and therefore is associated to a quasioperator on~$\K$ with
domain~$\K_0$. 
\end{proof}

An entirely analogous construction generalizes Wick powers of
Heisenberg operators,~$\Wick{\Phi(g)^n}$, from regular~$g\in\Hilb$ to
singular~$g\in\Hilb_0^\dagger$.

\begin{lemma}\label{lem:quasioperator2}
For every~$g\in\Hilb_0^\dagger$ there is a unique quasioperator~$\Wick{\Phi(g)^n}$
on~$\K$ with domain~$\K_0$ such that
$$
{\matElem{f^*}{\Wick{\Phi(g)^n}}{h^*}\over\bracket{f^*}{h^*}}
=\biggl({\matElem{f^*}{\Phi(g)}{h^*}\over\bracket{f^*}{h^*}}\biggr)^n
\qquad\hbox{for all}\quad f,h\in\Hilb_0.
$$
\end{lemma}

\begin{proof}
As before, the function
$$
\ket{f^*}\times\ket{h^*}\mapsto\biggl({i\over 2}\bigl[\langle
  f,g\rangle-\langle g,h\rangle\bigr]\biggr)^n\bracket{f^*}{h^*}
\qquad\hbox{for all}\quad
f,h\in\Hilb_0
$$
on~$\Psi_0\times\Psi_0$
is jointly continuous in the topology of~$\K_0$. Extending it to all of~$\K_0$
by linearity, it defines a quasioperator on~$\K$ with
domain~$\K_0$. 
\end{proof}

%

We can now extend the normal-ordering operation by linearity to the
algebra of polynomials on a Heisenberg operator~$\Phi(g)$, that is,
if~$P(x)=\sum_{k=0}^n p_kx^k$ we define
$$
\Wick{P\bigl(\Phi(g)\bigr)}=\sum_{k=0}^n p_k\Wick{\Phi(g)^k}.
$$
Then, it is easily checked that
$$
\Wick{(P+Q)\bigl(\Phi(g)\bigr)}=\Wick{P\bigl(\Phi(g)\bigr)}+\Wick{Q\bigl(\Phi(g)\bigr)}
$$
for all polynomials~$P,Q\in\C[x]$. This holds both at the level of
operators on~$\K$, if~$g\in\Hilb$, and as an equation between
quasioperators on~$\K_0$.

\begin{corollary}\label{cor:remarkable}
Let~$F\from\C^n\to\C$ be an entire function. Then, for
all~$g\in\Hilb_0^\dagger$, there is a unique
quasioperator~$\Wick{F(\Phi(g))}$ on~$\K$ with domain~$\K_0$ satisfying
$$
{\matElem{f^*}{\Wick{F(\Phi(g))}}{h^*}\over\bracket{f^*}{h^*}}=F\biggl({\matElem{f^*}{\Phi(g)}{h^*}\over\bracket{f^*}{h^*}}\biggr)
\qquad\hbox{for all}\quad
f,h\in\Hilb_0.
$$ 
\end{corollary}

We have proved this formula for single Heisenberg operators in
Equation~(\ref{eq:Heisenberg}), and for monomials of the Heisenberg
operators in Lemma~\ref{lem:Wick}; it also holds for Heisenberg
quasioperators (Lemma~\ref{lem:quasioperator1}) and their Wick powers (Lemma~\ref{lem:quasioperator2}). We have defined the normal-ordering
operator on the entire algebra of polynomials on the Heisenberg
(quasi)operator~$\Phi(g)$ by linearity from the normal-ordered monomials and, since
the operation
$$
X\mapsto{\matElem{f^*}{X}{h^*}\over\bracket{f^*}{h^*}}
$$
is complex linear, our desired formula holds for all polynomials of
Heisenberg operators. 

\begin{proof}
For the proof, we do as before and define a complex function
on~$\Psi_0\times\Psi_0$ by
$$
\ket{f^*}\times\ket{h^*}\mapsto F\biggl({\matElem{f^*}{\Phi(g)}{h^*}\over\bracket{f^*}{h^*}}\biggr)\bracket{f^*}{h^*}
$$
which is jointly  continuous in the topology of~$\K_0$, and extends by
linearity to a sesquilinear form on~$\K_0$ defining a
quasioperator with the required properties.
%
\end{proof}

By analogy with~$W(f)=e^{-i\Phi(f)}$, we can now define
$$
\Wick{W(g)}=\sum_{n\ge 0}{(-i)^n\over n!}\,\Wick{\Phi^n(g)}.
$$
and then
$$
{\matElem{f^*}{\Wick{W(g)}}{h^*}\over\bracket{f^*}{h^*}}=\exp{\matElem{f^*}{-i\Phi(g)}{h^*}\over\bracket{f^*}{g^*}}.
$$
This means that~$\Wick{W(g)}$ is defined as a quasioperator on the
span of the regular coherent states.
We can now deduce the following useful formula.

\begin{lemma}
$$
\Wick{W(g)}={W(g)\over\matElem{0}{W(g)}{0}}
\qquad\hbox{for all}\quad
g\in\Hilb
$$
as an equation between quasioperators on~$\K$ with domain~$\K_0$.
\end{lemma}

This shows that the definition of the normal-ordered Weyl
quasioperator~$W(g)$ for~$g\in\Hilb_0^\dagger$ is analogous to
resolving a singularity of the form~$0/0$ by taking a limit.

\begin{proof}
We particularize Equation~(\ref{eq:matElem}) 
$$
\matElem{f^*}{W(g)}{h^*}=e^{\omega(g,h+f)/2i}e^{\omega(h,f)/2i}e^{-\|h+g-f\|^2/4}
$$
to~$g=0$
$$
\bracket{f^*}{h^*}=e^{\omega(h,f)/2i}e^{-\|h-f\|^2/4}
$$
and, to~$f=h=0$
$$
\matElem{0}{W(g)}{0}=e^{-\|g\|^2/4}.
$$
Then,  
$$
{\matElem{f^*}{W(g)}{h^*}\over\bracket{f^*}{h^*}\matElem{0}{W(g)}{0}}=e^{(\langle f,g\rangle-\langle g,h\rangle)/2}.
$$
By Equation~(\ref{eq:Heisenberg}), the right-hand side is
$$
\exp{\matElem{f^*}{-i\Phi(g)}{h^*}\over\bracket{f^*}{h^*}}={\matElem{f^*}{\Wick{W(g)}}{h^*}\over\bracket{f^*}{h^*}}
$$
by Corollary~\ref{cor:remarkable} applied to~$W(g)=e^{-i\Phi(g)}$.
\end{proof}

\chapter{$p$-form Electromagnetism as a Free Boson Field}

\label{chap:qed}

In this chapter we show how the oscillating modes of~$p$-form
electromagnetism in~$(p+1)$-dimensions have a free boson field
representation, define certain physically interesting observables as
quasioperators, and prove that suitable analogues of the classical
equations of motion hold as quasioperator equations.

According to Section~\ref{sec:freeBosonField}, in order to construct a
free boson field representation we need a complex Hilbert
space~$\Hilb$ consisting of classical observables of~$p$-form
electomagnetism. The same vector space with its real structure will be
denoted~$\Phase^*$ since the space of observables is the dual of the
physical phase space~$\Phase$. The complex inner
product~$\langle~,~\rangle$ on~$\Hilb$ must have as its imaginary part
the classical symplectic structure~$\omega$ on~$\Phase^*$. The free
boson field on~$\Hilb$ is the representation of~$\W(\Phase^*,\omega)$
produced by the GNS construction applied to a state~$\langle~\rangle$
with characteristic functional~$\mu(f)=\exp(-\|f\|^2/4)$ for
all~$f\in\Hilb$. Now, in order for time evolution to be unitary, it is
sufficient that~$\mu$ be invariant under time evolution; in other
words,~$\mu$ and hence~$\|~\|$ must be constants of the motion.

However, the analysis of the classical theory produces a real phase
space, without a complex structure and not having necessarily even a
real Hilbert space structure. That is, the starting point for
quantization is a classical phase space~$\Phase$ whose
dual~$(\Phase^*,\omega)$ is a a real topological vector space
with a continuous symplectic structure~$\omega$. Time evolution acts
on phase space as a strongly continuous one-parameter group of
\emph{bounded} operators~$T(t)$ preserving the symplectic structure
on~$\Phase^*$. To quantize these symplectic dynamics involves
constructing from~$(\Phase^*,\omega)$ and~$T$ a complex Hilbert
space~$\Hilb$ on which~$T(t)$ is a strongly-continuous one-parameter
group of \emph{unitary} operators. 

Ordinarily, for instance when quantizing a massive linear field such
as the Klein--Gordon field,~$\Hilb$ would carry a weaker norm
than~$\Phase^*$, and so~$\Phase^*$ would be contained
in~$\Hilb$. However, as we shall see, when there are infrared
divergences (as is the case for massless fields such as the Maxwell
field) neither~$\Hilb$ nor~$\Phase^*$ contain each other. However,
there is a common subspace of both~$\Phase^*$ and~$\Hilb$ on which all
the mathematical objects we are discussing are well-defined. This
space is constructed as a subspace of~$\Phase^*$ in a well-prescribed
way and then completed to obtain~$\Hilb$.

An additional complication is the existence of a nontrivial
Aharonov--Bohm sector. We have seen that the dynamics in this sector
are analogous to those of a free particle. In the case of the
electromagnetic field, we will see that the definition of~$\Hilb$
involves negative powers of the Laplacian, and so the Aharonov--Bohm
sector must be quantized in a different way, if at all.  Accordingly,
although we set out to quantize~$\Phase\simeq\Phase_o\oplus \Phase_f$,
we really only achieve a Fock quantization of the oscillating
sector~$\Phase_o$. We do not attempt to determine whether a free boson
field representation of the free modes is possible; we expect this to
be the case only when~$\Phase_f$ is finite-dimensional. Moreover, only
on~$\Phase_o$ is it possible to find a (densely-defined) complex
structure preserved by time evolution. On the space~$\Phase_f$ of
Aharonov--Bohm modes, the time evolution operator~$T_f(t)$ is a shear,
and there is no way to make it unitary.

The plan of this short chapter is as follows. In
Section~\ref{sec:freeBosonEM} we construct the free boson field
representation of the oscillating sector of~$p$-form
electromagnetism. In Section~\ref{sec:punchline} we use our
quasioperator technology from Section~\ref{sec:quasioperators} to make
sense of Wilson loop operators and their higher-dimensional
generalizations, as well as electromagnetic field operators at a
point, which are then shown to satisfy the Maxwell equations as
quasioperator equations. Most importantly, we end with a description
of the dynamics of the electromagnetic field in terms of Wilson loops,
without any need for `regularizing' or `smearing' these loops as in
the work of Varadarajan~\cite{varadarajan00,varadarajan01}.

\section{Free boson field representation}

\label{sec:freeBosonEM}

Our Result~\ref{thm:n+1phys} associated to~$p$-form
electromagnetism in~$n+1$ dimensions a real Hilbert space~$\Phase$
consisting of pairs~$X=[A]\oplus E$, where~$[A]$ is an equivalence
class of~$p$-forms modulo~$D_{p-1}$-exact~$p$-forms and~$E$ is
a~$p$-form such that~$D_{p-1}^*E=0$ (what one might call
\emph{twisted-divergenceless}). 
The symplectic structure
on~$\Phase$ was
$$ 
\omega(X,X')=(E,A')-(E',A).
$$
We need to make this~$\Phase$ into a
complex Hilbert space~$\Hilb$, and put a time-independent complex
inner product on it whose imaginary part is the symplectic
structure~$\omega$.  
This is equivalent to putting a real inner product~$h$ on~$\Phase$ which
is time-independent and satisfies
$$
h(X,X')=\omega(X,JX')
\qquad\hbox{for all}\quad
X,X'\in\Phase
$$
where~$J\from\Phase\to\Phase$ is a densely-defined complex structure,
that is, a real-linear map such that~$J^2=-1$ on a dense domain
of~$\Phase$. Now, because of the appearance below of inverse powers of
the twisted Laplacian~$L_p$, we will be forced to restrict our
attention to the oscillating sector~$\Phase_o$, and ignore the `free'
sector~$\Phase_f$ which was the intersection of~$\Phase$ with the
kernel of~$L_p$. Recall that the time evolution in~$\Phase_o$ is given
by
$$
T_o(t) 
\left( \! \begin{array}{c} A \\ E \end{array} \! \right) = 
\left( \begin{array}{cc} 
\cos(t\sqrt{L_p}) & \sin(t\sqrt{L_p})\,/\,\sqrt{L_p}\\ 
-\sqrt L_p\,\sin(t\sqrt{L_p}) & \cos(t\sqrt{L_p})  
\end{array} \right) 
\left( \! \begin{array}{c} A \\ E \end{array} \! \right)
$$
Defining multiplication by~$i$ by the action of~$J$, we can
make~$\Phase$ into a complex vector space. The completion of the dense
domain of~$K$ in~$\Phase$ with respect to the norm
$$
\|X\|^2=(E,L_p^{-1/2}E)+(A,L_p^{1/2}A')
$$
is the complex Hilbert space~$\Hilb$. The key facts about~$\Hilb$ are
summarized in the following theorem.

\begin{theorem}\label{thm:complex}
Let~$\e_o$ be a real Hilbert space with inner
product~$(~\mid~)$, let~$L$ be a nonnegative self-adjoint operator
on~$\e_o$ with vanishing kernel, and consider the real Hilbert space
$$
\A_o\colon=\{A\in\e_o\colon\|A\|^2+\|L^{1/2}A\|^2<\infty\}.
$$ 
Define time evolution on $\Phase_o=\A_o\oplus\e_o$ by 
$$
\partial_t(A\oplus E)=E\oplus -LA,
$$
which preserves the canonical symplectic structure on~$\A_o\oplus\e_o$,
namely
$$
\omega(A\oplus E,A'\oplus E')=(A\mid E')-(A'\mid E). 
$$
Then, there is a densely-defined complex structure~$J\colon\Y\to\Y$
given by~$J=-L^{-1/2}K$, or
$$
J(A\oplus E)\colon=-L^{-1/2}E\oplus L^{1/2}A,
$$
commuting with~$K$ and whose domain
$$
\Y\colon =\{A\oplus
E\in\Phase_o\colon\|A\|^2+\|L^{1/2}A\|^2+\|E\|^2+\|L^{-1/2}E\|^2<\infty\} 
$$
is dense in~$\Phase_o$, preserved by time evolution and satisfying
$$
\|Jx\|_\Y=\|x\|_\Y
\qquad\hbox{and}\quad
\omega(Jx,Jy)=\omega(x,y)
\qquad\hbox{for all}\quad x,y\in\Y.
$$ 
Finally, the completion of~$\Y$ with respect to the norm
$$
\|x\|_{\Hilb}^2\colon =\omega(x,Jy)
$$
is a complex Hilbert space~$\Hilb$ with inner product
$$
\langle x,y\rangle\colon =\omega(x,Jy)+i\omega(x,y)
$$
Time evolution defined on~$\Y$ then extends to a strongly-continuous
one-parameter group of unitary operators on~$\Hilb$, with nonnegative,
self-adjoint generator~$H=L^{1/2}$.
\end{theorem}

\begin{proof} 
First we need to show that~$\Y$ is dense in~$\Phase_o$. Since $\|A\oplus
E\|_\Y^2=\|A\oplus E\|_\Phase^2+\|L^{-1/2}E\|^2$, $\Y$ is dense
in~$\A_o\oplus\ran L^{1/2}$. To show that~$\Y$ is dense in~$\Phase_o$ we
need to show that~$\ran L^{1/2}$ is dense in~$\e_o$. Now, $L^{1/2}$ is
self-adjoint on~$\e$ and has vanishing kernel so, by
lemma~\ref{lem:ran_ker}, $\{\ran L^{1/2}\}^\perp=\ker L^{1/2}=\{0\}$. But
this implies that~$\ran L^{1/2}$ is dense in~$\e_o$.

Next, we need to show that $\Y$ is preserved by the time evolution of
equation~(\ref{eq:time_evolution_twisted}) or, equivalently, that
$\|T(t)\|_{\Y}<\infty$ for all~$t$. It is not hard to check that, in
fact, $\|T(t)(A\oplus E)\|_\Y=\|A\oplus E\|_\Y$ for all~$t$. An even
easier calculation shows that $\|J\|_\Y=1$, so~$J$ maps~$\Y$ to
itself.

Then, we need to show that~$J$ is compatible with~$\omega$. With the
analytical subtleties out of the way, it requires only straightforward
algebraic calculations to check that
\begin{itemize}
\item{1)}$\omega(Jx,Jy)=\omega(x,y)$ for all $x,y\in\Y$
\item{2)}$\omega\bigl(A\oplus E,J(A\oplus
E)\bigr)=\|L^{1/4}A\|^2+\|L^{-1/4}E\|^2\ge 0$.
\end{itemize}
Also, $\|x\|_\Hilb^2=\omega(x,Jx)$ is clearly a Hilbert-space norm.

Another simple calculation shows that $\|T(t)(A\oplus
E)\|_\Hilb=\|A\oplus E\|_\Hilb$, so $T(t)$ is a one-parameter unitary
group. Strong continuity is also easily checked. Finally, the
self-adjoint generator of time-evolution is determined by the
condition~$J\partial_t(A\oplus E)=H(A\oplus E)$, that is, $JK=H$. It
is also a straightforward algebraic calculation to check that~$\langle
A\oplus E,H(A\oplus E)\rangle=(A|LA)+(E|E)$, which is
nonnegative. Since time evolution is unitary with respect
to~$\langle~,~\rangle$, it follows that~$H$ is also self-adjoint.
\end{proof}

We can now apply Theorem~\ref{thm:FreeBosField} to the complex dual
of~$\Hilb$ (denoted~$\Hilb^\dagger$) to obtain the free boson field
over~$\Hilb^\dagger$. Note that because of the mis-match
between~$\Hilb$ in Theorem~\ref{thm:FreeBosField} and~$\Hilb^\dagger$
now, there is a sign difference in the definition of the generator of
time evolution, which was~$U(t)=e^{-itA}$ then and is~$T_o(t)=e^{JtH}$
now.

Although this construction seems natural enough, one might worry that
there may be more than one complex structure with the given
properties, but in fact it is unique, as asserted in the following
theorem.

\begin{theorem}
Let~$T_o(t)$ be a one-parameter group of symplectic transformations on
the linear symplectic space~$(\Phase,\omega)$. Then there is at most one
complex structure~$J$ on~$\Phase$ which is invariant, positive, symplectic
and such that the self-adjoint generator~$H$ of~$T_o(t)$ in the
completion of~$\Phase$ as a complex Hilbert space,~$\Hilb$, is nonnegative and with
vanishing kernel.
\end{theorem}

\begin{proof}

The self-adjoint generator~$H$ commutes both with the complex
structure~$J$ and with each element~$T_o(t)$ of the unitary
group. Hence, the spectral projections (see~\cite[Section VIII.3]{RS})
associated to~$H$ also commute with them. We can use these spectral
projections to restrict the problem to the subspaces~$\Phase_n$
of~$\Phase$
where~$H\ge 1/n$. The hypothesis of the theorem hold, but now the
self-adjoint generator~$H$ is strictly positive (that is, bounded below by a
positive constant). That uniqueness holds in this case is proved
in~\cite[Scholium 3.3]{BSZ}. 
\end{proof}

At this point, we redefine the meaning of~$\Phase$. It is clear that,
while mathematically convenient at the classical level, the real
Hilbert space structure of Theorems~\ref{thm:N+1}
and~\ref{thm:n+1phys} is really not the right one for Fock
quantization, which is the one given in Theorem~\ref{thm:complex}. We
now give concrete electromagnetic counterparts for all the objects
appearing in the development of the abstract free boson field
representation of Section~\ref{sec:freeBosonField}.

\begin{itemize}
\item The classical phase space~$\Phase$ consists of pairs of the
form~$X=[A]\oplus E$ such that~$h(X,X)<\infty$. It has a continuous
symplectic structure
$$
\omega(X,X')=(E,A')-(E',A)
$$
and a continuous complex structure $J\from\Phase\to\Phase$ given by
$$
J([A]\oplus E)=L_p^{-1/2}E\oplus(-L_p^{1/2}A).
$$
We denote~$\Phase$ by~$\Hilb$ when we want to view it as a complex
Hilbert space. Multiplication by~$i$ in~$\Hilb$ corresponds to the
action of~$J$ on~$\Phase$.
\item Real observables $F\in\Phase^*$ are associated to phase space points~$F^*=[Q]\oplus
J\in\Phase$ where, if~$X=[A]\oplus E$,
$$
F(X)=(J,A)-(Q,E)=\omega(F^*,X)
$$
Each such real observable defines a complex-linear
observable~$iF+FJ\in\Hilb^\dagger$. We have
$$
(iF+FJ)(X)=\langle F^*,X\rangle=h(F^*,X)+i\omega(F^*,X).
$$
This is consistent with the symplectic structure on~$\Phase^*$
$$
\omega(F,G)=-\omega(F^*,G^*)
\qquad\hbox{for all}\quad
F,G\in\Phase^*.
$$
It is customary to refer to observables primarily by~$F^*=[Q]\oplus
J$. 
\item The free boson field representation of~$\Hilb^\dagger$---the
complex dual of~$\Hilb$---has
characteristic functional
$$
\mu(F)=e^{-{1\over 4}[(Q,L_p^{1/2}Q)+(J,L_p^{-1/2}J)]}.
$$
\item The coherent states of the electromagnetic field are of the form
$$
\ket{[Q]\oplus J}
\qquad\hbox{with}\quad
[Q]\oplus J\in\Phase. 
$$
The inner product of two coherent states is
$$
\bracket{[Q]\oplus J}{[Q']\oplus J'}=e^{[(J,Q')-(J',Q)]/2i}e^{-{1\over 4}[(Q-Q',L_p^{1/2}(Q-Q')+(J-J',L_p^{-1/2}(J-J'))]}.
$$
\item The Weyl operator~$W(F)$, where~$F^*=[Q]\oplus J$, is defined by the following
action on the coherent states:
$$
W(F)\ket{[Q']\oplus J'}=e^{[(J',Q)-(J,Q')]/2i}\ket{[Q+Q']\oplus(J+J')}.
$$
The Heisenberg operator~$\Phi(F)$
satisfying~$W(F)=e^{-i\Phi(F)}$ has diagonal matrix elements on
coherent states given by
$$
\matElem{[Q]\oplus J}{\Phi(F)}{[Q]\oplus J}=-\omega([Q']\oplus
J',[Q]\oplus J)=(Q',J)-(Q,J')
$$
where~$F^*=[Q']\oplus J'$. In other words, the interpretation of~$F$
in~$\Phi(F)$ and in~$\ket{F}$ is very different: since the coherent
state~$\ket{F}$ is a semiclassical state of the quantum theory which
is peaked about the value~$F$ of the field configuration, it follows
that~$\Phi(F)$ does not represent the quantization of the
observable~$F$, but of~$JF$. We will see this in more detail in the
next section.
\item Time evolution is handled as follows. We have
$$
T_o(t)=e^{tJL^{1/2}}\from\Phase\to\Phase
$$
on the phase space. The time evolution of the observables is
$$
U(t)=e^{-tJL^{1/2}}\from\Phase^*\to\Phase^*.
$$
Then,~$\Gamma\bigl(U(t)\bigr)\from\K\to\K$ is defined by extending the
following action on the coherent states:
$$
\Gamma\bigl(U(t)\bigr)\ket{[Q]\oplus J}=\ket{T_o(t)([Q]\oplus F)},
$$
and for all~$F\in\Phase^*$ the equation
$$
\Gamma\bigl(U(t)\bigr)W(F)\Gamma\bigl(U(-t)\bigr)=W\bigl(U(t)F\bigr).
$$
\end{itemize}

\section{Field quasioperators}

\label{sec:punchline}

First we try to define~$\hat A$ as an operator-valued $p$-form on~$S$
or, equivalently, an operator with matrix elements valued
in~$\Omega^p(S)$. It turns out~$\hat A$ exists as a quasioperator, and
we construct it as follows. First,~$\widehat{A(x)}$ can be defined for
all~$x\in S$ as a quasioperator by directly quantizing the classical
observable~$A(x)$. Then~$\hat A$ is defined so that~$\hat
A(x)=\widehat{A(x)}$ for all~$x\in S$. This technique is also used to
define~$\hat B(x)$ and~$\hat E(x)$, and the upshot is that, almost by
definition, the formulas
$$
\hat B(x)=\dd\hat A(x)
\qquad\hbox{and}\quad
\oint_\gamma\hat A=\widehat{\textstyle\oint_\gamma A}
$$
hold as equations between quasioperators. Because in the physics
literature one does not distinguish between~$\Phase$ and~$\Phase^*$,
and it would be extremely awkward to use notations such
as~$\delta_x^*$, we identify them by means of using the duality
map~$*$ related to the symplectic structure~$\omega$.

When the gauge group is~$U(1)$, the proper holonomy is not
$\oint_\gamma A$ but the exponentiated version~$e^{i\oint_\gamma A}\in
U(1)$. However, because~$\gamma$ is a curve and~$A$ is
square-integrable, we know that the na\"\i{}ve candidate
for~$e^{i\oint_\gamma\hat A}$ has vanishing matrix elements between
any two coherent states, which is a problem. However, the
normal-ordered version of this exponentiated hlonomy exists as a
nonzero quasioperator on~$\K$ with domain containing the span~$\K_0$
of the smooth coherent states~$\ket{X}$ where~$X=[Q]\oplus J$ is not
only in~$\Phase$, but it is also infinitely differentiable. We will
denote the space of~$C^\infty$ elements of~$\Phase$---called
\emph{smooth field configurations}---by~$\Phase_0$.

\subsection{Quantizing the classical fields}

The classical observable~$A(x)$ is the densely-defined linear
functional on~$\Phase$ given by
$$
X=[A]\oplus E\mapsto A(x)
$$
In fact,~$A\in\Phase$ is in the domain of this observable as long
as~$A$ is continuous. Since a more convenient sufficient condition is
that~$A$ be infinitely differentiable, we give the following definition.

\begin{definition}[smooth coherent states]
Let~$\Phase$ be the oscillating phase space of $p$-form
electromagnetism, and let~$\K$ be the associated Fock space. We say
that~$X=[A]\oplus E\in\Phase$ is a \emph{smooth field configuration}, and
write~$X\in\Phase_0$, if~$[A]$ and~$E$ are
infinitely-differentiable. A coherent state~$\ket{X}$
with~$X\in\Phase_0$ is called a \emph{smooth coherent state}. We denote
by~$\K_0$ the span of the smooth coherent states.
\end{definition} 

\begin{proof}[Note]
The space~$\Phase_0$ is a domain of essential self-adjointness of the
Laplacian~$L_p$ inside~$\Phase$, and is therefore dense. Hence,~$\K_0$
is also dense in~$\K$.
\end{proof}

Note that the observable~$A(x)$ takes values in~$\Lambda^pT^*_xS$. We
get a real-valued observable by contracting it with a multivector~$v_x\in \Lambda^pT_xS$. We denote this contraction by~$A_v(x)$. The
quantum observable~$\hat A_v(x)$ should be a Heisenberg
operator~$\Phi(F)$ such that
$$
\matElem{X}{\Phi(F)}{X}=(v\delta_x,A)=\omega(F^*,X)=F(X),
$$
where~$v\delta_x$ is the distributional~$p$-form defined by the
equation~$A_v(x)=(v\delta_x,A)$ for all smooth~$A$. In other words,
since~$A(x)=\omega(0\oplus v\delta_x,A\oplus E)$, one should
define
$$
\widehat{A_v(x)}\sim\Phi(0\oplus v\delta_x),
$$
as a quasioperator.

Now, it follows from Equation~\ref{eq:Heisenberg} that
$$
{\matElem{X'}{\widehat{A_v(x)}}{X}\over\bracket{X'}{X}}={A_v(x)+A_v'(x)\over 2}+iL_p^{-1/2}{E'_v(x)-E_v(x)\over 2}
$$
since~$h(0\oplus v\delta_x,A\oplus E)=(L_p^{-1/2}E)_v(x)$. Hence, defining
a quasioperator-valued $p$-form~$\hat A$ by
\begin{equation}\label{eq:AMatElem}
{\matElem{X'}{\hat A}{X}\over\bracket{X'}{X}}={A+A'\over 2}+iL_p^{-1/2}{E'-E\over 2},
\end{equation}
one has
$$
\hat A_v(x)=\widehat{A_v(x)}
\qquad\hbox{for all}\quad v_x\in T_xS
$$
as an equation between quasioperators. 

In a entirely analogous manner, one can quantize the electric
field. Indeed,~$\hat E_v(x)$ is the quantum counterpart of
$$
A\oplus E\mapsto E(x)=(v\delta_x,E),
$$
with
$$
E(x)=-\omega(v\delta_x\oplus 0,A\oplus E)
\qquad\hbox{and}\quad
h(v\delta_x\oplus 0,A\oplus E)=(L_p^{1/2}A)_v(x).
$$
This means that
$$
\widehat{E_v(x)}\sim-\Phi(v\delta_x\oplus 0)
$$
and, demanding~$\hat E_v(x)=\widehat{E_v(x)}$,
$$
{\matElem{X'}{\hat E}{X}\over\bracket{X'}{X}}={E+E'\over 2}+iL_p^{1/2}{A-A'\over 2}.
$$


In the same way one can derive
$$
\widehat{(L_p A)_v(x)}=(L_p\hat A)_v(x)
$$
as quasioperators with domain~$\K_0$.

\subsection{Wilson surfaces as quasioperators}

Now that~$\hat A$ is defined as a~$p$-form (albeit
quasioperator-valued), we can define its integral on a
compact, oriented~$p$-dimensional submanifold~$\gamma$ of space in
such a way that
$$
\oint_\gamma\hat A=\widehat{\textstyle\oint_\gamma A}.
$$
as a quasioperator equation. In the~$p=1$ case, these observables are
called Wilson loops in physics. We call them \emph{Wilson surfaces} in
general.

We do this by 
observing that the
classical observable~$\oint_\gamma\hat A$ can be written as
$$
A\oplus E\mapsto\oint_\gamma A=(\Gamma_\gamma,A)
$$
where~$\Gamma$ is a distributional~$p$-form analogous to Dirac's
delta, uniquely defined by this equation and satisfying
$$
\oint_\gamma A=\omega(0\oplus\Gamma_\gamma,A\oplus E)
\qquad\hbox{and}\quad
h(0\oplus\Gamma_\gamma,A\oplus E)=\oint_\gamma L_p^{-1/2}E.
$$
So, letting
$$
\widehat{\textstyle\oint_\gamma A}\sim\Phi(0\oplus\Gamma_\gamma)
$$
it follows that
showing that
$$
{\matElem{X'}{\textstyle\oint_\gamma\hat
A}{X}\over\bracket{X'}{X}}=\oint_\gamma{\matElem{X'}{\hat
A}{X}\over\bracket{X'}{X}}.
$$
so
$$
\widehat{\textstyle\oint_\gamma A}=\oint_\gamma\hat A,
$$
as a quasioperator equation on~$\K_0$. 
In particular, we find that
$$
\matElem{X}{{\textstyle\oint_\gamma\hat
A}}{X}=\oint_\gamma A,
$$
when~$X=[A]\oplus E$ is a smooth field configuration. In fact, this
follows from the less
obvious expression
$$
{\matElem{X'}{{\textstyle\oint_\gamma\hat A}}{X}\over\langle X'\mid X\rangle}=\oint_\gamma\biggl({A+A'\over
    2}\biggr)+i\oint_\gamma{1\over\sqrt{L_p}}\biggl({E-E'\over
    2}\biggr)
$$
when~$X,X'$ are smooth field configurations, which is an easy
consequence of Equation~(\ref{eq:AMatElem}).

Finally, in order to extend our work to the case of a $U(1)$
connection which is much more common in the physics literature, we
would need to define the operator
$$
e^{i\oint_\gamma \hat A}
$$
which quantizes the holonomy
$$
e^{i\oint_\gamma A}\in U(1).
$$
As we know, there is a serious problem coming from the fact that~$\oint_\gamma\hat
A\sim\Phi(0\oplus\Gamma_\gamma)$,
and~$\|0\oplus\Gamma_\gamma\|=\infty$. However, we have shown that the
normal-ordered
$$
\Wick{e^{i\oint_\gamma\hat A}}=\Wick{W(0\oplus\Gamma_\gamma)}
$$ 
does exist as a quasioperator on~$\K$ with domain~$\K_0$. In fact, 
$$
{\matElem{X'}{e^{i\oint_\gamma\hat
A}}{X}\over\bracket{X'}{X}}=\exp{i\matElem{X'}{\oint_\gamma\hat A}{X}\over\bracket{X'}{X}}
$$
whenever~$X,X'$ are smooth coherent states. 

\subsection{The vacuum Maxwell equations}

We are now ready to show that the field quasioperators that we have
defined satisfy the vacuum Maxwell
equations in the following sense:
\begin{theorem}
Let~$\ket{X(t)}=\Gamma\bigl(U(t)\bigr)\ket{X}$ for all~$X\in\Phase$. Then,
\begin{eqnarray*}
{\partial\over\partial t}\matElem{X'(t)}{\hat
A}{X(t)}
&=&
\matElem{X'(t)}{\hat E}{X(t)}\\
{\partial\over\partial t}\matElem{X'(t)}{\hat
E(x)}{X(t)}
&=&
-\matElem{X'(t)}{L_p\hat A}{X(t)}\\
\end{eqnarray*}
\end{theorem}

\begin{proof}
First, recall that
$$
\Gamma(U(t))\ket X=\ket{T_o(t)X},
$$
so that
$$
\ket{X(t)}=\ket{T_o(t)X}.
$$
Therefore, if~$X=[A]\oplus E$, we have~$\ket{X(t)}=\ket{[A](t)\oplus
E(t)]}$, where~$[A](t)$ and~$E(t)$ are the solutions of the classical Maxwell
equations with initial data~$[A]\oplus E$. 

Now, from the known
expression for the matrix elements of~$\hat A$
$$
{\matElem{X'(t)}{\hat A}{X(t)}\over\bracket{X'(t)}{X(t)}}={A(t)+A'(t)\over
2}+{i\over\sqrt{L_p}}{E(t)-E'(t)\over 2}.
$$
On the left-hand side,~$\hat A$ and~$\hat E$ are~$p$-form-valued
operators on~$\K$ (quantum observables), while on the right-hand side
we have the classical solutions of the Maxwell equations evaluated at
time~$t$. Since the quantities on the right-hand side satisfy the
Maxwell equations, and~$\bracket{X'(t)}{X(t)}$ is independent of~$t$
because~$\Gamma(U(t))$ is unitary, we have
$$
{1\over\bracket{X'(t)}{X(t)}}{\partial\over\partial t}\matElem{X'(t)}{\hat A(x)}{X(t)}={E+E'\over
2}-i\sqrt{L_p}\Bigl({A-A'\over 2}\Bigr),
$$
but the right-hand side is precisely~$\matElem{X'(t)}{\hat
E(x)}{X(t)}\over\bracket{X'(t)}{X(t)}$.

Similarly, 
$$
{\matElem{X'(t)}{\hat E(x)}{X(t)}\over\bracket{X'(t)}{X(t)}}={E+E'\over
2}-i\sqrt{L_p}\Bigl({A-A'\over 2}\Bigr)
$$
implies that
\begin{eqnarray*}
{1\over\bracket{X'(t)}{X(t)}}{\partial\over\partial t}\matElem{X'(t)}{\hat E(x)}{X(t)}
&=&
-L_p\Bigl({E+E'\over
2}\Bigr)-i\sqrt{L_p}\Bigl({A-A'\over 2}\Bigr)\\
&=&
-L_p\Bigl({A+A'\over
2}+{i\over\sqrt{L_p}}{E-E'\over 2}\Bigr),
\end{eqnarray*}
and the result follows.
\end{proof}

The calculations involved in the proof of this fact are deceptively
simple. The point is that these would be purely formal had we not
developed a framework where objects such as~$\hat A(x)$ are
well-defined. All the hard work is hidden in
Chapter~\ref{chap:linear}.

Finally, here is the promised formula for the time evolution of
electromagnetism in terms of Wilson loop quasioperators:

\begin{corollary}
$$
{\partial\over\partial t}{\matElem{X'}{e^{i\oint_\gamma\hat
A}}{X}\over\bracket{X'}{X}}={i\matElem{X'}{{\textstyle \oint_\gamma\hat E}}{X}\over\langle X'\mid X\rangle}\exp{i\matElem{X'}{{\textstyle \oint_\gamma\hat A}}{X}\over\langle X'\mid X\rangle}.
$$
\end{corollary}

\begin{proof}
Differentiating
$$
{\matElem{X'}{e^{i\oint_\gamma\hat
A}}{X}\over\bracket{X'}{X}}=\exp{i\matElem{X'}{{\textstyle \oint_\gamma\hat A}}{X}\over\langle X'\mid X\rangle}
$$
we get
$$
{\partial\over\partial t}{\matElem{X'}{e^{i\oint_\gamma\hat
A}}{X}\over\bracket{X'}{X}}={i\matElem{X'}{{\textstyle \oint_\gamma\hat E}}{X}\over\langle X'\mid X\rangle}\exp{i\matElem{X'}{{\textstyle \oint_\gamma\hat A}}{X}\over\langle X'\mid X\rangle}.
$$
\end{proof}


\ssp
\bibliographystyle{alpha}
\bibliography{dissertation}
\dsp


\end{document}